\documentclass[11pt,a4paper]{ltxdoc}
\usepackage[italian,english]{babel}
\usepackage[ansinew]{inputenc}
\usepackage[T1]{fontenc}
\usepackage[svgnames]{xcolor}
\usepackage{graphicx}
\usepackage{polski}
\usepackage{amsmath,eqnarray}
\usepackage{amssymb}
\usepackage{rotating}
\usepackage{placeins}
\usepackage{physics}
\usepackage[left=2.5cm,top=2.5cm,right=2.5cm,bottom=2.5cm]{geometry}
\usepackage[labelformat=simple]{subfig}

\usepackage{setspace}
\usepackage{hyperref}
\usepackage{lscape}
\providecommand{\keywords}[1]
{
    \small
    \textbf{Keywords:} #1
}
\hypersetup{%
    pdfauthor={Gustavo Cevolani},%
    pdftitle={Towards quantization Conway Game of Life},%
    pdfsubject={LaTeX},%
    bookmarksnumbered=true,%
    bookmarksopen=true,%
    bookmarksopenlevel=0,%
    urlcolor=NavyBlue,%
    linkcolor=NavyBlue,%
    citecolor=black,%
    colorlinks=true,%
    pdfhighlight=/P,%
    pdfstartview=FitH,%
    plainpages=false,
}
\EnableCrossrefs
\CodelineIndex
\RecordChanges

\begin{document}

    \title{Towards quantization Conway Game of Life}
    \author{Krzysztof Pomorski \\Faculty of Electrical and Computer Engineering \\ Cracow University of Technolog (Cracow)\\E-mail: kdvpomorski@gmail.com \and%
    Dariusz Kotula \\Faculty of Computer Science and Telecommunication \\ Cracow University of Technology (Cracow) \\E-mail: dariusz.kotula@student.pk.edu.pl}
    \maketitle

    \begin{abstract}
    Classical stochastic Conway Game of Life is expressed by the dissipative Schr\"{o}dinger equation and dissipative tight-binding model. This is conducted at the prize of usage of time dependent anomalous non-Hermitian Hamiltonians as with occurrence of complex value potential that do not preserve the normalization of wave-function and thus allows for mimicking creationism or annihilationism of cellular automaton. Simply saying time-dependent complex value eigenenergies are similar to complex values of resonant frequencies in electromagnetic resonant cavities reflecting presence of dissipation that reflects energy leaving the system or being pumped into the system. At the same time various aspects of thermodynamics were observed in cellular automata that can be later reformulated by quantum mechanical pictures. The usage of Shannon entropy and mass equivalence to energy points definition of cellular automata temperature. Contrary to intuitive statement the system dynamical equilibrium is always reflected by negative temperatures. Diffusion of mass, energy and temperature as well as phase of proposed wave function is reported and can be directly linked with second
    thermodynamics law approximately valid for the system, where neither mass nor energy is conserved. The concept of complex-valued mass mimics wave-function behavior. Equivalence an anomalous second Fick law and dissipative Schr\"{o}dinger equation is given. Dissipative Conway Game of Life tight-binding Hamiltonian is given using phenomenological justification.
    \end{abstract}
    \newpage
    \tableofcontents
    \newpage

    \section{Stochastic Conway Game of Life}
    \label{sec:stochastic-conway-game-of-life}

    \subsection{Introduction to Classical Conway's Game of Life (CCGoL)}
    \label{subsec:introduction-to-classical-conway's-game-of-life-(ccgol)}
\begin{figure}
        \centering
        \includegraphics[height=.3 \linewidth]{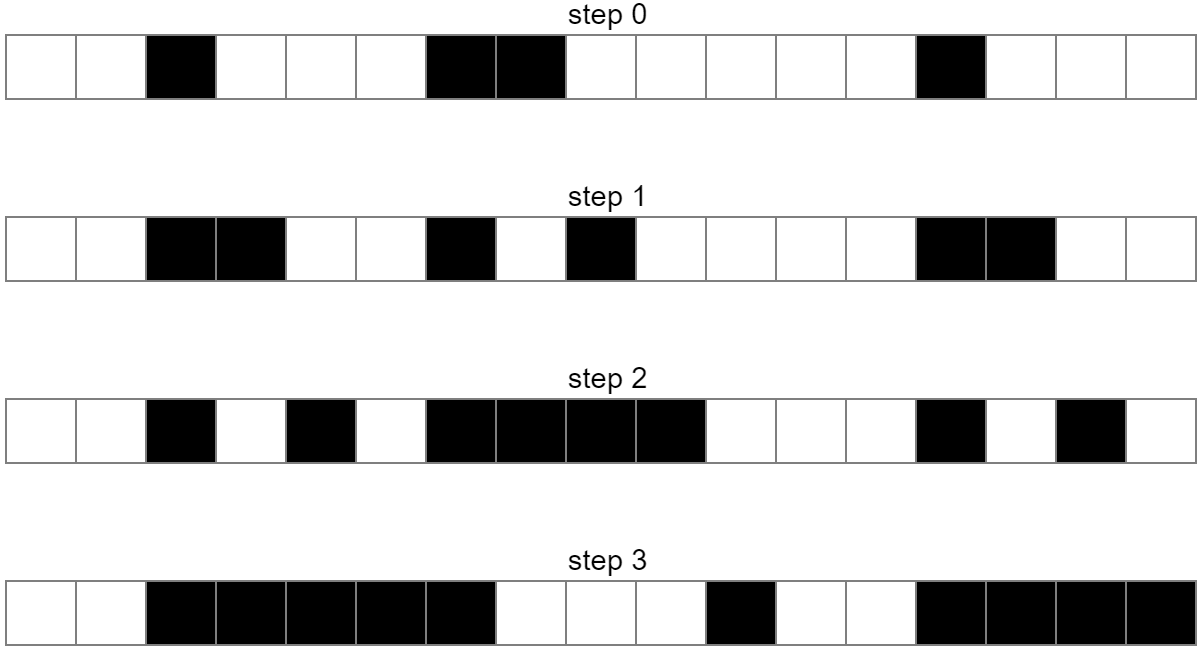}
        \caption{Evolution of a one-dimensional cellular automaton in successive cycles with left side partial logical negation rule (if the state of nearest left cell is alive, then the state of a given cellular automaton will change to its opposite).}
        \label{fig:1D_automaton}
    \end{figure}

    Sharp logic circuit can be expressed by deterministic finite state machine.
    There is common believe that deterministic Conway Game of Life can be equivalent to any logical system with properly taken boundary condition for cellular automaton at the beginning of simulation and by proper measurement of system state after certain simulation time.
    We follow the reasoning described in~\cite{MDPIKotula}.
    A cellular automaton~\cite{cite15} is a system consisting of cells with sharp 0 or 1 values arranged most often on a one-, two-, or three-dimensional regular lattice. 
    The dynamics of the model depends on the definition of individual cell states and the rules of transitions between them~\cite{cite4}.
    The transition table can be deterministic or based on probabilistic rules.
    One of the simplest examples is a one-dimensional cellular automaton.
    Suppose that the cells placed on the lattice can be in one of two states, which are marked with white (default assigned to dead state or logical zero) or black color (default assigned to alive state or logical one).
    We define a rule that if a given cell is black, then the cell to the right of it will change its state.
    This situation is depicted in a Figure~\ref{fig:1D_automaton}.
    \FloatBarrier
    We can spot how the system changes in the subsequent steps of the simulation.
    The parameter determining the change of the cell state is the state of the left neighbor of a given cell.
    There are many other possible parameters to choose from, e.g., the condition of the state of both neighboring cells or having nearest neighbors with opposite states.
    In order to determine the system dynamics, we must have defined the initial cellular automaton states (information about the initial system dynamical state) and a specific set of deterministic or probabilistic rules.

\begin{figure}
        \centering
        \includegraphics[height=0.6 \linewidth]{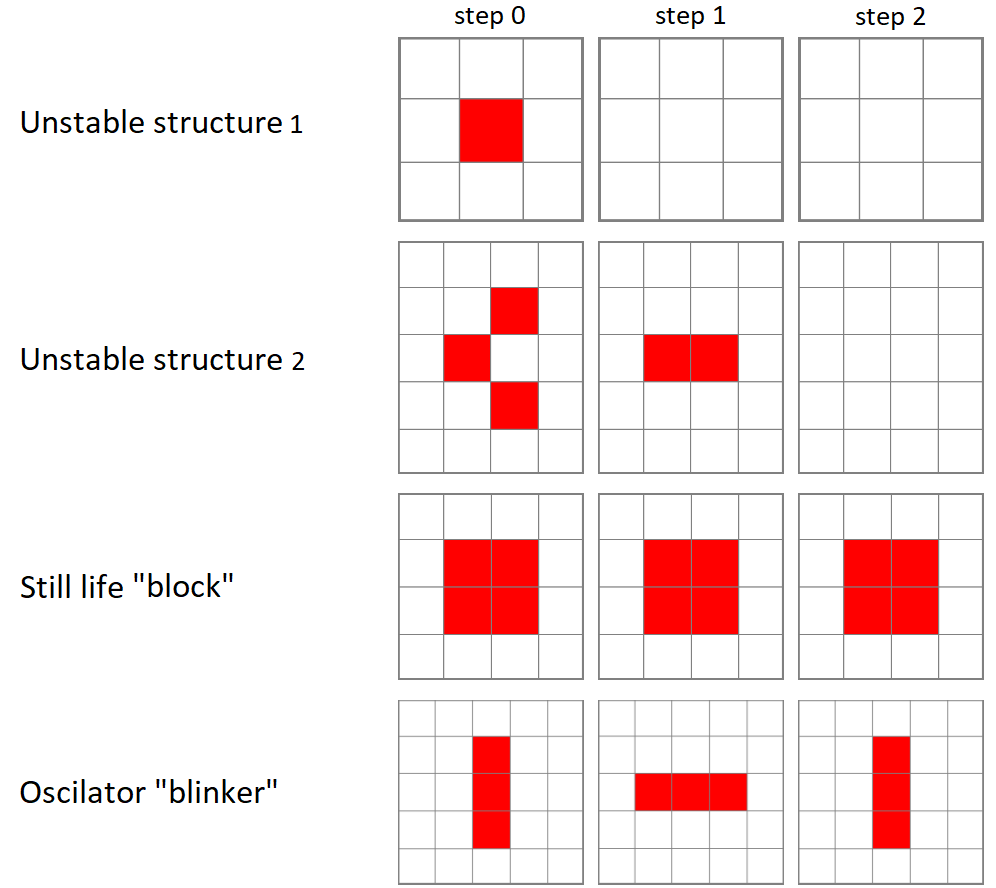}
        \caption{Evolution of various topologies of cellular automaton structures over time with deterministic rules of CCGoL. Two dynamically unstable structures and two structures can be identified that have dynamical stability over time.}
        \label{fig:structures}
    \end{figure}

    Conway's Game of Life~\cite{cite1,cite16} is an example of a cellular automaton with deterministic rules.
    The cellular automaton system consists of cells located on a two-dimensional lattice, which can be in one of two states: alive or dead.
    The rules specify the required number of neighbors and cell states that are taken into account to determine their states in the next cycle (time index).
    Given the nearest neighborhood of a given cell expressed through the state of 8 closest cells, we can write the \mbox{3 main rules} of the Classical Conway's Game of Life (CCGoL) as follows:
    \begin{enumerate}
        \item If a dead cell has exactly 3 neighbors, it comes alive in the next cycle.
        \item If a living cell has 2 or 3 neighbors, it survives in the next cycle.
        \item If a cell has a different number of neighbors than stated above, it will be dead
        in the next cycle.
    \end{enumerate}
    The rules defined in this way allow for the generation of various types of structure topologies with automaton states set to 1, as shown in Figure~\ref{fig:structures}.
    The most common type of structures are ``unstable structures'', which change in successive cycles but do not return to their initial state.
    A single cell cannot survive on the lattice because it has fewer than 2 neighbors.
    A dead cell surrounded by live cells cannot come to life because it has a number of neighbors different from 3.
    If the simulation is continued for a sufficiently long time, structures usually remain on the lattice that are unchanging over time - ``still lifes'' (an example is the ``block'' shown in Figure~\ref{fig:structures}) - or change over time in a periodic way, so they return to their original shape after $k$ cycles - ``oscillators'' (an example is the ``blinker'' shown in Figure~\ref{fig:structures}).

\begin{figure}
        \centering
        \subfloat[\centering]{{\includegraphics[height=0.15 \linewidth]{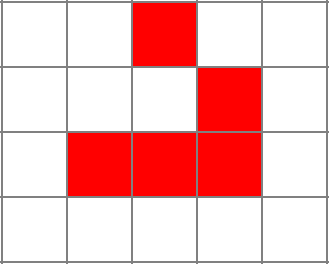} }}
        \subfloat[\centering]{{\includegraphics[height=0.15 \linewidth]{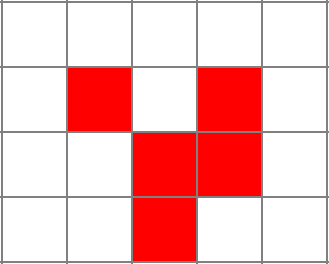} }}
        \subfloat[\centering]{{\includegraphics[height=0.15 \linewidth]{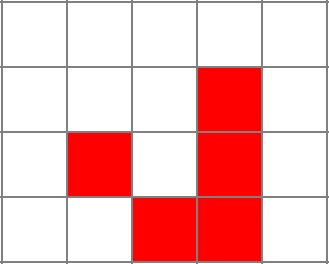} }}
        \subfloat[\centering]{{\includegraphics[height=0.15 \linewidth]{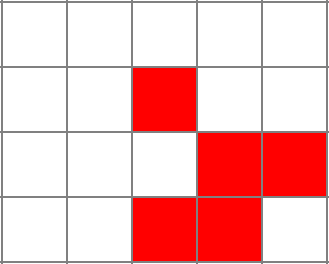} }}
        \subfloat[\centering]{{\includegraphics[height=0.15 \linewidth]{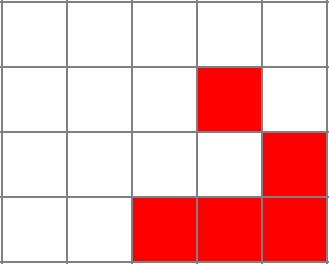} }}
        \caption{Evolution of the ``glider'' configuration (\textbf{a}--\textbf{e}) of cellular automata propagating over time in deterministic CCGoL.}
        \label{fig:glider}
    \end{figure}

\begin{figure}
        \centering
        \includegraphics[height=.28 \linewidth]{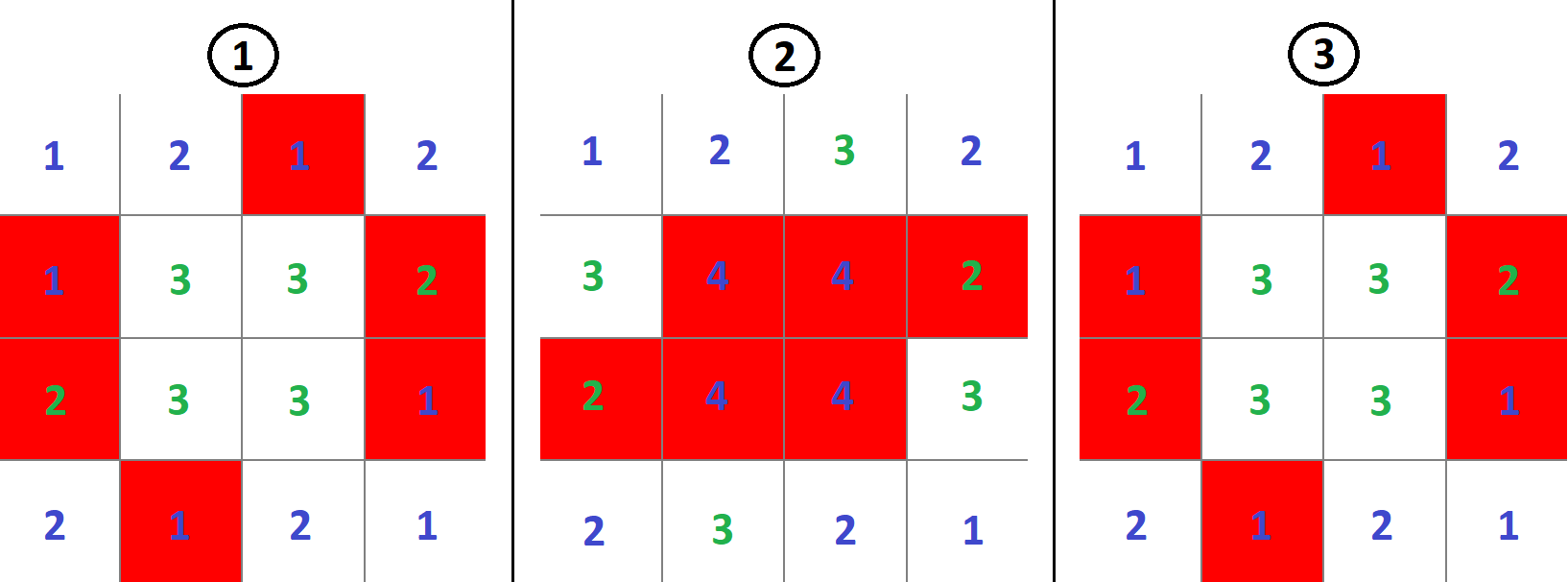}
        \caption{Evolution of the ``toad'' configuration of cellular automaton acting as an oscillator in successive cycles in CCGoL, where numbers correspond to number of living neighbors of the given cell. The blue color of digits means that in the next cycle the given cell will be dead, while in case of the green ones, that in the next cycle the given cell will be alive.}
        \label{fig:oscilator}
    \end{figure}

    There are also structures that move in a certain direction named ``gliders'', as depicted in Figure~\ref{fig:glider}.
    In a very real way, these gliders behave in accordance with Newton's first law of dynamics, preserving momentum (speed and direction of propagation in this case).
    We can also identify ``blinkers'', ``star ships'', and objects which periodically eject \mbox{``gliders'' - ``guns''~\cite{cite5}}.
    Figure~\ref{fig:oscilator} shows one of the many oscillators in Conway's Game of Life during successive iterations of the system simulation.
    The ``toad'' oscillator has a period of 2, which means that it switches continuously between two different fixed configurations.
    Each 2-dimensional discrete lattice field has a specified number that indicates the number of neighbors of the given cell.
    If the number is green, the cell will be alive in the next cycle.
    If the number is blue, the cell will be dead in the next cycle.
    It shall be underlined that conservation of momentum and angular momentum takes place in rare situation in Conway Game of Life and in most cases it is not preserved.

    \FloatBarrier

    \subsection{Introduction to Stochastic Classical Conway's Game of Life (SCCGoL)}
    \label{subsec:introduction-to-stochastic-classical-conway's-game-of-life-(sccgol)}
\begin{figure}
        \centering
        \includegraphics[width=0.7\linewidth]{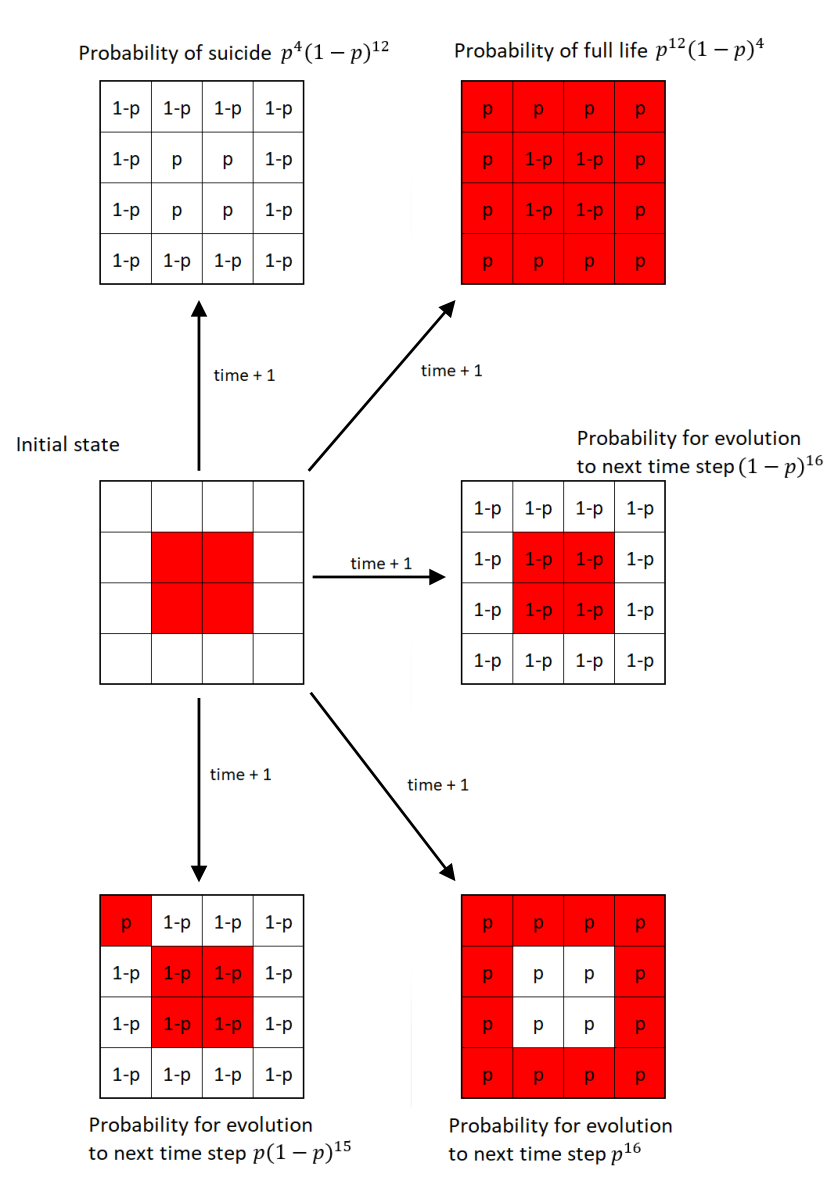}
        \caption{Possible scenarios during one time step in Stochastic Classical Conway's Game of Life (D/C) with Discrete $\{0,1\}$ or Continuous lattice point values~\cite{MDPIKotula}.}
        \label{fig:scenario}
    \end{figure}

    The modification of the fully deterministic Classical Conway's Game of Life by adding probability to this simulator can be done in a variety of ways~\cite{cite17,cite18}.
    The Stochastic Classical Conway's Game of Life (SCCGoL) was developed from the Deterministic Classical Conway's Game of Life by adding the probability of spontaneous change to the initially deterministic rules describing the states of cells.
    With a given prefixed spontaneous probability value \emph{p}, the state of the cell can change regardless of the number of neighbors resulting only in 0 or 1 (Stochastic Conway's Game of Life in Discrete mode = \textbf{SCCGoL(D)}).
    One shall consider all possible scenarios (cell configurations) or subsets of it characterized by a given probability that might take place during the next time step as depicted in Figure~\ref{fig:scenario}.
    This is somewhat similar to the case of Quantum Mechanics, where one evolves from one point in space to another point in space over a large class of trajectories formally recognized as path integral approach.
    Instead of discrete values of 0 and 1, we introduce cell states that have continuous values between 0 and 1, which are called mass that will be later assigned to Stochastic Conway's Game of Life in Continuous mode = \textbf{SCCGoL(C)}.
    Due to the fact that SCCGoLs have different rules from CCGoLs, cells almost never have exactly two or three neighbors.
    A condition for a given cell to come to life from a dead cell state (creationism of a live cell) is that it has a number of neighbors in a certain range of values.
    Similar rules apply to a living cell, justifying its live or dead state in the next time iteration.
    By setting standard intervals of allowed/forbidden numbers of neighboring values, in which the cell is alive/dead, and by adding the additional spontaneous rule probability for the cell to change in the next iteration (probability of changing the state of a cell regardless of the number of neighbors), we are able to create a simulation where the cells almost never die, since it is very difficult from a probabilistic point of view for a given cell to stay alive.

    \FloatBarrier

    \section{Incorporation of complex value stochastic finite state machine into Schr\"{o}dinger equation in one dimensional stochastic Conway Game of Life}
    \label{sec:incorporation-of-classical-stochastic-finite-state-machine-into-schroedinger-equation}
    We obtain stochastic dynamics coming from stochastic Conway game of life~\cite{MDPIKotula} or any other classical stochastic game or process expressed by $\rho(x,t)$ or in more general way by 2 dimensional case $\rho(x,y,t)$ or 3 dimensional situation $\rho(x,y,z,t)$ or even in case of N dimensional space $\rho(x_1,x_2,..,x_N,t)$.
    It is instructive to consider one dimensional complex value Conway Game of Life that is semi-classical stochastic game with built-in
    interference effect that mimics quantum mechanics to certain way, but it is still classical game.

    \subsection{Description of complex value stochastic Conway Game of Life in one dimension}
    \label{subsec:description-of-stochastic-conway-game-of-life}
    Let us consider one dimensional classical stochastic Conway Game of Life, where mass of cell is complex value and has continuous values with no upper or lower constrains.
    We set probability of spontaneous change of cell $p$ from existing alive state of $\lvert m(k,t_s)\rvert \neq 0$ into $\lvert m(k,t_s+1)\rvert= 0$ (probability of spontaneous cell death on the precondition that two nearest neighbors have mass sum greater than simulation numerical zero or more precisely $10^{-5}$).
    Given cell with mass 0 (dead cell) or not bigger than $10^{-5}$ can become alive with probability $p$ and in such case the given cell will acquire the mass that average mass of its neighbors.
    At the same time with probability of $1-p$ there is occurrence of rules similar to those known from classical deterministic Conway Game of Life.
    We have following rules:
    \newline \newline
    1. If the sum of given live cell with mass $m(k,t_s)=\lvert m(k,t_s)\rvert e^{\sqrt{-1}\Theta(k,t_s)}$ (at position $k$) nearest neighbors mass ($m(k-1,t_s)+m(k+1,t_s)=m_{all}(t_{s})$) is in interval of magnitude values between 0.3 and 1.5 ($\lvert m_{all}\rvert \in [0.3,1.5]$) then $m(k,t_s+1)=\lvert m(k,t_s)\rvert e^{\sqrt{-1}*(\Theta(k,t_s)+0.1)}$.
    It is noticeable here that phase is increasing its value in quite similar way as in case of wavefunction for eigenenergy state for Quantum Mechanics characterized by Schr\"{o}dinger equation.
    \newline \newline
    2. If the sum of masses of nearest neighbors of given dead cell is in interval of magnitude values between 0.1 and 1.5 ($\lvert m_{all}(t_{s}\rvert) \in [0.1,1.5]$) then ($m(k,t_{s})\rightarrow m(k,t_s+1)=\frac{1}{2}m_{all}(t_{s})=m(k-1,t_s)+m(k+1,t_s)$).
    \newline \newline
    3. If a cell does not fulfill any of the rules mentioned above ($\lvert m(k,t_s)\rvert>1.5$ or $\lvert m(k,t_s)\rvert<0.1$ for a dead cell or $\lvert m(k,t_s)\rvert<0.3$ for a live cell), then in the next time step that cell will be dead ($m(k,t_s+1)=0$).
    \newline \newline
    In case of a two-dimensional cellular automaton we set probability of spontaneous change of cell $p$ from existing alive state of $\lvert m(k,l,t_s)\rvert \neq 0$ into $\lvert m(k,l,t_s+1)\rvert= 0$ on condition that the mass of its neighbors is in interval of magnitude values between 0.3 and 1.
    In other case given alive cell will acquire mass that is equal to $m(k,l,t_s)=\frac{1}{2}m_{all}=\frac{1}{2}(m(k-1,l-1,t_s)+m(k,l-1,t_s)+m(k+1,l-1,t_s)+m(k-1,l,t_s)+m(k+1,l,t_s)+m(k-1,l+1,t_s)+m(k,l+1,t_s)+m(k+1,l+1,t_s))$.
    Given dead cell with mass 0 or not bigger than $10^{-5}$ can become alive with probability $p$ and in such case the given cell will acquire the mass that is mass of its neighbors divided by two ($m(k,l,t_s)=\frac{1}{2}m_{all}$).
    At the same time with probability of $1-p$ there is occurrence of rules similar to those known from classical deterministic Conway Game of Life.
    We have following rules:
    \newline \newline
    1. If the sum of given live cell with mass $m(k,l,t_s)=\lvert m(k,l,t_s)\rvert e^{\sqrt{-1}\Theta(k,l,t_s)}$ (at position $k,l$) nearest neighbors mass ($m_{all}$) is in interval of magnitude values between 0.3 and 1 ($m_{all} \in [0.3,1]$) then $m(k,l,t_s+1)=\lvert m(k,l,t_s)\rvert e^{\sqrt{-1}*(\Theta(k,l,t_s)+0.1)}$.
    \newline \newline
    2. If the sum of given dead cell nearest neighbors mass is in interval of magnitude values between 0.45 and 1 ($m_{all} \in [0.45,1]$) then $m(k,l,t_s+1)=\frac{1}{2}m_{all}$.
    \newline \newline
    3. If a cell does not fulfill any of the rules mentioned above, then in the next time step that cell will be dead ($m(k,l,t_s+1)=0$).
    \newline \newline
    Situation of various scenarios of complex value Game of Life are given by Figures \ref{fig:potential}, \ref{fig:modulzfazy}, \ref{fig:konf14_average_characteristics}.
    Especially interesting is observing difference of Conway Game of Life in one dimension with presence and absence of phase.
    The real value Conway Game of Life always ends up with a kind of uniform probability distribution, while complex value Conway Game of Live ends up with Gaussian distribution what is shown in Figures \ref{fig:wavefunction_without_phase} and \ref{fig:modulzfazy}.
    \newline

    Due to certain conceptional analogies from quantum mechanics in Complex value stochastic Conway Game of Life it is instructed to derive such Schr\"{o}dinger potential that we parameterize one-dimensional complex value Game of Life.

    \subsection{Derivation of Schr\"{o}dinger potential mimicking second Fick Law dynamics in one and in two dimensions}
    \label{subsec:derivation-of-schroedinger-potential-mimicking-second-fick-law-dynamics-in-one-and-in-2-dimensions}
    It shall not be surprising that quantum mechanics can mimic statistical physics.
    After all Schr\"{o}dinger equation has classical limit of its solutions.
    Furthermore, there are deep analogies between quantum mechanics and classical statistical physics.
    We start from second Fick law in one dimensional geometric space expressing probability distribution dynamics with time $\rho(x,y)$ by means of $D(x,t)$ diffusion function summarized by equation or operator equation given as
    \begin{align}
        D(x,t)\frac{d^2}{dx^2}\rho(x,t)=\frac{d}{dt}\rho(x,t), \frac{d^2}{dx^2}\rho(x,t)=\frac{1}{D(x,t)}\frac{d}{dt}\rho(x,t), \nonumber \\
        \left(\frac{d}{dx}\rho(x,t)\right)\left(\frac{d^2}{dx^2}\rho(x,t)\right)=\frac{1}{2}\frac{d}{dx}\left(\frac{d}{dx}\rho(x,t)\right)^2=\left(\frac{d}{dx}\rho(x,t)\right)\frac{1}{D(x,t)}\frac{d}{dt}\rho(x,t)
        \label{eq:fick}.
    \end{align}
    \begin{align}
        \rho_{,x} = \left(\int h(x,t)dx \right) =\left( \int_{x_0}^{x} dx'  \frac{1}{D(x',t)}\frac{d}{dt}\rho(x',t) \right).
    \end{align}
    We can always write it by stating operator equality $\frac{d}{dt}=D(x,t)\frac{d^2}{dx^2}$, what implies
    that $(\frac{d}{dt})^n=\left(D(x,t)\frac{d^2}{dx^2}\right)^n$ in case of operator acting on $\rho$, where n is integer value.
    We can obtain the same probability distribution $\rho$ by means of Schr\"{o}dinger equation
    \begin{align}
        \left[-\frac{\hbar^2}{2m}\frac{d^2}{dx^2}+V(x,t)\right]\psi(x,t)=i\hbar\frac{d}{dt}\psi(x,t)
        \label{eq:schroedinger}
    \end{align}
    with notations
    \begin{align}
        \psi = \psi(x,t), \rho= \rho(x,t),\frac{d}{dx}\rho=\rho_{,x} ,\frac{d^2}{dx^2}\rho=\rho_{,x,x}, \\
        \Theta  = \Theta(x,t),\frac{d}{dx}\Theta(x,t)=\Theta(x,t)_{,x},\frac{d^2}{dx^2}\Theta(x,t)=\Theta(x,t)_{,x,x},   \\
        D = D(x,t), V = V(x,t) \nonumber
    \end{align}
    and by expressing wave function as
    \begin{align}
        \psi = \sqrt{\rho}e^{i\Theta}, \rho = \psi^2 e^{-2i\Theta}.
    \end{align}
    Incorporation of probability distribution in Schr\"{o}edinger equation dynamics leads to equation
    \begin{align}
        \left(-\frac{\hbar^2}{2m}\frac{d^2}{dx^2}+V\right)\sqrt{\rho}e^{i\Theta}=i\hbar\frac{d}{dt}\left(\sqrt{\rho}e^{i\Theta}\right)
    \end{align}
    that results in following steps
    \begin{align}
        -\frac{\hbar^2}{2m}\frac{d^2}{dx^2}\left(\sqrt{\rho}e^{i\Theta}\right)+V\sqrt{\rho}e^{i\Theta}=i\hbar\left(\frac{\rho_{,t}}{2\sqrt{\rho}}e^{i\Theta}+i\sqrt{\rho}\Theta_{,t}e^{i\Theta}\right)
    \end{align}
    \begin{align}
        -\frac{\hbar^2}{2m}\frac{d^2}{dx^2}\left(\sqrt{\rho}e^{i\Theta}\right)+V\sqrt{\rho}e^{i\Theta}=\frac{i\hbar\rho_{,t}}{2\sqrt{\rho}}e^{i\Theta}-\hbar\sqrt{\rho}\Theta_{,t}e^{i\Theta}
    \end{align}
    \begin{align}
        \frac{2\sqrt{\rho}}{i\hbar}e^{-i\Theta}\left(\hbar\sqrt{\rho}\Theta_{,t}e^{i\Theta}-\frac{\hbar^2}{2m}\frac{d^2}{dx^2}\left(\sqrt{\rho}e^{i\Theta}\right)+V\sqrt{\rho}e^{i\Theta}\right)=\rho_{,t}
    \end{align}
    \begin{align}
        \frac{2\rho}{i}\Theta_{,t}-\frac{\hbar\sqrt{\rho}}{im}e^{-i\Theta}\frac{d^2}{dx^2}\left(\sqrt{\rho}e^{i\Theta}\right)+\frac{2\rho}{i\hbar}V=\rho_{,t}
    \end{align}
    \begin{align}
        \frac{i\hbar\sqrt{\rho}}{m}e^{-i\Theta}\frac{d^2}{dx^2}\left(\sqrt{\rho}e^{i\Theta}\right)-2i\rho\Theta_{,t}-\frac{2i\rho}{\hbar}V=\rho_{,t}
    \end{align}
    Second derivative from classical inspired wave-function with respect to position is
    \begin{align}
        \frac{d^2}{dx^2}\left(\sqrt{\rho}e^{i\Theta}\right)=\frac{d}{dx}\left(\frac{\rho_{,x}}{2\sqrt{\rho}}e^{i\Theta}+i\sqrt{\rho}\Theta_{,x}e^{i\Theta}\right)
    \end{align}
    and results in
    \begin{align}
        \frac{d^2}{dx^2}\left(\sqrt{\rho}e^{i\Theta}\right)=e^{i\Theta}\left(\frac{\rho_{,x,x}}{2\sqrt{\rho}}-\frac{(\rho_{,x})^2}{4\sqrt{\rho^3}}+i\frac{\rho_{,x}}{\sqrt{\rho}}\Theta_{,x}+i\sqrt{\rho}\Theta_{,x,x}-\sqrt{\rho}(\Theta_{,x})^2\right)
    \end{align}
    Finally we obtain
    \begin{align}
        \frac{i\hbar}{m}\left(\frac{\rho_{,x,x}}{2}-\frac{(\rho_{,x})^2}{4\rho}+i\rho_{,x}\Theta_{,x}+i\rho\Theta_{,x,x}-\rho(\Theta_{,x})^2\right)-2i\rho\Theta_{,t}-\frac{2i\rho}{\hbar}V=\rho_{,t}
    \end{align}
    \begin{align}
        \frac{i\hbar}{2m}\rho_{,x,x}+\frac{i\hbar}{m}\left(-\frac{(\rho_{,x})^2}{4\rho}+i\rho_{,x}\Theta_{,x}+i\rho\Theta_{,x,x}-\rho(\Theta_{,x})^2\right)-2i\rho\Theta_{,t}-\frac{2i\rho}{\hbar}V=\rho_{,t}
    \end{align}
    that allow us to determine effective Schr\"{o}dinger potential $V$ mimicking second Fick law in the form:
    \begin{align}
        \frac{i\hbar}{m}\left(\frac{\rho_{,x,x}}{2}-\frac{(\rho_{,x})^2}{4\rho}+i\rho_{,x}\Theta_{,x}+i\rho\Theta_{,x,x}-\rho(\Theta_{,x})^2\right)-2i\rho\Theta_{,t}-\rho_{,t}=\frac{2i\rho}{\hbar}V
    \end{align}
    \begin{align}
        \frac{\hbar}{2i\rho}\left(\frac{i\hbar}{m}\left(\frac{\rho_{,x,x}}{2}-\frac{(\rho_{,x})^2}{4\rho}+i\rho_{,x}\Theta_{,x}+i\rho\Theta_{,x,x}-\rho(\Theta_{,x})^2\right)-2i\rho\Theta_{,t}-\rho_{,t}\right)=V
    \end{align}
    \begin{align}
        \frac{\hbar^2}{2m\rho}\left(\frac{\rho_{,x,x}}{2}-\frac{(\rho_{,x})^2}{4\rho}+i\rho_{,x}\Theta_{,x}+i\rho\Theta_{,x,x}-\rho(\Theta_{,x})^2\right)-\hbar\Theta_{,t}+\frac{i\hbar}{2\rho}\rho_{,t}=V
    \end{align}
    Now we state the quantum phase in the following form using some analogy between gradient of phase and velocity known from London equations \cite{London} describing the superconducting current relation being the same as Schr\"{o}dinger probability current and vector potential.
    Due to existence of Aharonov-Bohm effect (postulated in phenomenology constructed wave-function coming from probability map generated by classical stochastic process) vector potential is gradient of wave-function phase.
    Such considerations lead to formula given as
    \begin{align}
        \Theta=g\sqrt{\rho}\frac{d}{dx}\sqrt{\rho}=\frac{1}{2}g\rho_{,x},
        \label{eq:phase}
    \end{align}
    where g is some phenomenological constant.
    \begin{align}
        \frac{\hbar^2}{2m\rho}\left(\frac{\rho_{,x,x}}{2}-\frac{(\rho_{,x})^2}{4\rho}+i\rho_{,x}\frac{1}{2}g\rho_{,x,x}+i\rho\frac{1}{2}g\rho_{,x,x,x}-\rho\frac{1}{4}g^2\left(\rho_{,x,x}\right)^2\right)-\hbar\frac{1}{2}g\rho_{,x,t}+\frac{i\hbar}{2\rho}\rho_{,t}=V
    \end{align}
    \begin{align}
        \frac{\hbar^2\rho_{,x,x}}{4m\rho}-\frac{\hbar^2(\rho_{,x})^2}{8m\rho^2}-\frac{\hbar^2 g^2\left(\rho_{,x,x}\right)^2}{8m}-\frac{\hbar g\rho_{,x,t}}{2}+i\left(\frac{\hbar^2\rho_{,x}g\rho_{,x,x}}{4m\rho}+\frac{\hbar^2 g\rho_{,x,x,x}}{4m}+\frac{\hbar\rho_{,t}}{2\rho}\right)=V
        \label{eq:potential}
    \end{align}
    We have real part of potential in the form
    \begin{align}
        Re[V(t)]= \frac{\hbar^2\rho_{,x,x}}{4m\rho}-\frac{\hbar^2(\rho_{,x})^2}{8m\rho^2}-\frac{\hbar^2 g^2\left(\rho_{,x,x}\right)^2}{8m}-\frac{\hbar g\rho_{,x,t}}{2}= \nonumber \\
        = \frac{\hbar^2}{4m\rho}\frac{1}{D(x,t)}\frac{d}{dt}\rho(t)-\frac{\hbar^2(\rho_{,x})^2}{8m\rho^2}-\frac{\hbar^2 g^2\left(\frac{1}{D(x,t)}\rho_{,t}\right)^2}{8m}-\frac{\hbar}{2}g\frac{d}{dt}\rho_{,x} = \nonumber \\
        =\frac{\hbar^2}{4m\rho}\frac{1}{D(x,t)}\frac{d}{dt}\rho(t)-\frac{\hbar^2\left(\left( \int_{x_0}^{x} dx' \frac{1}{D(x',t)}\frac{d}{dt}\rho(x',t)  \right)^2\right)}{8m\rho^2} 
        -\frac{\hbar^2 g^2\left(\frac{1}{D(x,t)}\rho_{,t}\right)^2}{8m}-\frac{\hbar}{2}g\frac{d}{dt}\left( \int_{x_0}^{x} dx' \frac{1}{D(x',t)}\frac{d}{dt}\rho(x',t) \right).
    \end{align}
    Now we can extract imaginary part of Schr\"{o}dinger potential in the form
    \begin{align}
        Im[V(t)]=\left(\frac{\hbar^2\rho_{,x}g\rho_{,x,x}}{4m\rho}+\frac{\hbar^2 g\rho_{,x,x,x}}{4m}+\frac{\hbar\rho_{,t}}{2\rho}\right)   =\left(\frac{\hbar^2}{8m\rho}g\frac{d}{dx}(\rho_{,x})^2+\frac{\hbar^2 g}{4m}\frac{d}{dx}\left(\frac{1}{D(x,t)}\rho_{,t}\right)+\frac{\hbar\rho_{,t}}{2\rho}\right) = \nonumber \\
        =\left(\frac{\hbar^2}{8m\rho}g\frac{d}{dx}\left(\int_{x_0}^{x}\frac{1}{D(x',t)}\rho_{,t}dx'\right)^2+\frac{\hbar^2 g}{4m}\frac{d}{dx}\left(\frac{1}{D(x,t)}\rho_{,t}\right)+\frac{\hbar\rho_{,t}}{2\rho}\right)=\nonumber \\
        =\left(\frac{\hbar^2}{8m\rho}g\frac{d}{dx}\left(\int_{x_0}^{x}\frac{d^2}{dx'^2}\rho(x',t)dx'\right)^2+\frac{\hbar^2 g}{4m}\frac{d^3}{dx^3}(\rho(x,t))+\frac{\hbar D(x,t)\frac{d^2}{dx^2}\rho}{2\rho}\right).
    \end{align}
    We have obtained integro-differential equations for effective Schr\"{o}dinger potential in function of diffusion parameter $D(x,t)$.
    Now we can write effective Schrodinger potential by angle $\Theta_p(x,t)$ and potential modulus $V_m(x,t)=\lvert V(x,t)\rvert$.
    It is given by following formulas

    \begin{align}
        V(x,t)=V(x,t)_m e^{i \Theta_p(x,t)}, V(x,t)_m=\sqrt{(Re[V(t)])^2+(Im[V(t)])^2}=\nonumber \\
        =  \Bigg[ \Bigg[ \Bigg(\frac{\hbar^2}{8m\rho}g\frac{d}{dx}\Bigg(\int_{x_0}^{x}\frac{d^2}{dx'^2}\rho(x',t)dx'\Bigg)^2+\frac{\hbar^2 g}{4m}\frac{d^3}{dx^3}(\rho(x,t))+\frac{\hbar D(x,t)\frac{d^2}{dx^2}\rho}{2\rho}\Bigg) \Bigg] ^2 +    \nonumber \\
        + \Bigg[  \frac{\hbar^2}{4m\rho}\frac{1}{D(x,t)}\frac{d}{dt}\rho(t)-\frac{\hbar^2\left(\left( \int_{x_0}^{x} dx' \frac{1}{D(x',t)}\frac{d}{dt}\rho(x',t)  \right)^2\right)}{8m\rho^2} \nonumber  \\
        -\frac{\hbar^2 g^2\left(\frac{1}{D(x,t)}\rho_{,t}\right)^2}{8m}-\frac{\hbar}{2}g\frac{d}{dt}\left( \int_{x_0}^{x} dx' \frac{1}{D(x',t)}\frac{d}{dt}\rho(x',t) \right) \Bigg]^2 \Bigg]^{\frac{1}{2}}.
    \end{align}
        Formula for phase of effective Schr\"{o}dinger potential is given as
    \begin{align}
        \Theta_p(x,t)=ArcCos \Bigg[  \frac{\hbar^2}{4m\rho}\frac{1}{D(x,t)}\frac{d}{dt}\rho(t)-\frac{\hbar^2\left(\left( \int_{x_0}^{x} dx' \frac{1}{D(x',t)}\frac{d}{dt}\rho(x',t)  \right)^2\right)}{8m\rho^2} \nonumber  \\
        -\frac{\hbar^2 g^2\left(\frac{1}{D(x,t)}\rho_{,t}\right)^2}{8m}-\frac{\hbar}{2}g\frac{d}{dt}\left( \int_{x_0}^{x} dx' \frac{1}{D(x',t)}\frac{d}{dt}\rho(x',t) \right) \Bigg] \times \nonumber \\
        \Bigg[
            \Bigg[ \Bigg[ \Bigg(\frac{\hbar^2}{8m\rho}g\frac{d}{dx}\Bigg(\int_{x_0}^{x}\frac{d^2}{dx'^2}\rho(x',t)dx'\Bigg)^2+\frac{\hbar^2 g}{4m}\frac{d^3}{dx^3}(\rho(x,t))+\frac{\hbar D(x,t)\frac{d^2}{dx^2}\rho}{2\rho}\Bigg) \Bigg] ^2 +    \nonumber \\
            + \Bigg[  \frac{\hbar^2}{4m\rho}\frac{1}{D(x,t)}\frac{d}{dt}\rho(t)-\frac{\hbar^2\left(\left( \int_{x_0}^{x} dx' \frac{1}{D(x',t)}\frac{d}{dt}\rho(x',t)  \right)^2\right)}{8m\rho^2} \nonumber  \\
            -\frac{\hbar^2 g^2\left(\frac{1}{D(x,t)}\rho_{,t}\right)^2}{8m}-\frac{\hbar}{2}g\frac{d}{dt}\left( \int_{x_0}^{x} dx' \frac{1}{D(x',t)}\frac{d}{dt}\rho(x',t) \right) \Bigg]^2 \Bigg]^{\frac{1}{2}}
            \Bigg]^{-1} \Bigg].
    \end{align}


    Basing on the both two last formulas we see that quantum mechanics can simulate classical world expressed by second Fick law.
    The effective Schr\"{o}dinger potential depends on non-local way of diffusion constant from Fick law.
    Indeed quantum mechanics can mimick classical world expressed by 2nd Fick law.
    It is understandable since Quantum Turing Machine can simulate Classical Turing Machine.
    Therefore nanochip basing on the laws of quantum mechanics can simulate classical statistical physics process.

    \subsection{Numerical computation of Schr\"{o}dinger complex value potential}
    \label{subsec:numerical-computation-of-schroedinger-complex-value-potential}
\begin{figure}
        \centering
        \includegraphics[width=0.5 \linewidth]{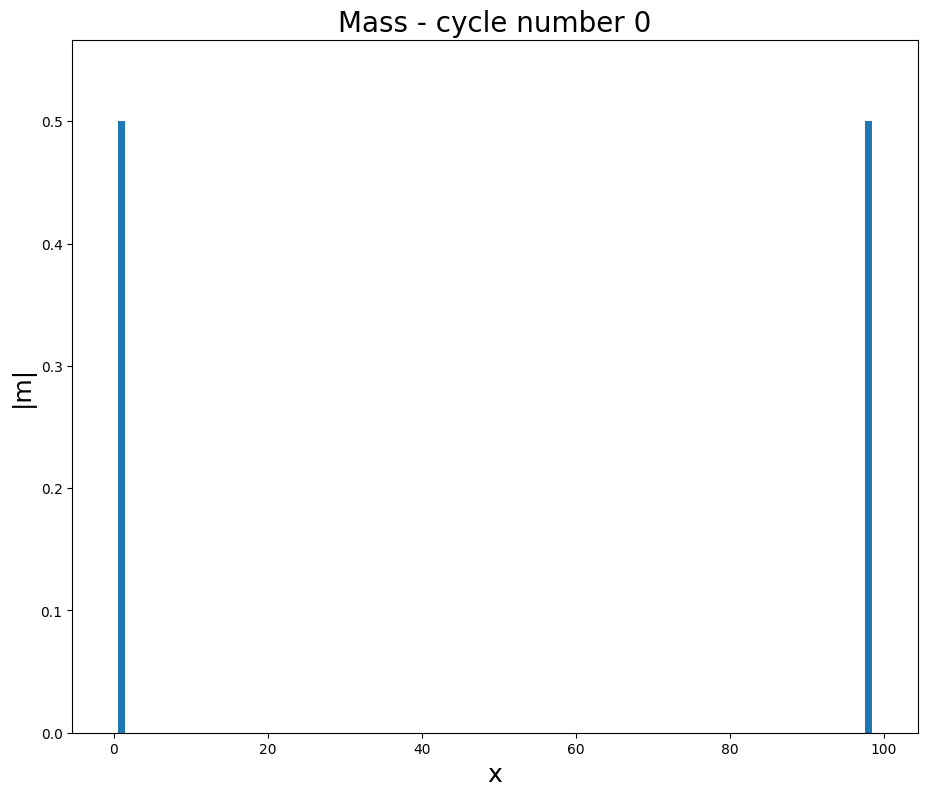}
        \caption{Initial structure of cell distribution in real or complex value Conway Game of Life. Given initial conditions correspond to cell evolution given by Figures \ref{fig:potential}, \ref{fig:potential2}, \ref{fig:modulzfazy} and \ref{fig:wavefunction_without_phase}.}
        \label{fig:sc5}
    \end{figure}

\begin{figure}
        \centering
        \subfloat[Mass at t=30\centering]{{\includegraphics[width=.3\linewidth]{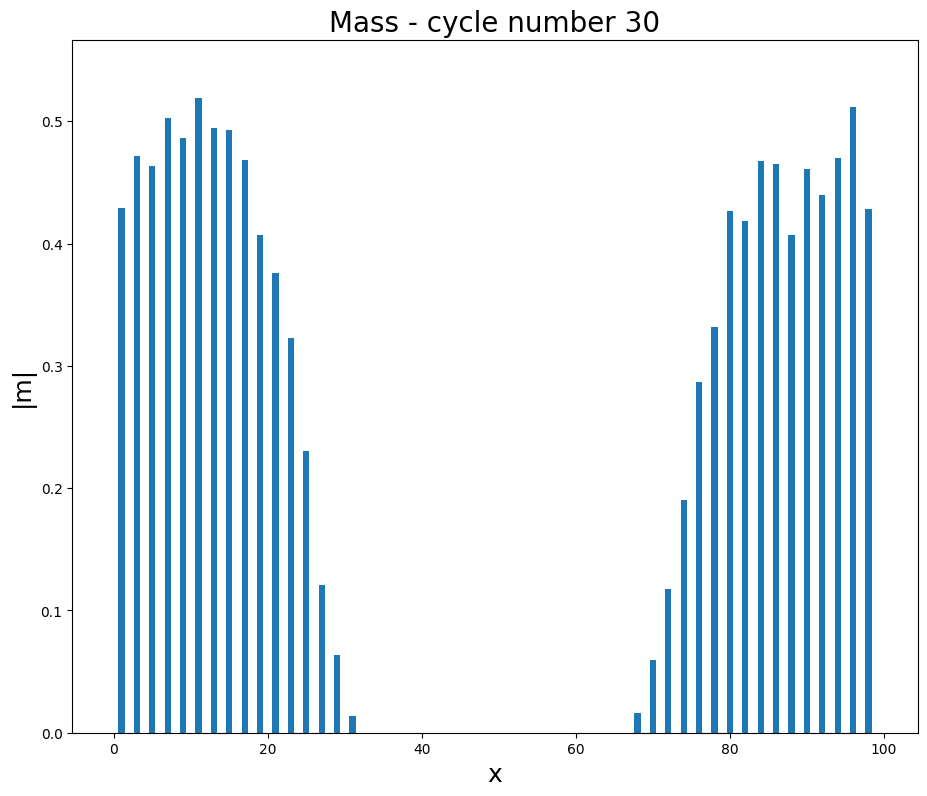} }}
        \subfloat[Mass at t=60\centering]{{\includegraphics[width=.3\linewidth]{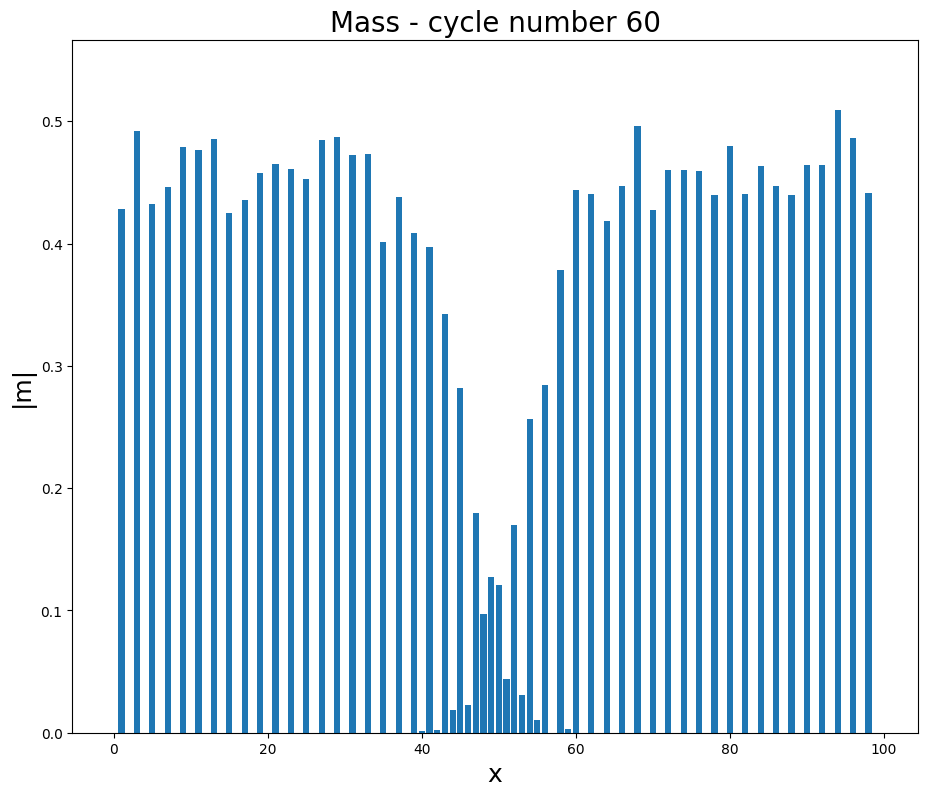} }}
        \subfloat[Mass at t=150\centering]{{\includegraphics[width=.3\linewidth]{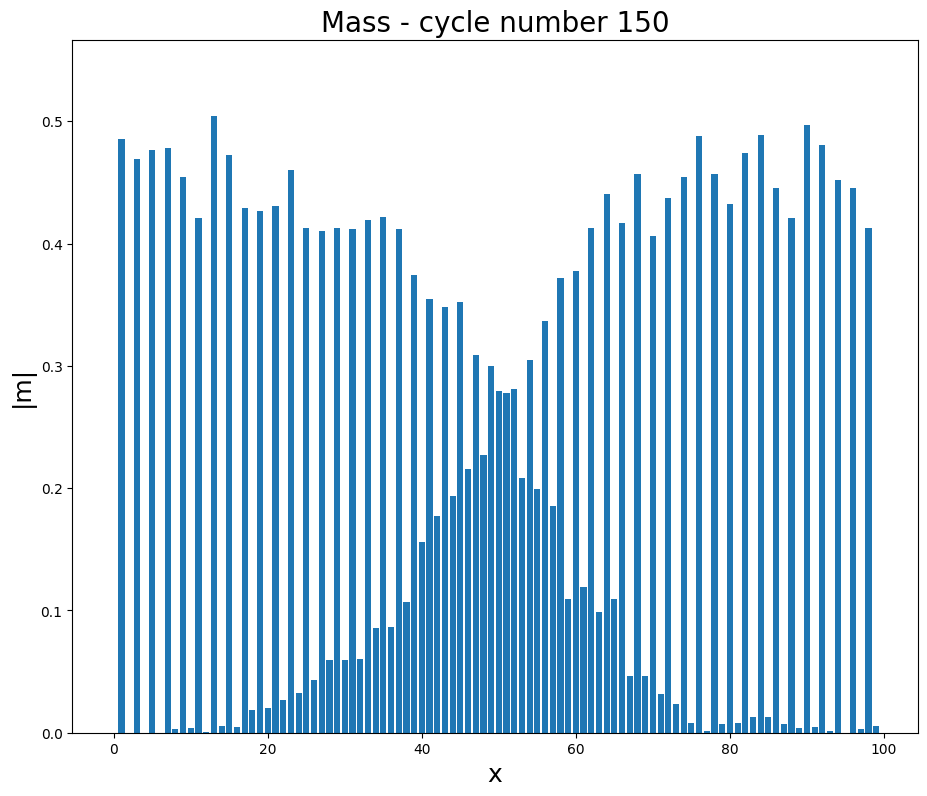} }}
        \quad
        \subfloat[$\mathfrak{R}(V)$ at t=30\centering]{{\includegraphics[width=.3\linewidth]{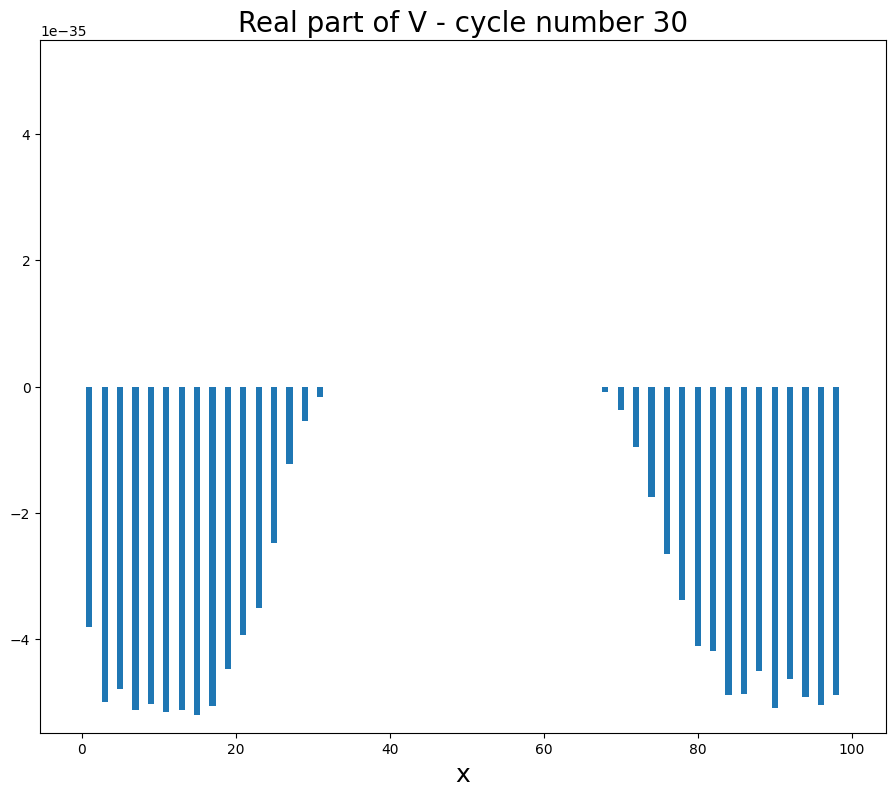} }}
        \subfloat[$\mathfrak{R}(V)$ at t=60\centering]{{\includegraphics[width=.3\linewidth]{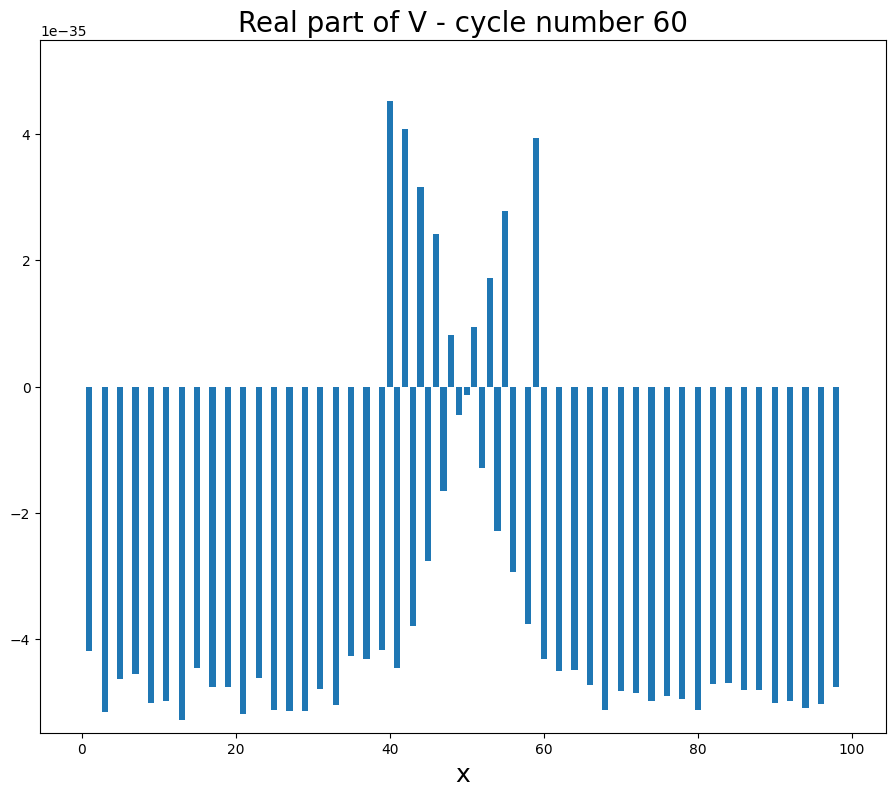} }}
        \subfloat[$\mathfrak{R}(V)$ at t=150\centering]{{\includegraphics[width=.3\linewidth]{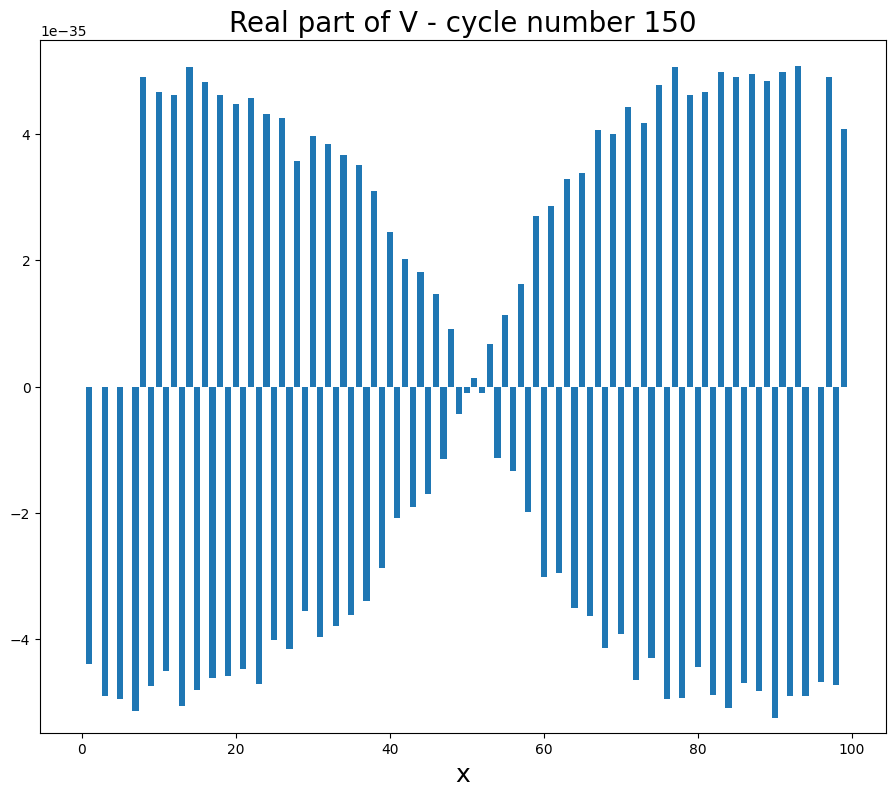} }}
        \quad
        \subfloat[$\mathfrak{I}(V)$ at t=30\centering]{{\includegraphics[width=.3\linewidth]{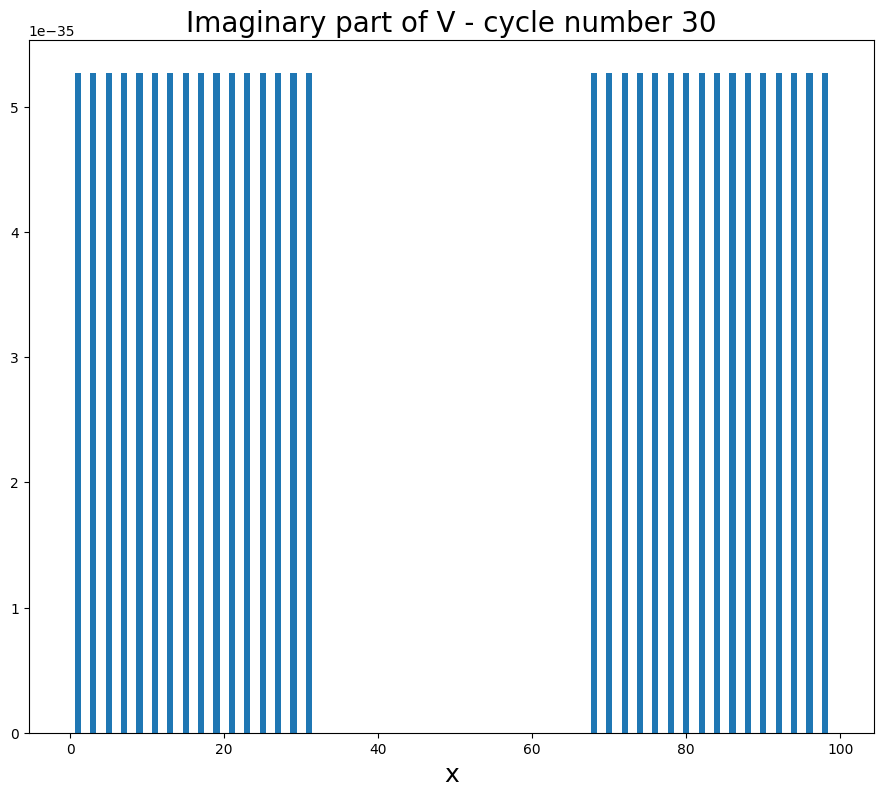} }}
        \subfloat[$\mathfrak{I}(V)$ at t=60\centering]{{\includegraphics[width=.3\linewidth]{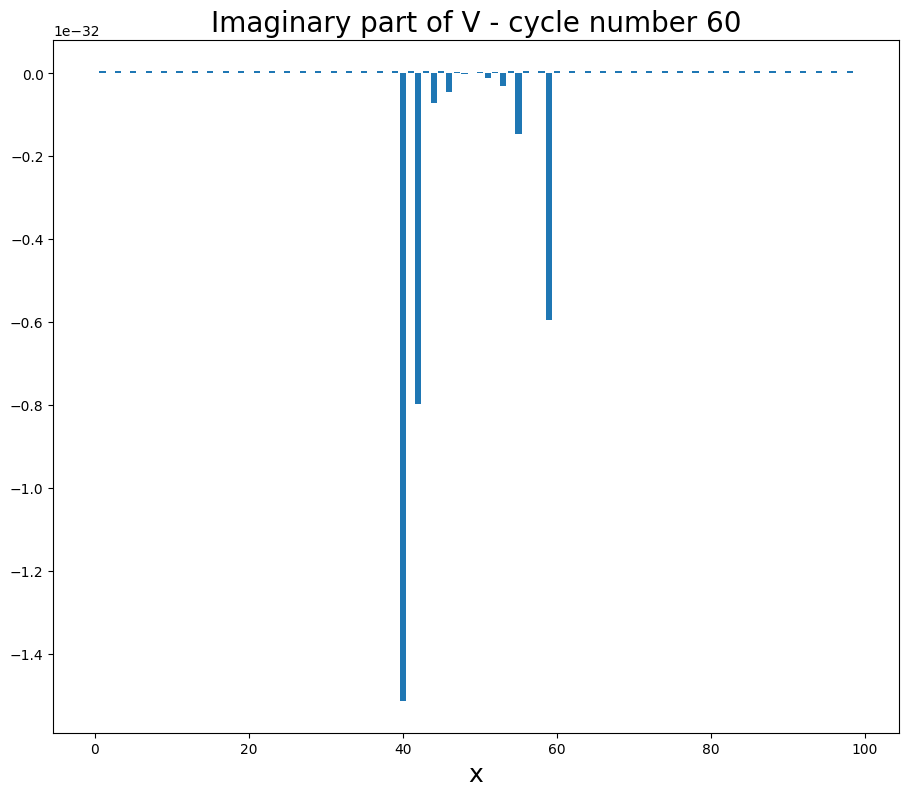} }}
        \subfloat[$\mathfrak{I}(V)$ at t=150\centering]{{\includegraphics[width=.3\linewidth]{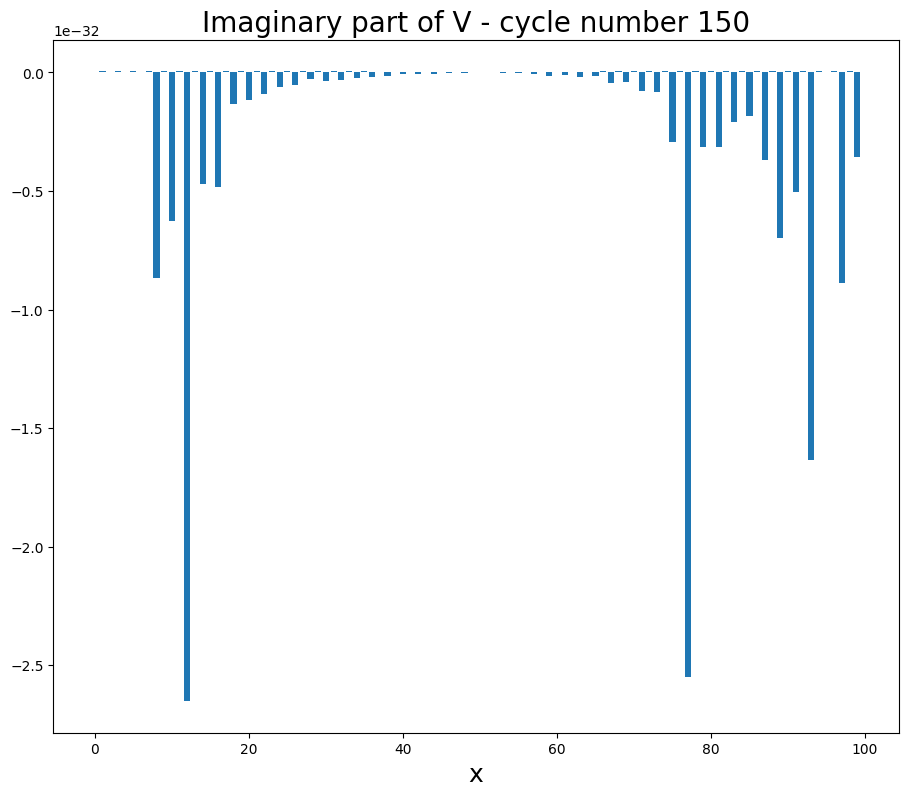} }}
        \caption{Dependence of mass (or more precisely probability density) on simulation in one dimensional Conway Game of Life at different time steps. Real and imaginary values of complex value potential from Schr\"{o}dinger equation mimicking preimposed probability distribution depicted above is given (numerical results according to equation~\ref{eq:potential}, $g=m=1$).}
        \label{fig:potential}
    \end{figure}

\begin{figure}
        \centering
        \subfloat[Mass at t=30\centering]{{\includegraphics[width=.3\linewidth]{figures/sc5_mass_1} }}
        \subfloat[Mass at t=60\centering]{{\includegraphics[width=.3\linewidth]{figures/sc5_mass_2} }}
        \subfloat[Mass at t=150\centering]{{\includegraphics[width=.3\linewidth]{figures/sc5_mass_3} }}
        \quad
        \subfloat[$\mathfrak{R}(V_{\hbar})$ at t=30\centering]{{\includegraphics[width=.3\linewidth]{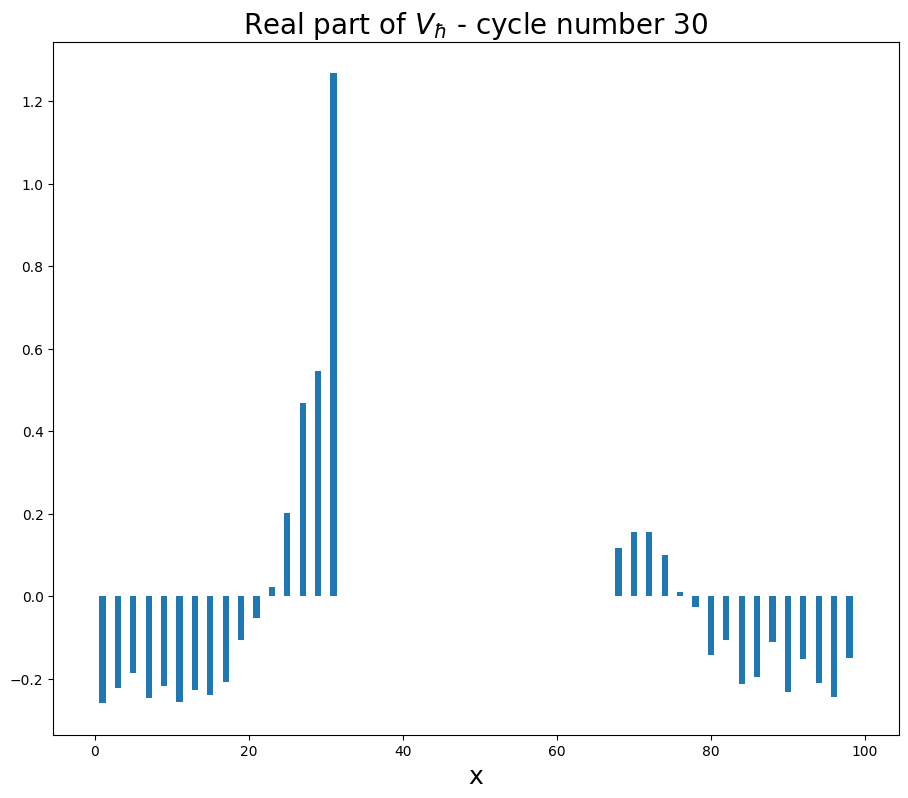} }}
        \subfloat[$\mathfrak{R}(V_{\hbar})$ at t=60\centering]{{\includegraphics[width=.3\linewidth]{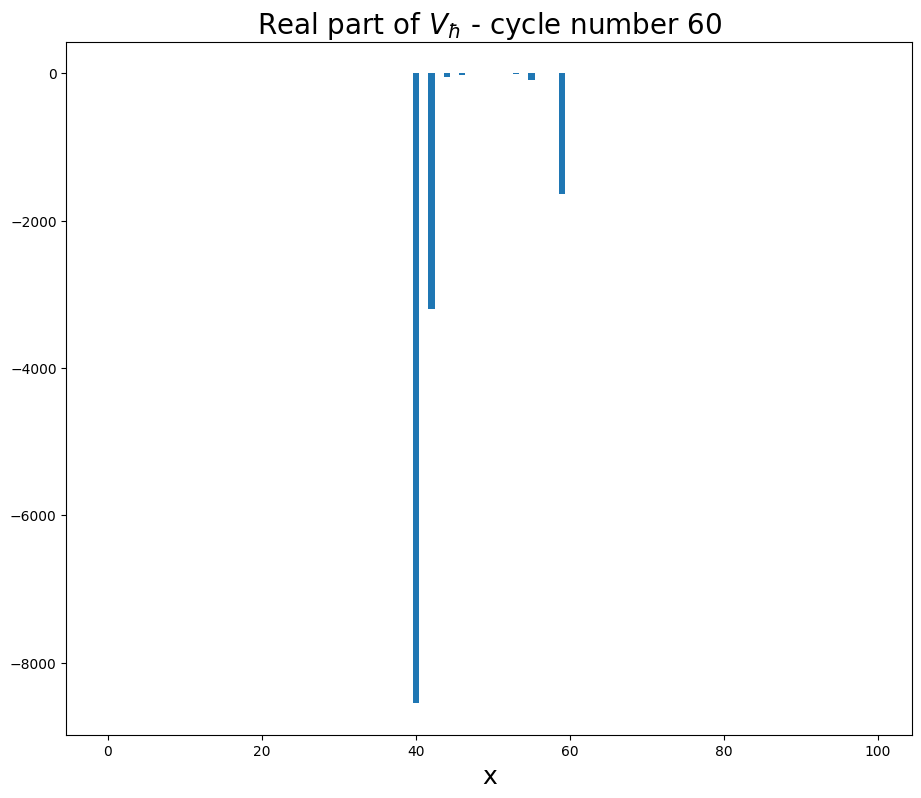} }}
        \subfloat[$\mathfrak{R}(V_{\hbar})$ at t=150\centering]{{\includegraphics[width=.3\linewidth]{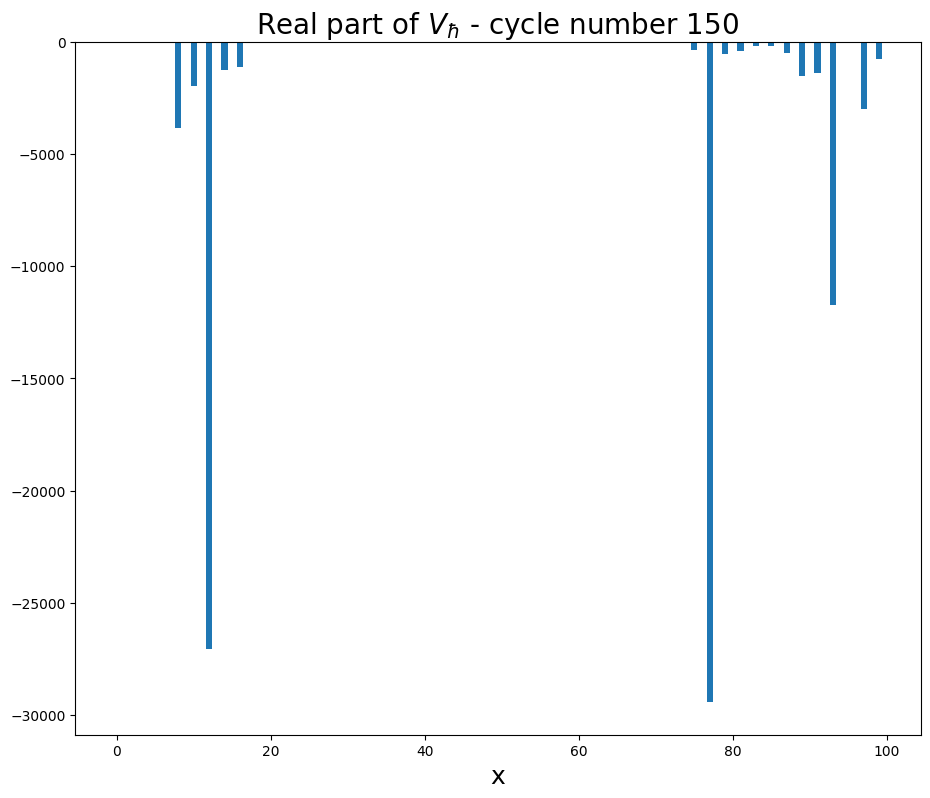} }}
        \quad
        \subfloat[$\mathfrak{I}(V_{\hbar})$ at t=30\centering]{{\includegraphics[width=.3\linewidth]{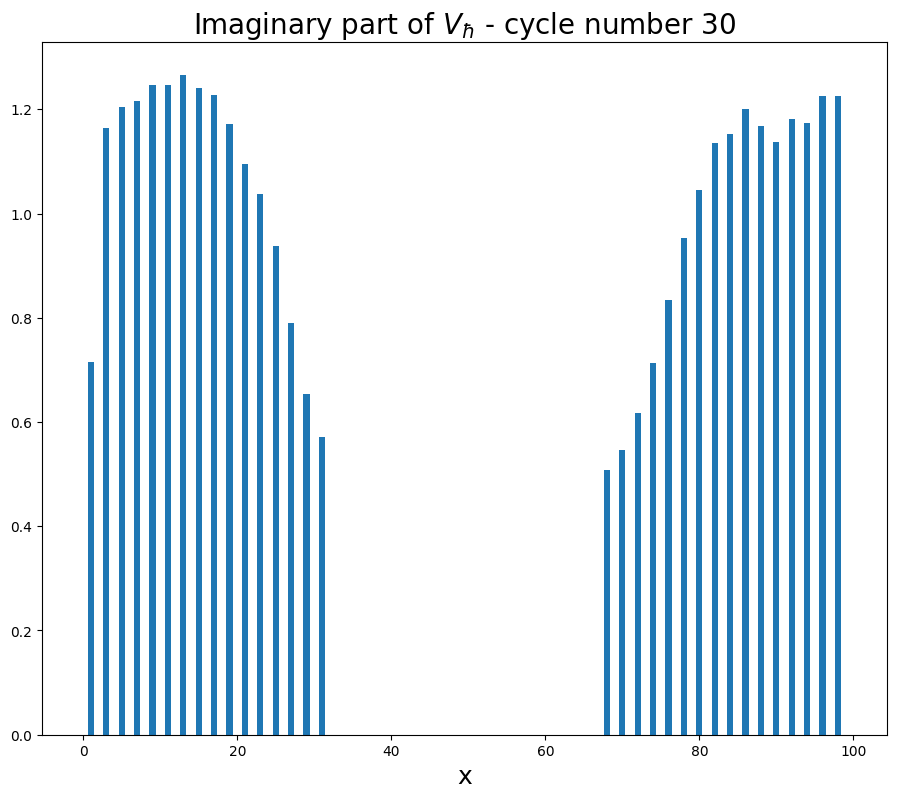} }}
        \subfloat[$\mathfrak{I}(V_{\hbar})$ at t=60\centering]{{\includegraphics[width=.3\linewidth]{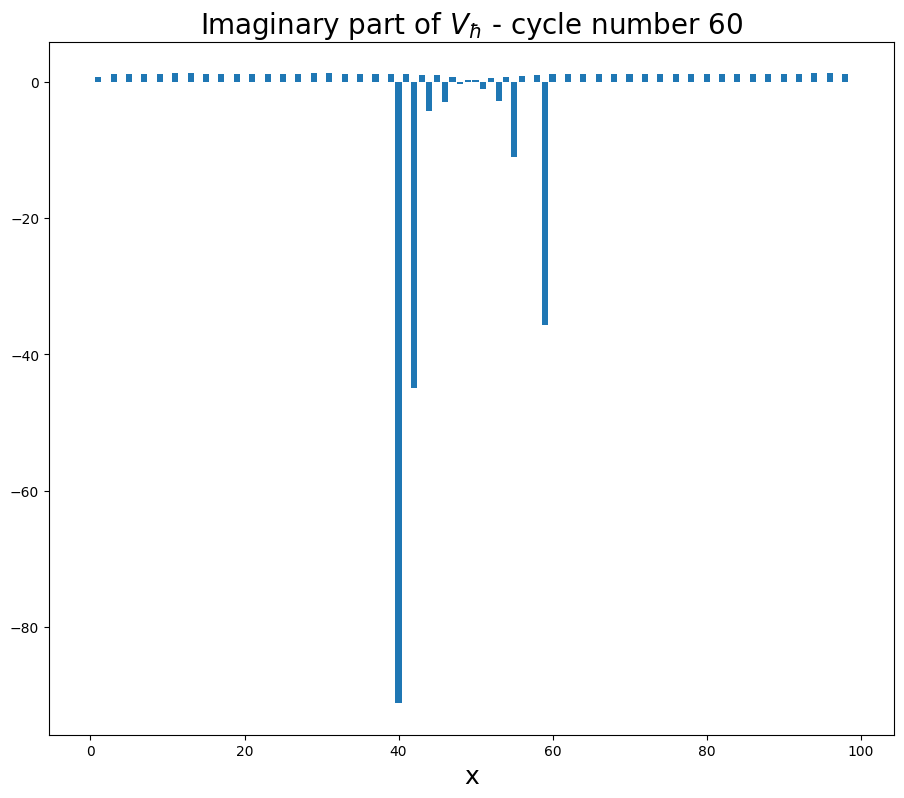} }}
        \subfloat[$\mathfrak{I}(V_{\hbar})$ at t=150\centering]{{\includegraphics[width=.3\linewidth]{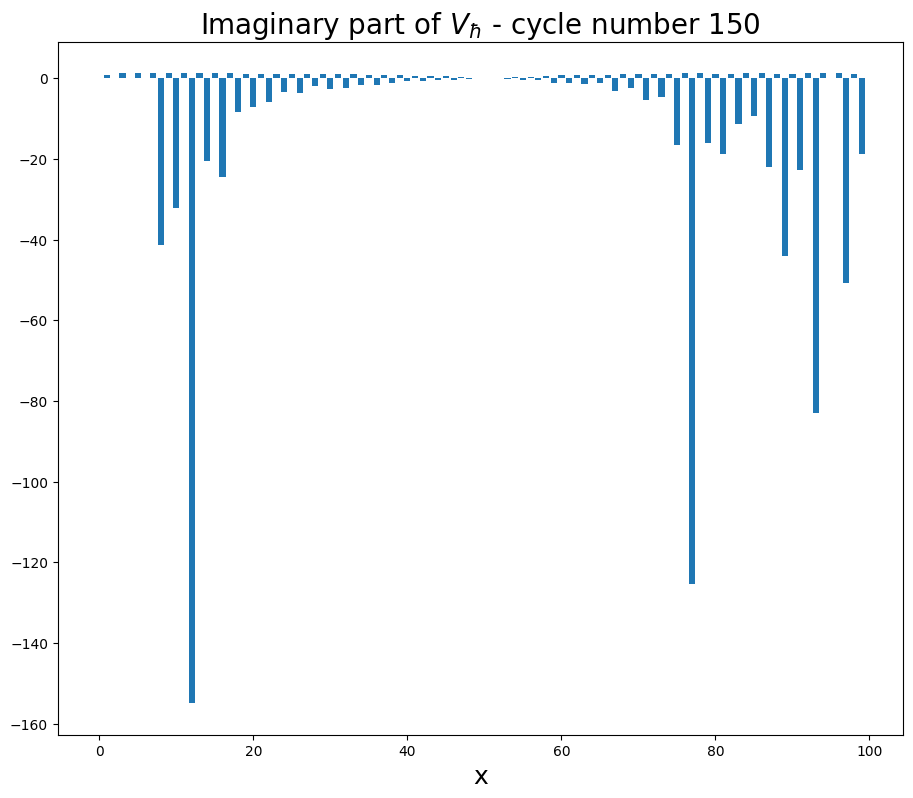} }}
        \caption{Dependence of mass (or more precisely probability density) on simulation in one dimensional Conway Game of Life at different time steps. Real and imaginary values of complex value potential from Schr\"{o}dinger equation divided by reduced Planck constant mimicking preimposed probability distribution depicted above is given (numerical results according to equation~\ref{eq:potential2} with $g=1$).}
        \label{fig:potential2}
    \end{figure}

\begin{figure}[ht]
        \centering
        \subfloat[$\Theta$ at t=30\centering]{{\includegraphics[width=.3\linewidth]{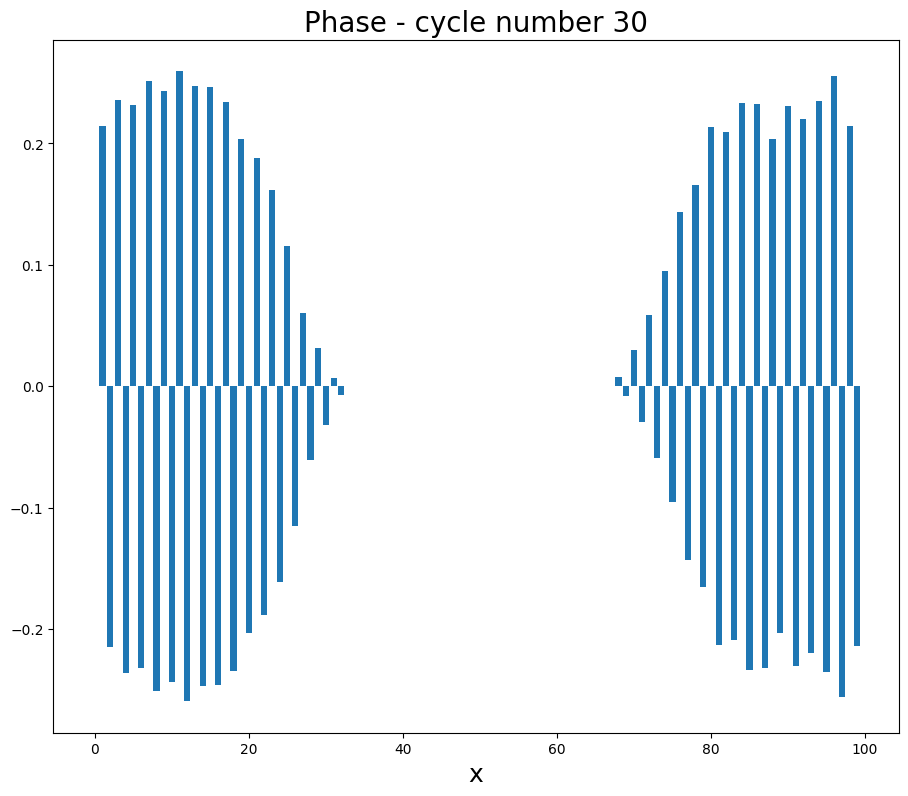} }}
        \subfloat[$\Theta$ at t=60\centering]{{\includegraphics[width=.3\linewidth]{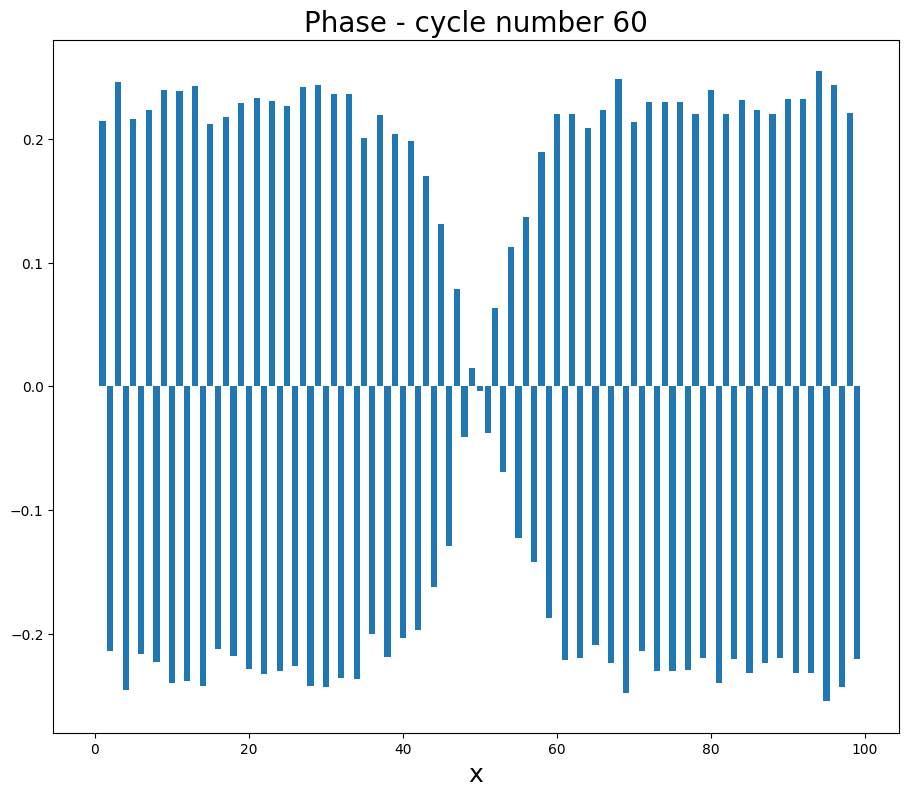} }}
        \subfloat[$\Theta$ at t=150\centering]{{\includegraphics[width=.3\linewidth]{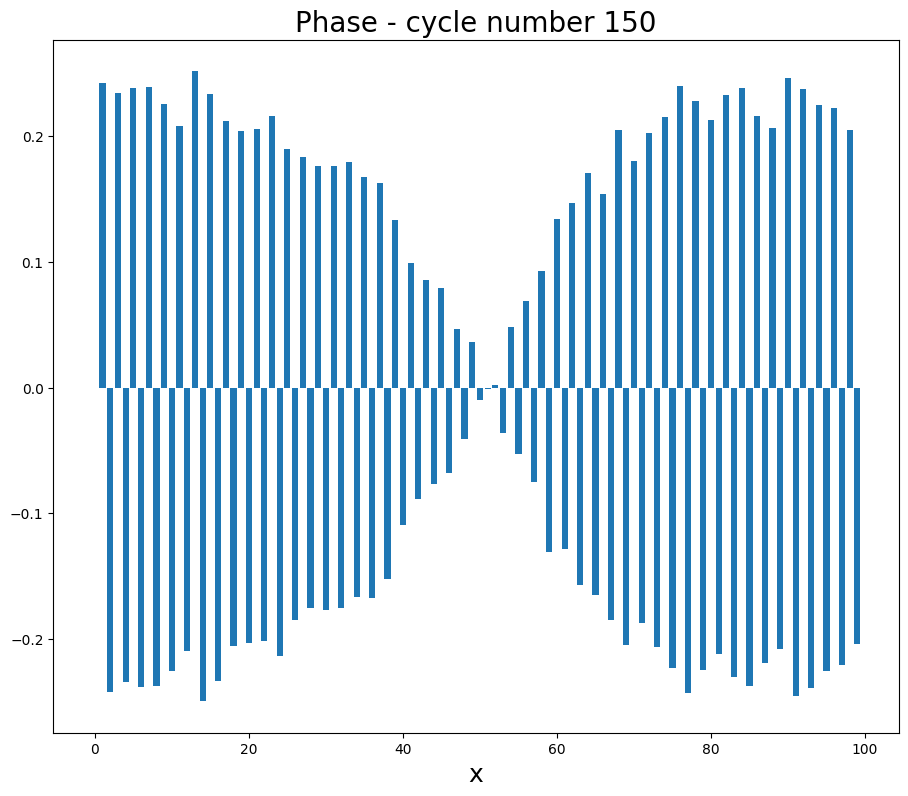} }}
        \quad
        \subfloat[$\lvert \psi\rvert$ at t=30\centering]{{\includegraphics[width=.3\linewidth]{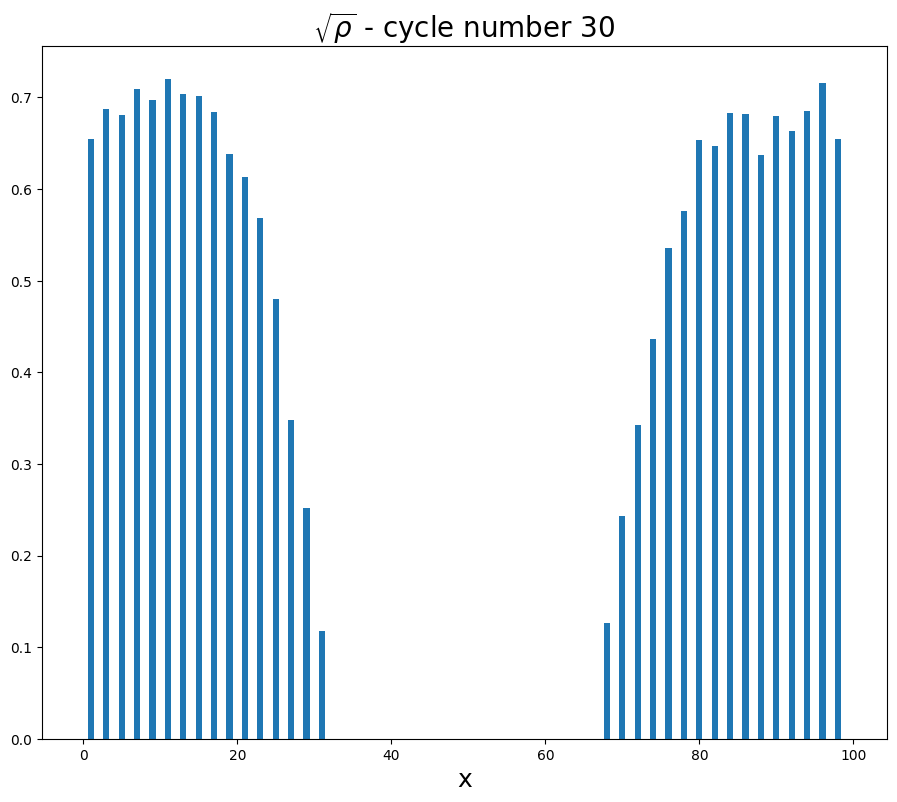} }}
        \subfloat[$\lvert \psi\rvert$ at t=60\centering]{{\includegraphics[width=.3\linewidth]{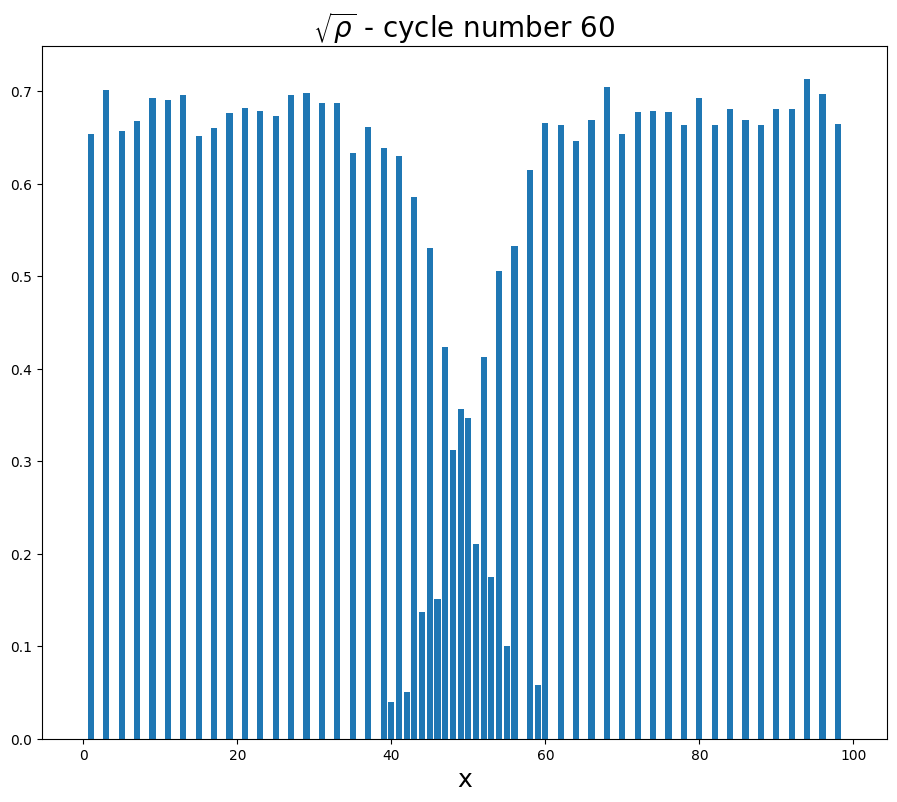} }}
        \subfloat[$\lvert \psi\rvert$ at t=150\centering]{{\includegraphics[width=.3\linewidth]{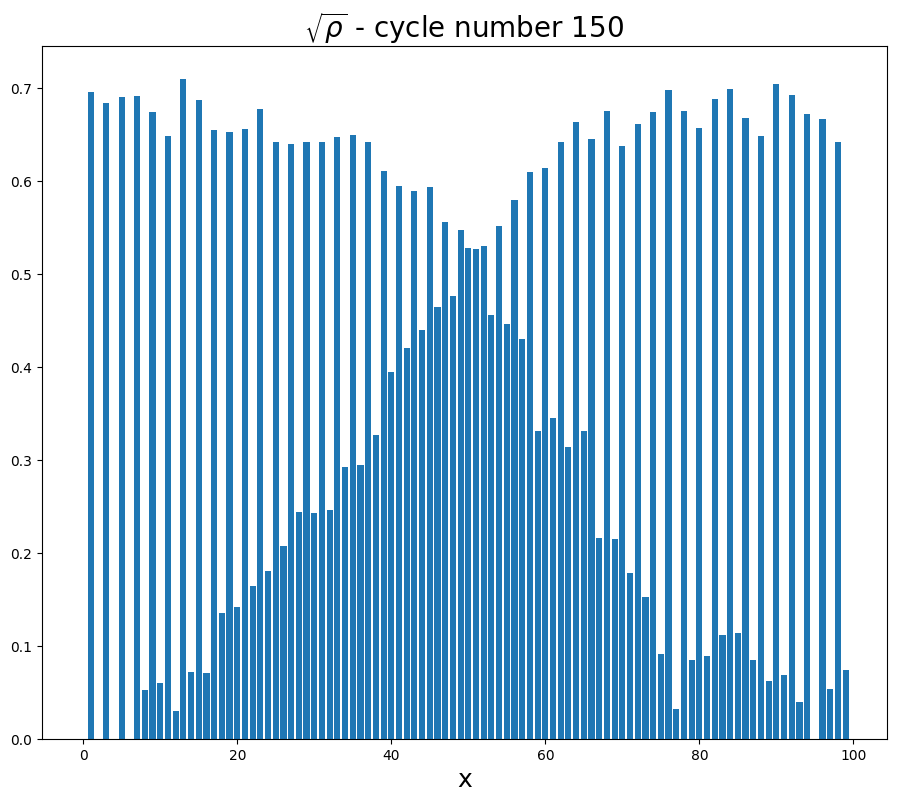} }}
        \caption{Dependence of phase of wave function on simulation in one dimensional Conway Game of Life at different time steps. Modulus of the wave function as equivalence between the square root of the mass. Simulation was conducted 200 times for the purpose of probability density distribution.}
        \label{fig:modulzfazy}
    \end{figure}

\begin{figure}
        \centering
        \subfloat[Mass at t=30\centering]{{\includegraphics[width=.2\linewidth]{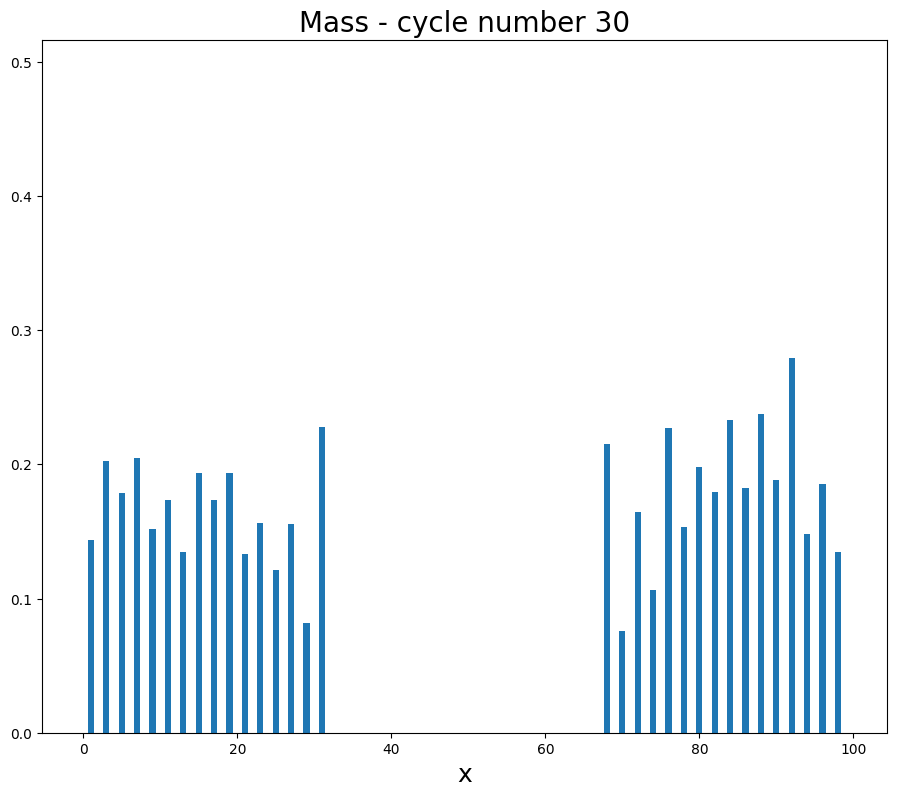} }}
        \subfloat[Mass at t=60\centering]{{\includegraphics[width=.2\linewidth]{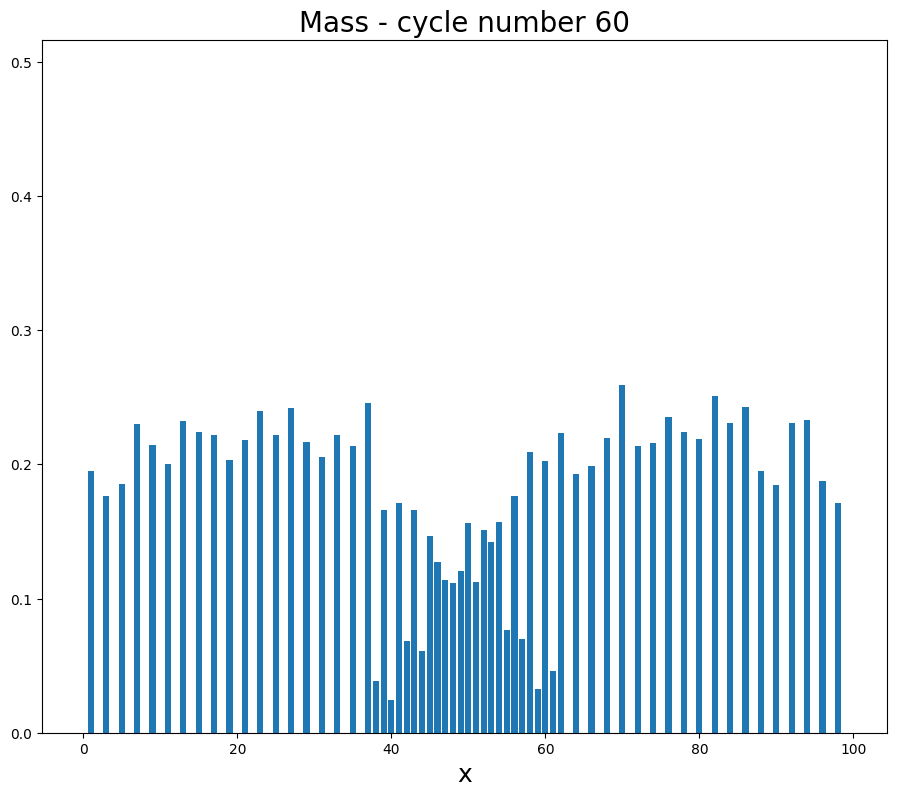} }}
        \subfloat[Mass at t=150\centering]{{\includegraphics[width=.2\linewidth]{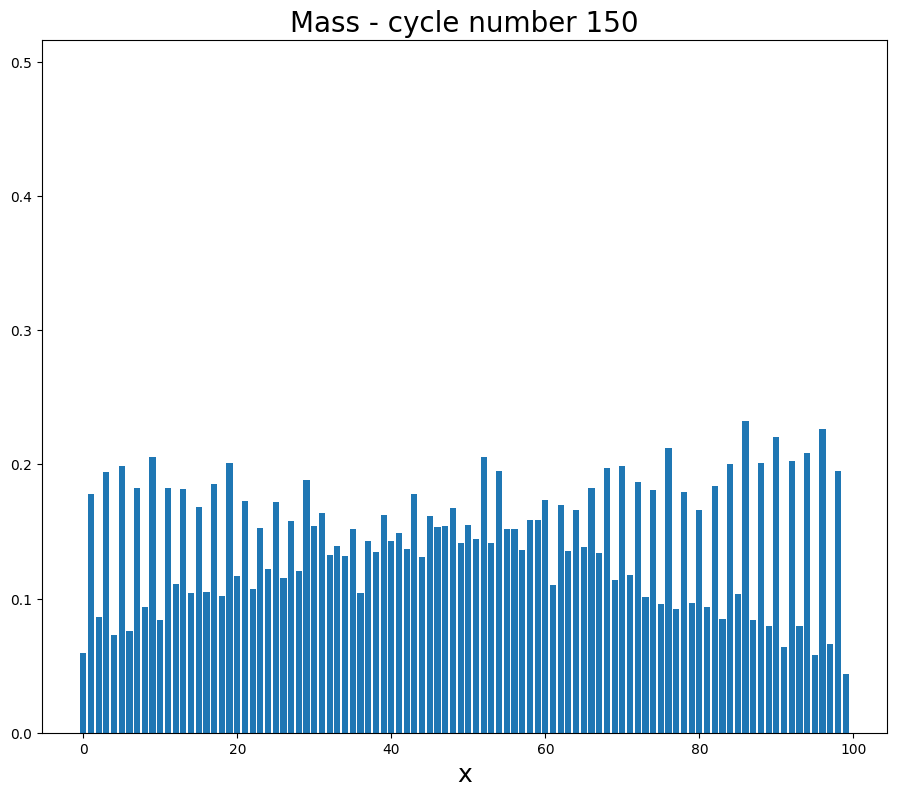} }}
        \quad
        \subfloat[$\Theta$ at t=30\centering]{{\includegraphics[width=.2\linewidth]{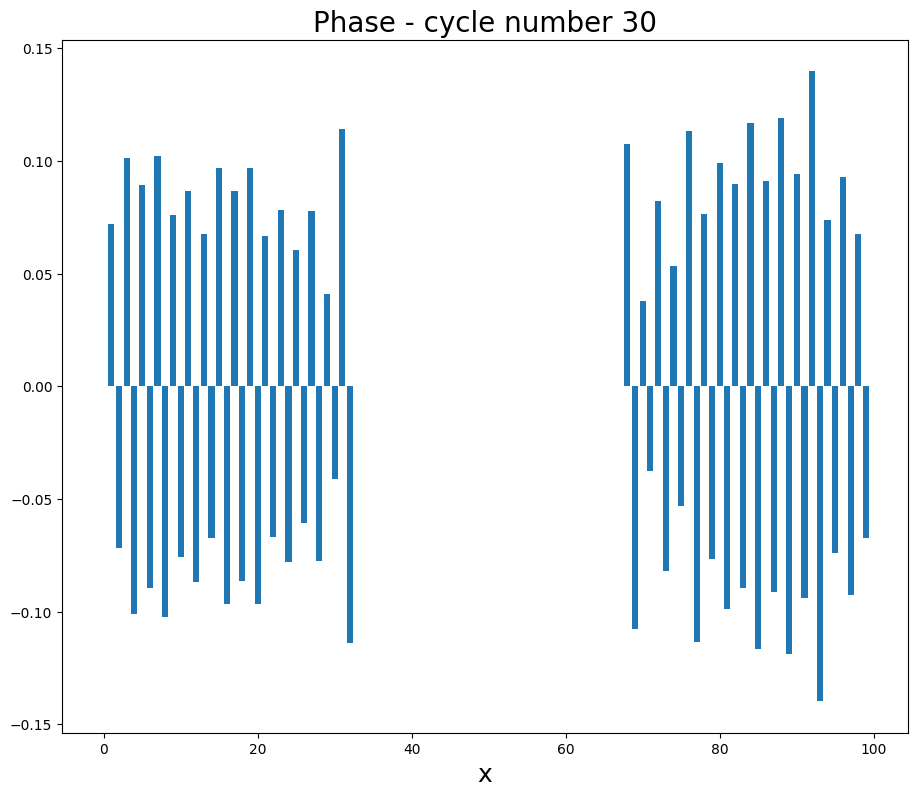} }}
        \subfloat[$\Theta$ at t=60\centering]{{\includegraphics[width=.2\linewidth]{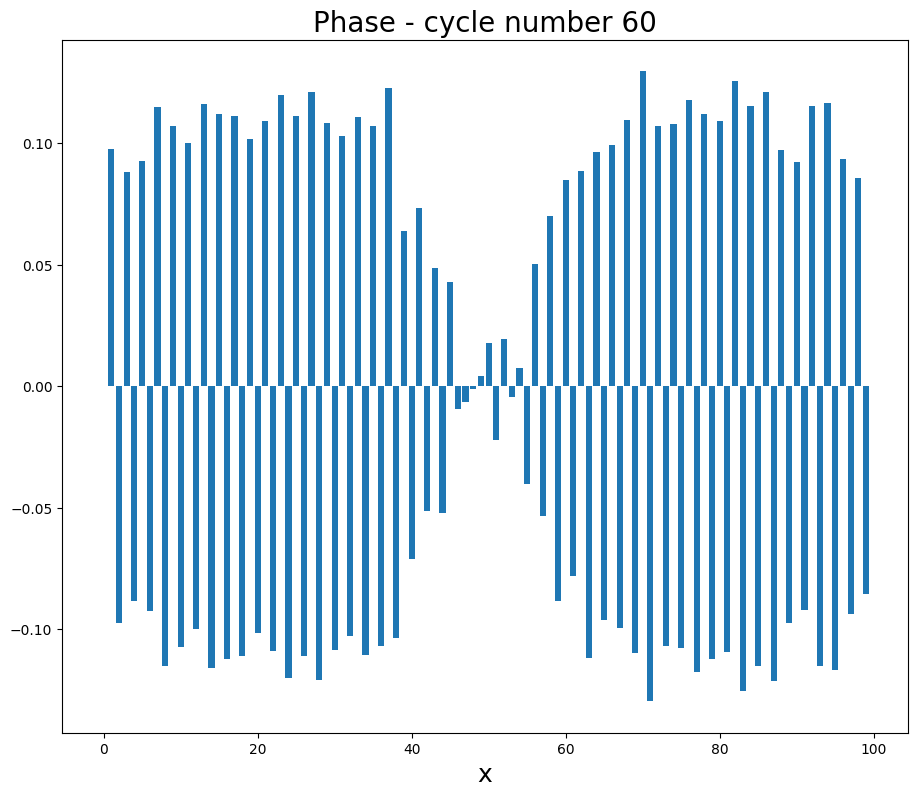} }}
        \subfloat[$\Theta$ at t=150\centering]{{\includegraphics[width=.2\linewidth]{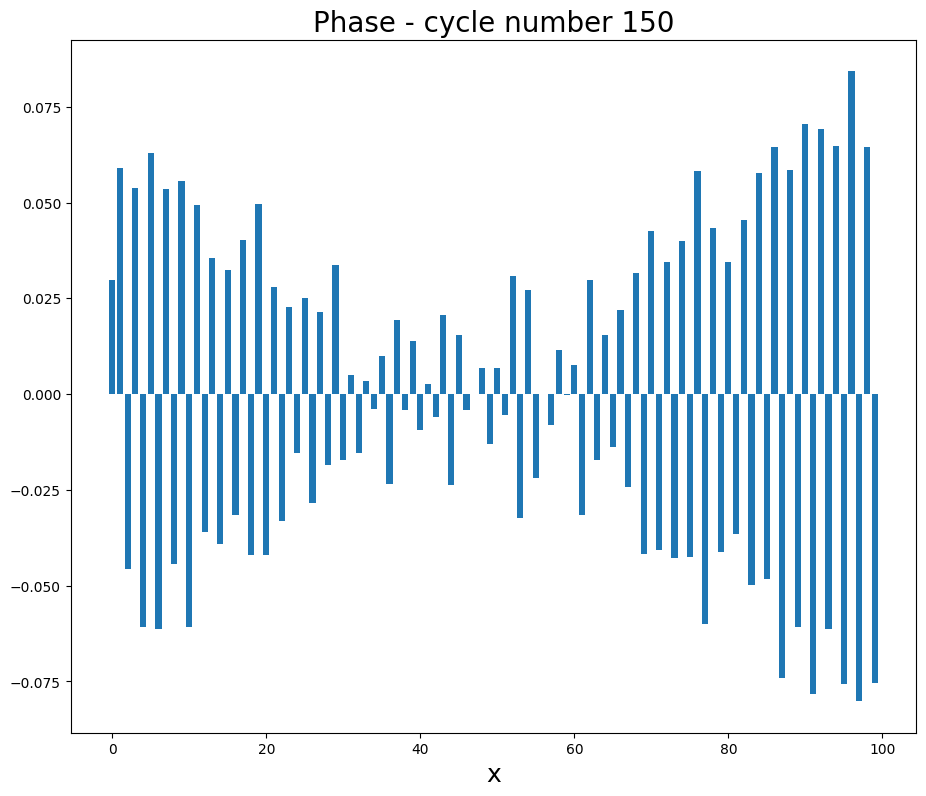} }}
        \quad
        \subfloat[$\lvert \psi\rvert$ at t=30\centering]{{\includegraphics[width=.2\linewidth]{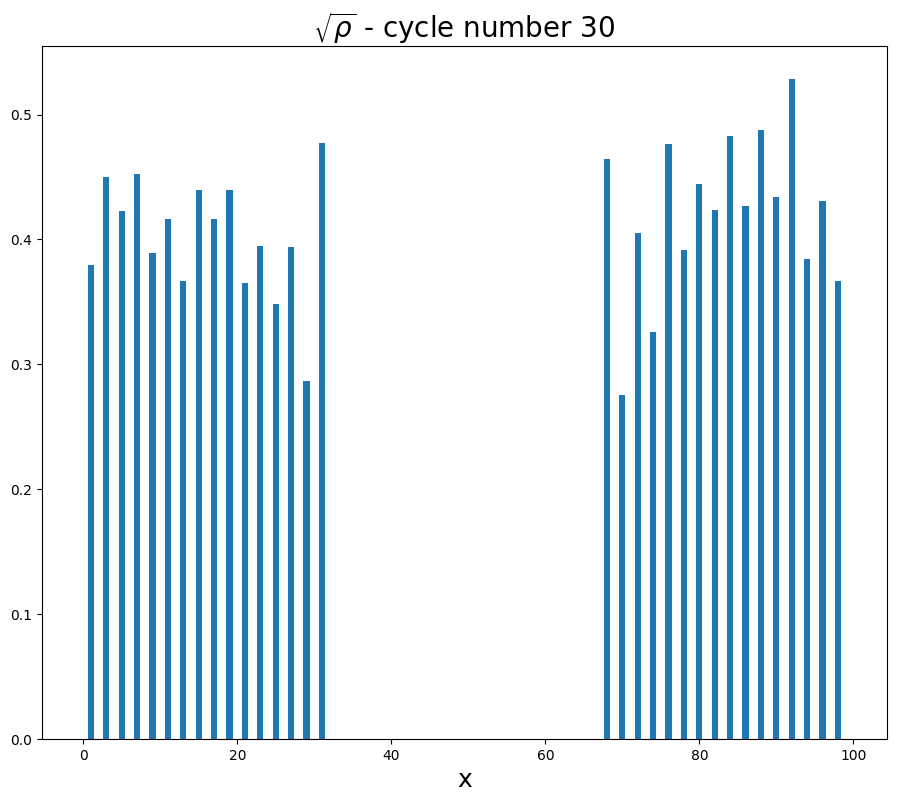} }}
        \subfloat[$\lvert \psi\rvert$ at t=60\centering]{{\includegraphics[width=.2\linewidth]{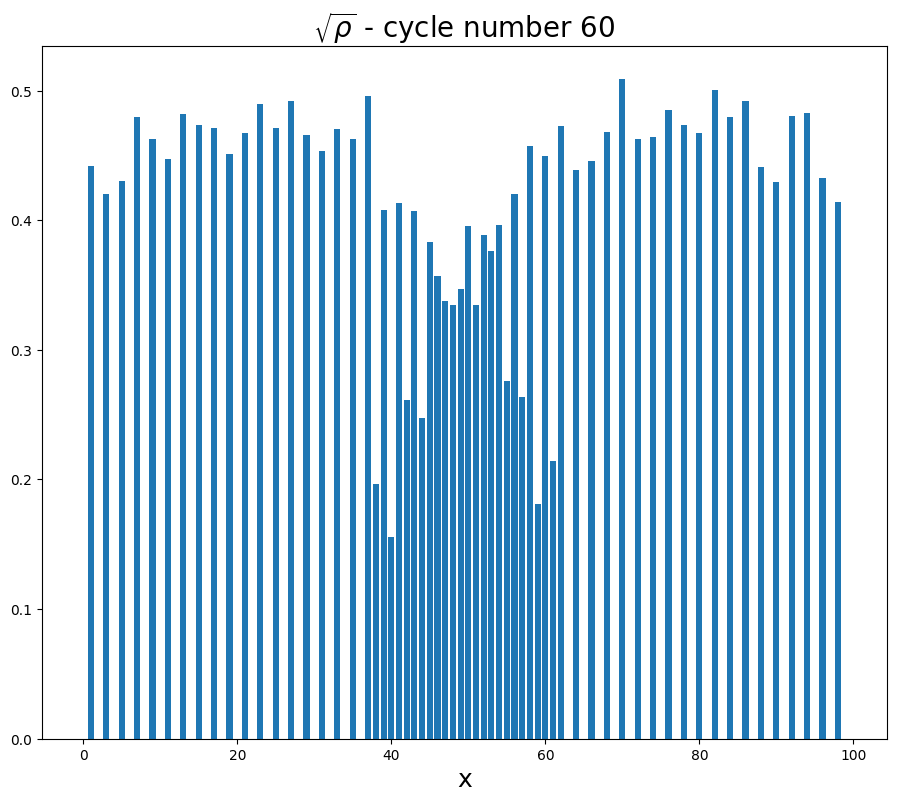} }}
        \subfloat[$\lvert \psi\rvert$ at t=150\centering]{{\includegraphics[width=.2\linewidth]{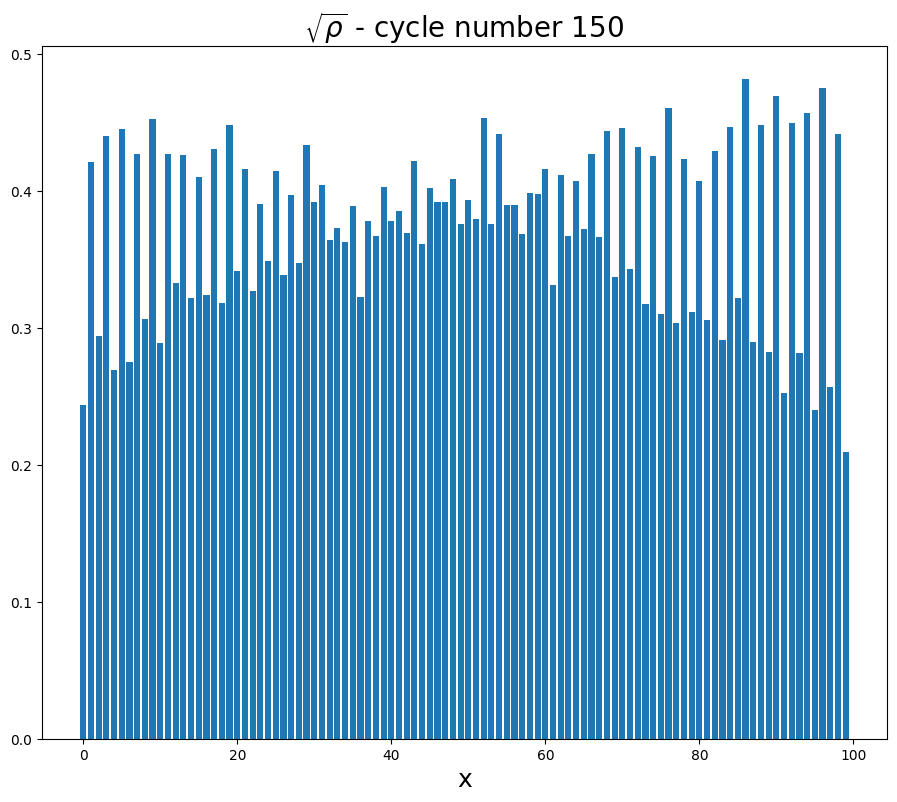} }}
        \quad
        \subfloat[$Re(\psi)$ at t=30\centering]{{\includegraphics[width=.2\linewidth]{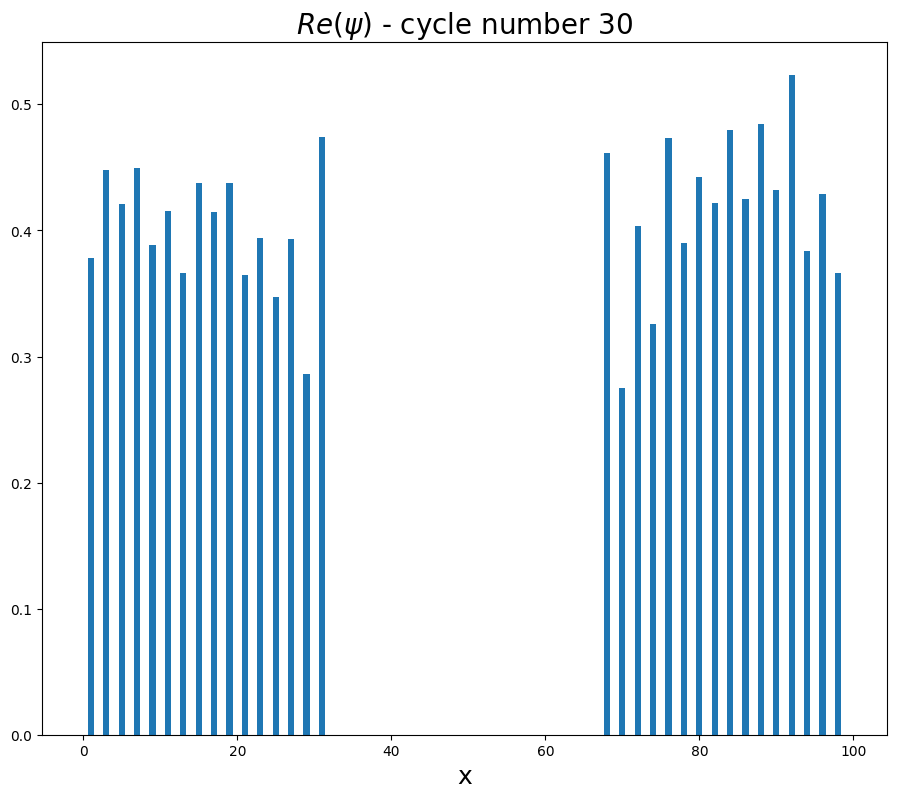} }}
        \subfloat[$Re(\psi)$ at t=60\centering]{{\includegraphics[width=.2\linewidth]{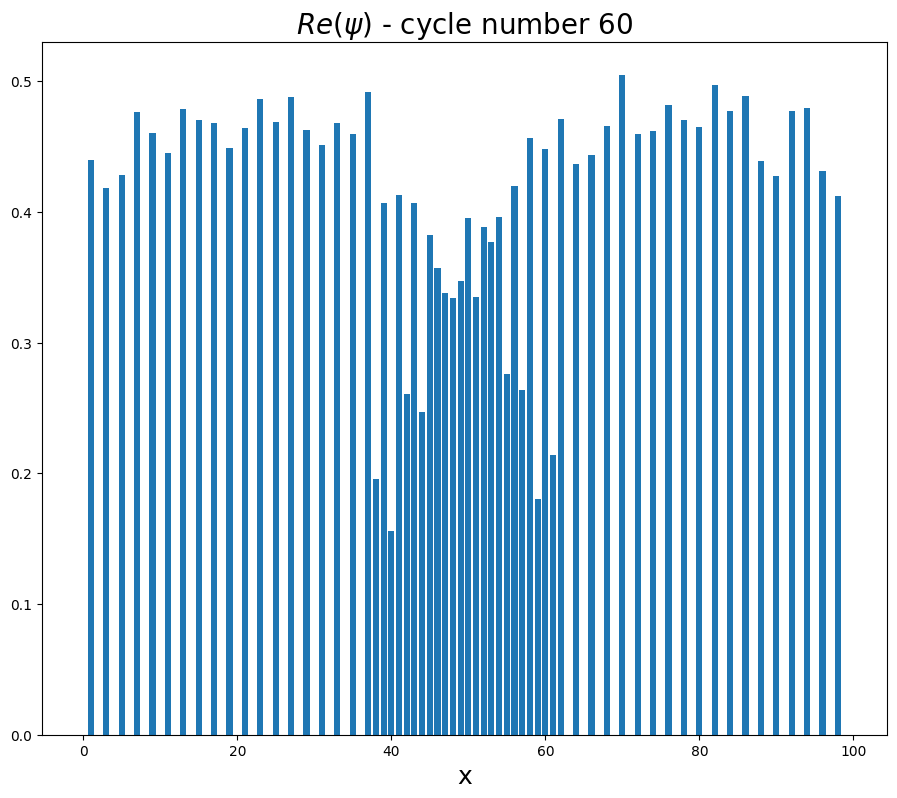} }}
        \subfloat[$Re(\psi)$ at t=150\centering]{{\includegraphics[width=.2\linewidth]{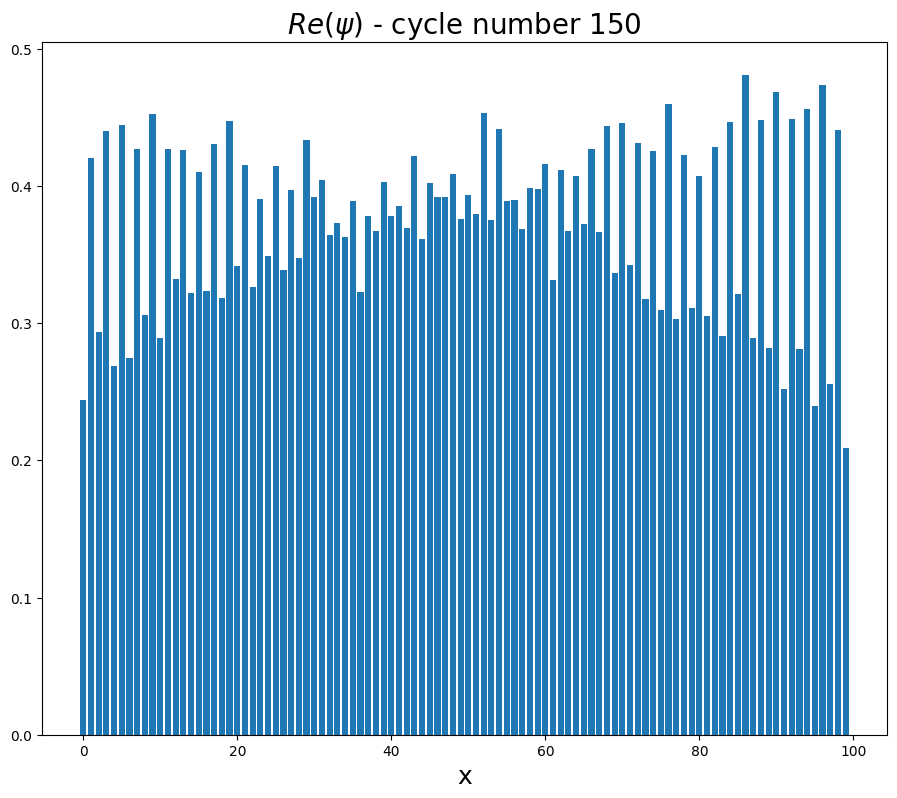} }}
        \quad
        \subfloat[$Im(\psi)$ at t=30\centering]{{\includegraphics[width=.2\linewidth]{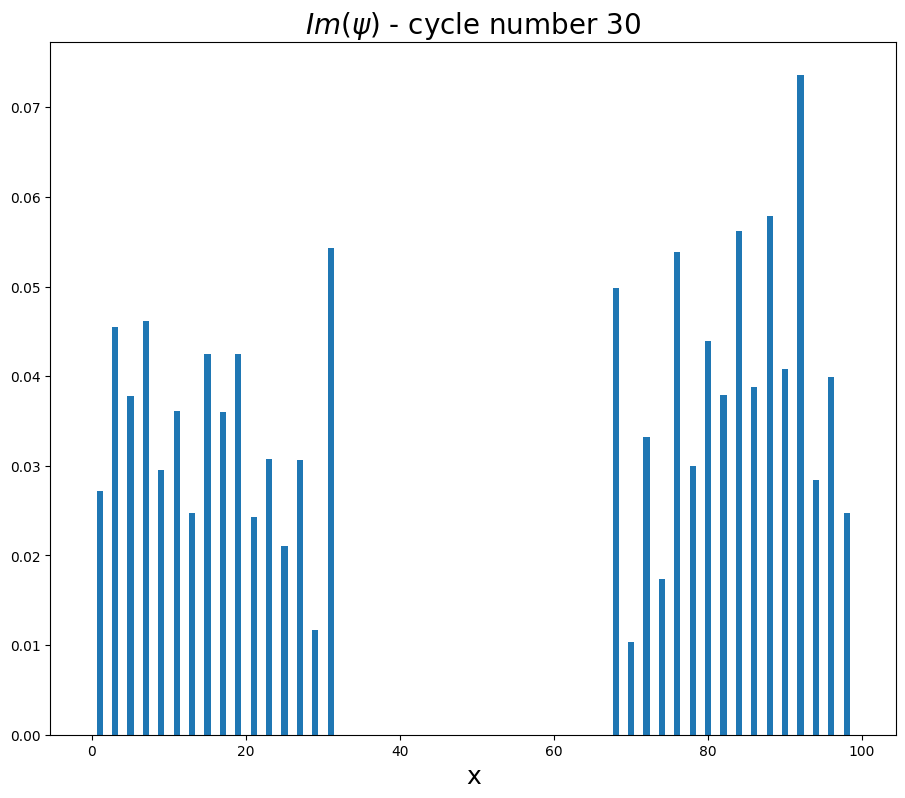} }}
        \subfloat[$Im(\psi)$ at t=60\centering]{{\includegraphics[width=.2\linewidth]{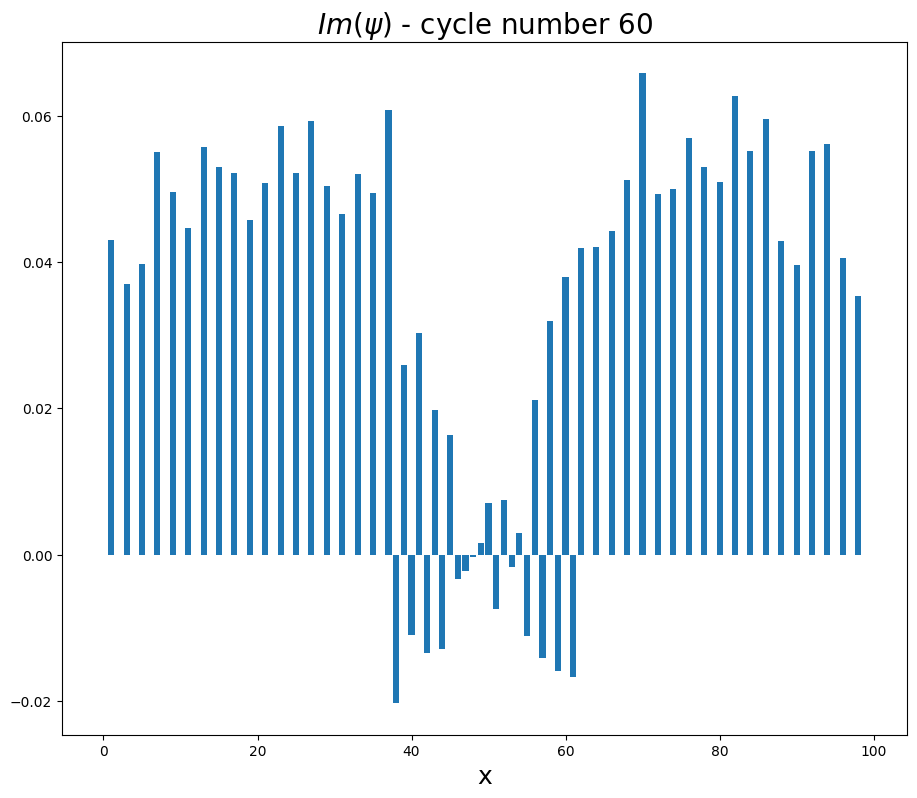} }}
        \subfloat[$Im(\psi)$ at t=150\centering]{{\includegraphics[width=.2\linewidth]{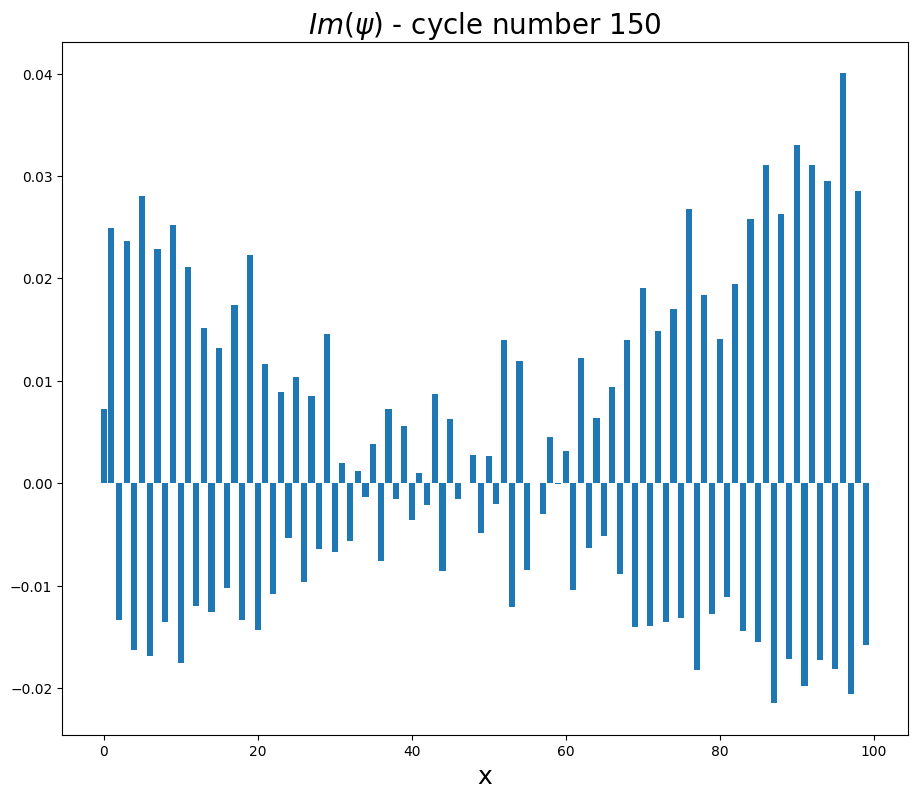} }}
        \caption{Evolution of mass distribution in real value one-dimensional Game of Life as given by (a),(b),(c) with initial situation given by Figure~\ref{fig:sc5}.}
        \label{fig:wavefunction_without_phase}
    \end{figure}

    Given equation~\ref{eq:potential}, we can numerically calculate the real and imaginary parts of the potential (for an initial structure shown in Fig.~\ref{fig:sc5}), as depicted in Fig.~\ref{fig:potential} for parameters $g=m=1$.
    Following previous analytical computation of complex value potential we arrive to equations that can be written as
    \begin{align}
        \frac{\hbar\rho_{,x,x}}{4m\rho}-\frac{\hbar(\rho_{,x})^2}{8m\rho^2}-\frac{\hbar g^2\left(\rho_{,x,x}\right)^2}{8m}-\frac{g\rho_{,x,t}}{2}+i\left(\frac{\hbar\rho_{,x}g\rho_{,x,x}}{4m\rho}+\frac{\hbar g\rho_{,x,x,x}}{4m}+\frac{\rho_{,t}}{2\rho}\right)=\frac{V}{\hbar},
    \end{align}
    where we use units such as
    \begin{align}
        \nonumber
        \frac{\hbar}{2m}\approx\frac{1}{2},
    \end{align}
    so we obtain simplified equation in the form as
    \begin{align}
        \frac{\rho_{,x,x}}{4\rho}-\frac{(\rho_{,x})^2}{8\rho^2}-\frac{g^2\left(\rho_{,x,x}\right)^2}{8}-\frac{g\rho_{,x,t}}{2}+i\left(\frac{\rho_{,x}g\rho_{,x,x}}{4\rho}+\frac{g\rho_{,x,x,x}}{4}+\frac{\rho_{,t}}{2\rho}\right)=V_{\hbar}
        \label{eq:potential2}
    \end{align}
    where $V_{\hbar}=\frac{V}{\hbar}$.
    Numerically we obtain the following dependence of complex value potential as depicted in Fig.~\ref{fig:potential2}.
    To complete the quantum picture of Stochastic Conway Game of Life, we present the evolution of the phase (calculated according to the equation~\ref{eq:phase}) and the modulus of the wave function ($\sqrt{\rho}=\lvert\psi\rvert$) of the system as shown in the Figure~\ref{fig:modulzfazy}.
    We recognize an explicit equivalence between the mass and the square of the wave function.
    \FloatBarrier

    \section{From quantum mechanics to generalized second Fick Law in one dimension (classical system mimicking quantum system)}
    \label{sec:from-quantum-mechanics-to-generalized-second-fick-law-in-one-dimension}
    We will assume second Fick law to be valid as expressed by formula~\ref{eq:fick}.
    We preassume existence of quantum mechanical process expressed by dissipative Schr\"{o}dinger equation (existence of complex value potential that is time-dependent) and we search for its implication on expected second Fick law that shall suppose to built bridge between $\frac{d}{dt}\rho$ and $\frac{d^2}{dx^2}\rho$.
    We start from Schr\"{o}dinger equation by plugging direct probability distribution $\rho$ accompanied with wave-function phase dependence $\Theta$ and we obtain
%

    \begin{align}
        -\frac{\hbar^2}{2m}\frac{d^2}{dx^2}(\sqrt{\rho}\exp(i\Theta))+V(x)\sqrt{\rho}\exp(i\Theta)=i\hbar \frac{d}{dt}[\sqrt{\rho}\exp(i\Theta)]= \nonumber \\
        =i\hbar \frac{d}{dt}\left[\rho \frac{\exp(i\Theta)}{\sqrt{\rho}}\right]= \frac{\exp(i\Theta)}{\sqrt{\rho}} i\hbar \frac{d}{dt} \rho - e^{i\Theta} \frac{1}{2}\frac{\rho}{(\rho)^{3/2}} i\hbar \frac{d}{dt}\rho -\hbar e^{i \Theta} \sqrt{\rho} \frac{d}{dt}\Theta
    \end{align}
    In next step we replace $\frac{d}{dt}\rho$ with $D(x,t)\frac{d^2}{dx^2}$ that leads to the equation of the form
    \begin{align}
        -\frac{\hbar^2}{2m}\frac{d^2}{dx^2}(\sqrt{\rho}e^{i\Theta})+V(x)\sqrt{\rho}e^{i\Theta}=i\hbar \frac{d}{dt}\left[\sqrt{\rho}e^{i\Theta}\right]=i\hbar \frac{d}{dt}\left[\rho \frac{1}{\sqrt{\rho}}e^{i\Theta}\right]= \nonumber \\
        \frac{e^{i\Theta}}{\sqrt{\rho}} i\hbar D \frac{d^2}{dx^2} \rho - i\hbar e^{i\Theta} \frac{1}{2}\frac{\rho}{(\rho)^{3/2}} D\frac{d^2}{dx^2}\rho -\hbar e^{i \Theta} \sqrt{\rho} \frac{d}{dt}\Theta
    \end{align}
    and finally we obtain equation incorporating diffusion coefficient $D(x,t)=\frac{\frac{d}{dt}\rho}{\frac{d^2}{dx^2}\rho}$ in linear way as
    \begin{align}
        -\frac{\hbar^2}{2m}\frac{d^2}{dx^2}(\sqrt{\rho}e^{i\Theta})+V(x)\sqrt{\rho}e^{i\Theta} +\hbar e^{i \Theta} \sqrt{\rho} \frac{d}{dt}\Theta= 
        \frac{e^{i\Theta}}{\sqrt{\rho}} i\hbar D \frac{d^2}{dx^2} \rho - i\hbar e^{i\Theta} \frac{1}{2}\frac{\rho}{(\rho)^{3/2}} D\frac{d^2}{dx^2}\rho
    \end{align}
    that brings
    \begin{align}
        -\frac{\hbar^2}{2m}\frac{d^2}{dx^2}(\sqrt{\rho}e^{i\Theta})+V(x)\sqrt{\rho}e^{i\Theta} +\hbar e^{i \Theta} \sqrt{\rho} \frac{d}{dt}\Theta= 
        D(x,t) \left[ \frac{e^{i\Theta}}{\sqrt{\rho}} i\hbar
        - i\hbar e^{i\Theta} \frac{1}{2}\frac{\rho}{(\rho)^{3/2}}  \right]  \frac{d^2}{dx^2}\rho
    \end{align}
    and thus we have
    \begin{align}
        -e^{-i\Theta}\frac{\hbar^2}{2m}\frac{d^2}{dx^2}(\sqrt{\rho}e^{i\Theta})+V(x)\sqrt{\rho} +\hbar  \sqrt{\rho} \frac{d}{dt}\Theta= 
        D \Big[ \frac{1}{\sqrt{\rho}} i\hbar
        - i\hbar \frac{1}{2}\frac{\rho}{(\rho)^{3/2}}  \Big]  \frac{d^2}{dx^2}\rho
    \end{align}
    and finally we have
    \begin{align}
        \frac{-e^{-i*\Theta}\frac{\hbar^2}{2m}\frac{d^2}{dx^2}(\sqrt{\rho}e^{i\Theta})+V(x)\sqrt{\rho} +\hbar  \sqrt{\rho} \frac{d}{dt}\Theta }{\Big[ \frac{1}{\sqrt{\rho}} i\hbar
        - i\hbar \frac{1}{2}\frac{\rho^{2/2}}{(\rho)^{3/2}}  \Big]  \frac{d^2}{dx^2}\rho}=D
    \end{align}
    Potential V can be complex but coefficient D is real value.
    Finally we have
    \begin{align}
        \frac{-(\rho)^{1/2}e^{-i\Theta}\frac{\hbar^2}{2m}\frac{d^2}{dx^2}(\sqrt{\rho}e^{i\Theta})+V(x)\rho +\hbar  \rho \frac{d}{dt}\Theta }{\frac{1}{2} i\hbar \frac{d^2}{dx^2}\rho}=D
    \end{align}
    and with
    \begin{align}
        \nonumber
        \frac{\hbar}{2m}\approx\frac{1}{2}
    \end{align}
    we obtain
    \begin{align}
        \frac{-(\rho)^{1/2}e^{-i\Theta}\frac{1}{2}\frac{d^2}{dx^2}(\sqrt{\rho}e^{i\Theta})+V_{\hbar}(x)\rho +\rho \frac{d}{dt}\Theta }{\frac{1}{2} i \frac{d^2}{dx^2}\rho}=D
    \end{align}
Consequently we have

    \begin{align}
        \frac{i(\frac{1}{2}\rho_{,x,x}+\frac{1}{4\rho}(\rho_{,x})^2-\rho(\Theta_{re,x}+i\Theta_{im,x})^2 -2(V_{\hbar, re}(x)+iV_{\hbar,im}(x))\rho -\rho \frac{d}{dt}(\Theta_{re}+i\Theta_{im})) }{ \frac{d^2}{dx^2}\rho}
        \nonumber \\
        -
        \frac{(\frac{1}{2}\rho_{,x}(\Theta_{re,x}+i\Theta_{im,x}) +2\rho(\Theta_{re,x,x}+i\Theta_{im,x,x}))}{ \frac{d^2}{dx^2}\rho}
        =D
    \end{align}

    We recognize the existence of two phase components. First component is real value $\Theta_{re}(t)$ and second component has $\Theta_{im}(t)$.

   The equation for phase evolution with time is given as
    \begin{align}
        i \hbar \frac{d}{dt}(e^{i(\Theta_{re}(t)+i\Theta_{im}(t))}\sqrt{\rho(x,t_0)})=(E_{re}(t)+iE_{im}(t))(\sqrt{\rho(x,t_0)}e^{i(\Theta_{re}(t)+i\Theta_{im}(t))})
    \end{align}
    and consequently
    \begin{align}
        i \hbar \frac{d}{dt}(e^{i(\Theta_{re}(t)+i\Theta_{im}(t))}\sqrt{\rho(x,t_0)})=(-\frac{\hbar^2}{2m}\frac{d^2}{dx^2}+V_{re}(t)+iV_{im}(t))(\sqrt{\rho(x,t_0)}e^{i(\Theta_{re}(t)+i\Theta_{im}(t))})
    \end{align}
    or

    \begin{align}
        i \hbar \frac{d}{dt}(e^{i(\Theta_{re}(t)+i\Theta_{im}(t))}\sqrt{\rho(x,t_0)})=
        -(e^{i(\Theta_{re}(t)+i\Theta_{im}(t))}\sqrt{\rho(x,t_0)})(\frac{d}{dt}\Theta_{re}(x,t)+\frac{d}{dt}\Theta_{im}(t))
        =
        \nonumber \\
        =(-\frac{\hbar^2}{2m}\frac{d^2}{dx^2}+V_{re}(x,t)+iV_{im}(x,t))(\sqrt{\rho(x,t_0)}e^{i(\Theta_{re}(x,t)+i\Theta_{im}(t))}).
    \end{align}
    We finally arrive to
    \begin{align}
        -\hbar (\frac{d}{dt}\Theta_{re}(t)+i\frac{d}{dt}\Theta_{im}(t))
        =
        -(e^{-i(\Theta_{re}(t)+i\Theta_{im}(t))}\frac{1}{\sqrt{\rho(x,t_0)}})(\frac{d}{dt}\Theta_{re}(x,t)+\frac{d}{dt}\Theta_{im}(t)) \nonumber \\
        (-\frac{\hbar^2}{2m}\frac{d^2}{dx^2}+V_{re}(x,t)+iV_{im}(x,t))
        \nonumber \\ (\sqrt{\rho(x,t_0)}e^{i(\Theta_{re}(x,t)+i\Theta_{im}(t))})
    \end{align}
    or
    \begin{align}
        -\hbar i\frac{d}{dt}\Theta_{im}(t)
        =
        -(e^{-i(\Theta_{re}(t)}\frac{1}{\sqrt{\rho(x,t_0)}})(\frac{d}{dt}\Theta_{re}(x,t)+\frac{d}{dt}\Theta_{im}(t)) \nonumber \\
        (-\frac{\hbar^2}{2m}\frac{d^2}{dx^2}+V_{re}(x,t)+iV_{im}(x,t))
        \nonumber \\ (\sqrt{\rho(x,t_0)}e^{i(\Theta_{re}(x,t))})+\hbar (\frac{d}{dt}\Theta_{re})
    \end{align}
    and
      $  -\hbar i\frac{d}{dt}\Theta_{im}(t) = -(e^{-i(\Theta_{re}(t)}\frac{1}{\sqrt{\rho(x,t_0)}})(\frac{d}{dt}\Theta_{re}(x,t)+\frac{d}{dt}\Theta_{im}(t)) $ 
       $ (-\frac{\hbar^2}{2m}\frac{d^2}{dx^2}+V_{re}(x,t)+iV_{im}(x,t))$  
     $   (\sqrt{\rho(x,t_0)}e^{i(\Theta_{re}(x,t))})+\hbar (\frac{d}{dt}(\sqrt{\rho}\frac{d}{dx}\sqrt{\rho}))) $.
    We set $\Theta_{im}(t_0)=0$ at initial time and $\Theta_{im}(t_0)$ accounts for cell creationism or annilationism.
    Alternatively always setting $\Theta_{im}(t)=0$ we deal with time-dependent $\rho$.
    In such a way we obtain
    \begin{align}
        -\frac{-\frac{i\rho_{,x,x}}{2}+\frac{i(\rho_{,x})^2}{4\rho}+\rho_{,x}\Theta_{,x}+\rho\Theta_{,x,x}-i\rho(\Theta_{,x})^2+V_{\hbar}(x)\rho +\rho \frac{d}{dt}\Theta }{ \frac{d^2}{dx^2}\rho}=D,
    \end{align}
    and this expression is used in conducted simulations.
    We can extract real values from proposed diffusion coefficient given by equation
    \begin{align}
        \frac{-\rho_{,x}\Theta_{,x}-\rho\Theta_{,x,x}+Im\left[V_{\hbar}(x)\right]\rho}{\frac{d^2}{dx^2}\rho}=Re[D]
    \end{align}
    and imaginary values from proposed diffusion coefficient given by equation
    \begin{align}
        \frac{\frac{\rho_{,x,x}}{2}-\frac{(\rho_{,x})^2}{4\rho}-\rho(\Theta_{,x})^2-Re\left[V_{\hbar}(x)\right]\rho -\rho \frac{d}{dt}\Theta }{\frac{d^2}{dx^2}\rho}=Im[D].
    \end{align}
    \FloatBarrier
    \begin{figure}
        \centering
        \subfloat[Mass at t=30\centering]{{\includegraphics[width=.3\linewidth]{figures/sc5_mass_1} }}
        \subfloat[Mass at t=60\centering]{{\includegraphics[width=.3\linewidth]{figures/sc5_mass_2} }}
        \subfloat[Mass at t=150\centering]{{\includegraphics[width=.3\linewidth]{figures/sc5_mass_3} }}
        \quad
        \subfloat[$\mathfrak{R}(D)$ at t=30\centering]{{\includegraphics[width=.3\linewidth]{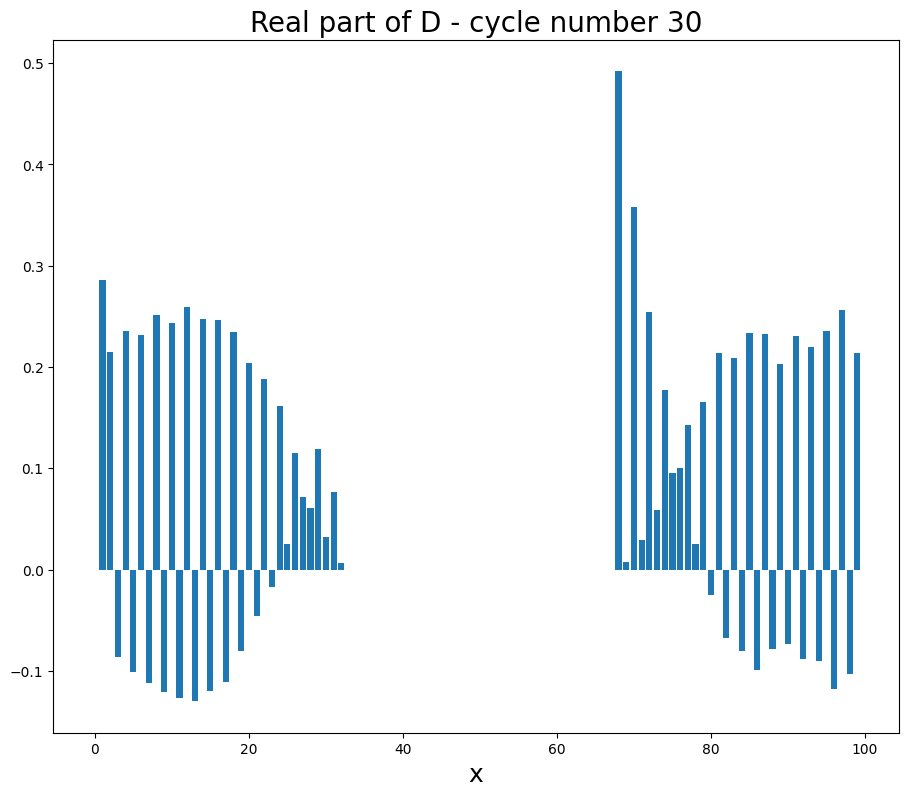} }}
        \subfloat[$\mathfrak{R}(D)$ at t=60\centering]{{\includegraphics[width=.3\linewidth]{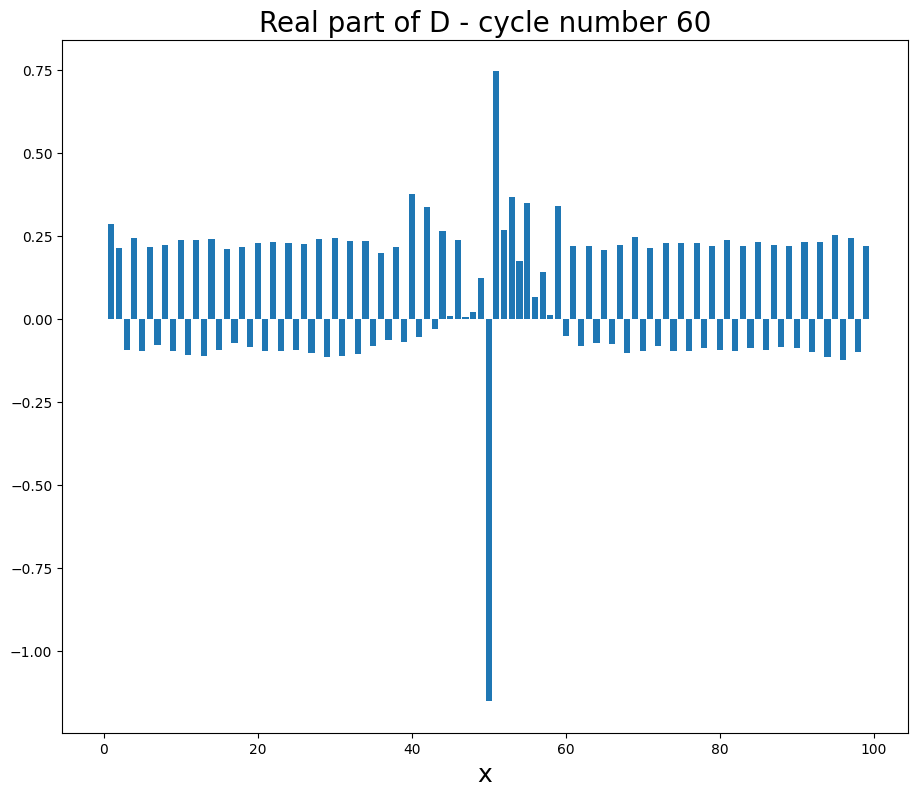} }}
        \subfloat[$\mathfrak{R}(D)$ at t=150\centering]{{\includegraphics[width=.3\linewidth]{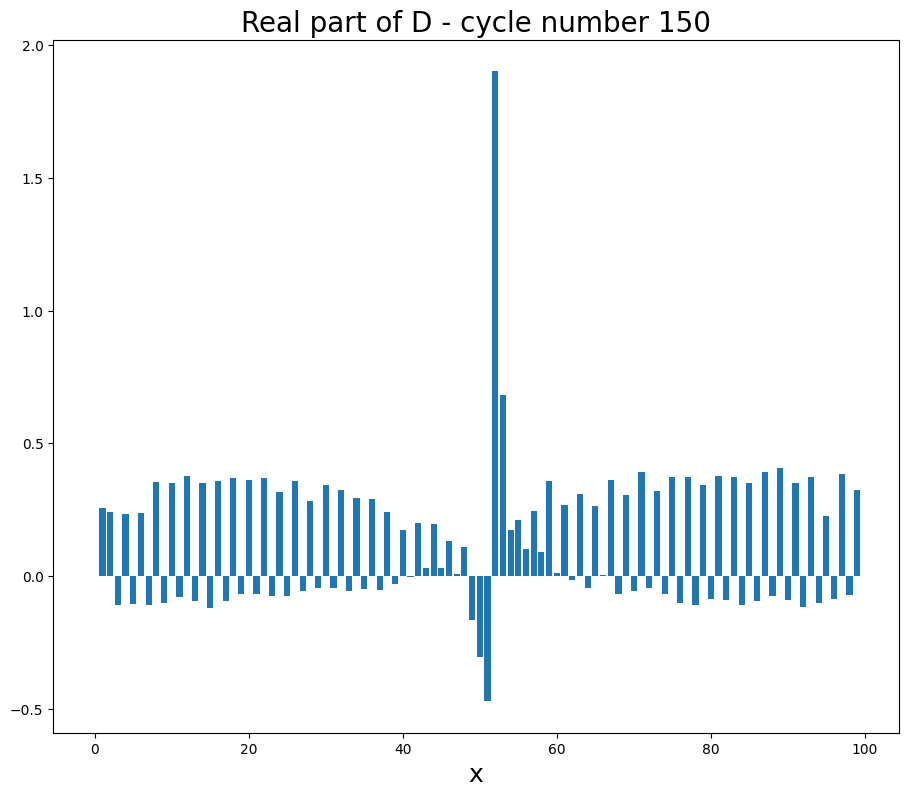} }}
        \quad
        \subfloat[$\mathfrak{I}(D)$ at t=30\centering]{{\includegraphics[width=.3\linewidth]{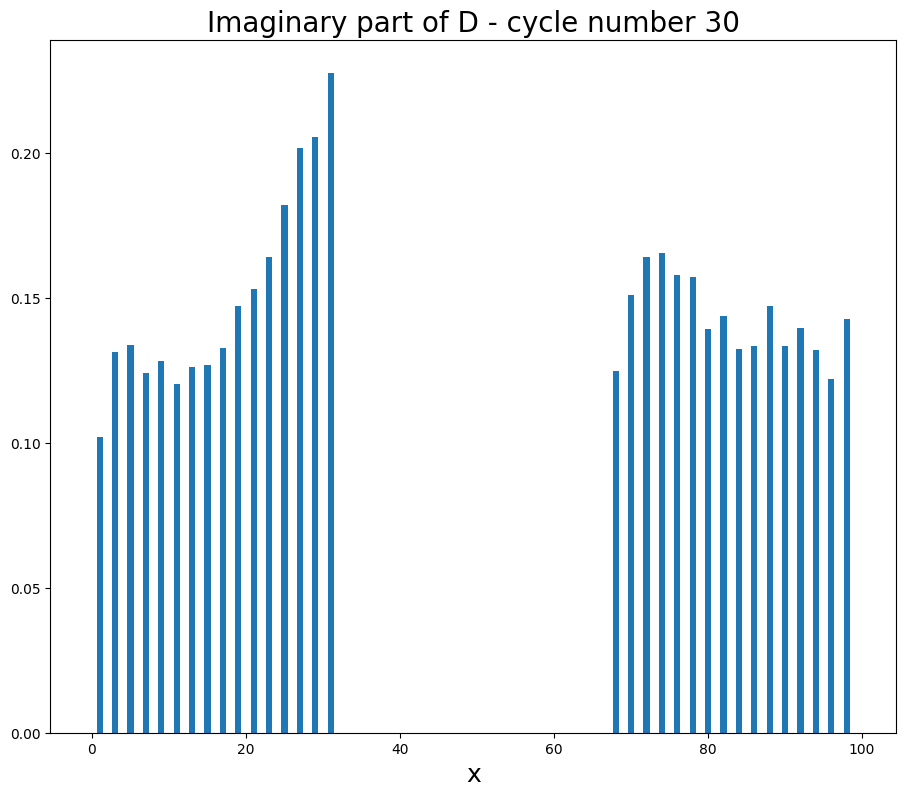} }}
        \subfloat[$\mathfrak{I}(D)$ at t=60\centering]{{\includegraphics[width=.3\linewidth]{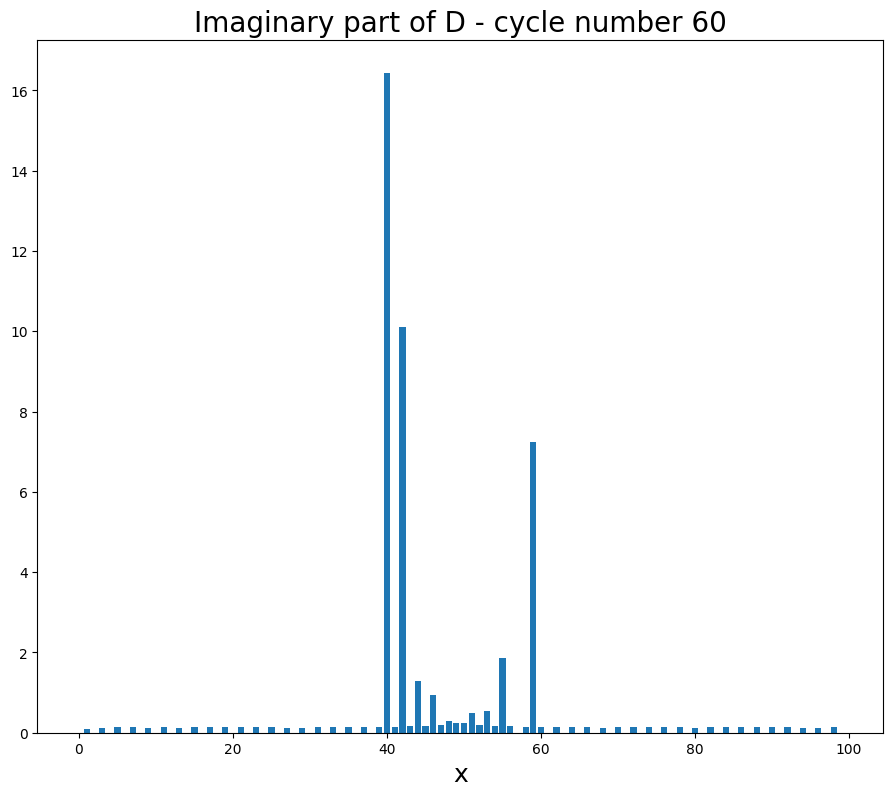} }}
        \subfloat[$\mathfrak{I}(D)$ at t=150\centering]{{\includegraphics[width=.3\linewidth]{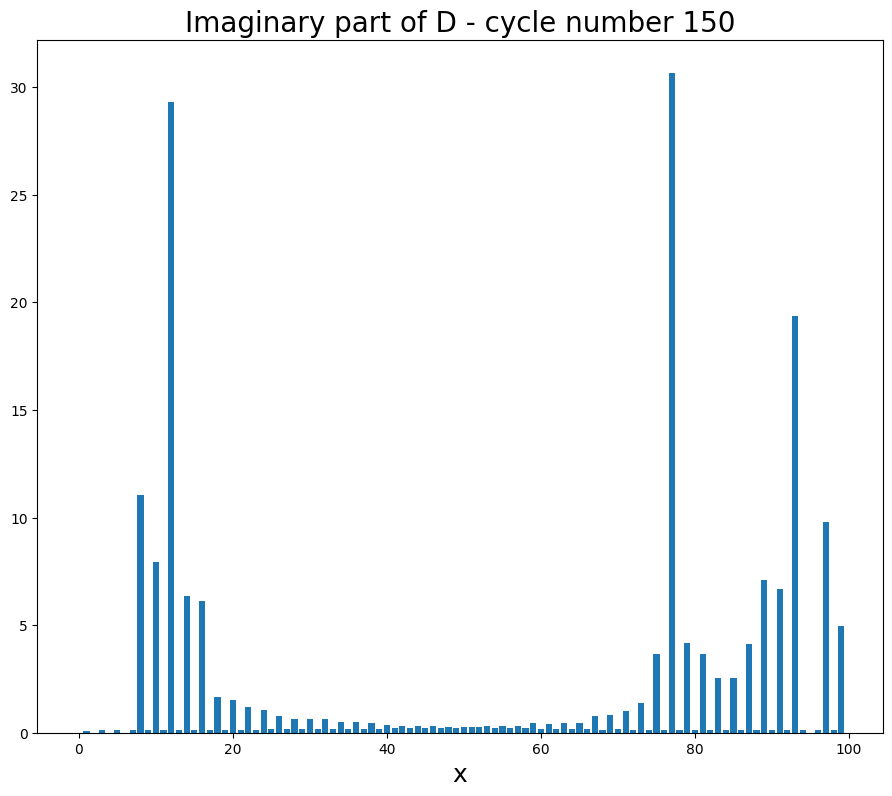} }}
        \caption{Anomalous dependence of diffusion coefficient from Fig.~\ref{fig:potential}.}
        \label{fig:D}
    \end{figure}
    \begin{figure}
        \centering
        \subfloat[Mass at t=30\centering]{{\includegraphics[width=.3\linewidth]{figures/sc5_mass_1_no_phase} }}
        \subfloat[Mass at t=60\centering]{{\includegraphics[width=.3\linewidth]{figures/sc5_mass_2_no_phase} }}
        \subfloat[Mass at t=150\centering]{{\includegraphics[width=.3\linewidth]{figures/sc5_mass_3_no_phase} }}
        \quad
        \subfloat[$\mathfrak{R}(D)$ at t=30\centering]{{\includegraphics[width=.3\linewidth]{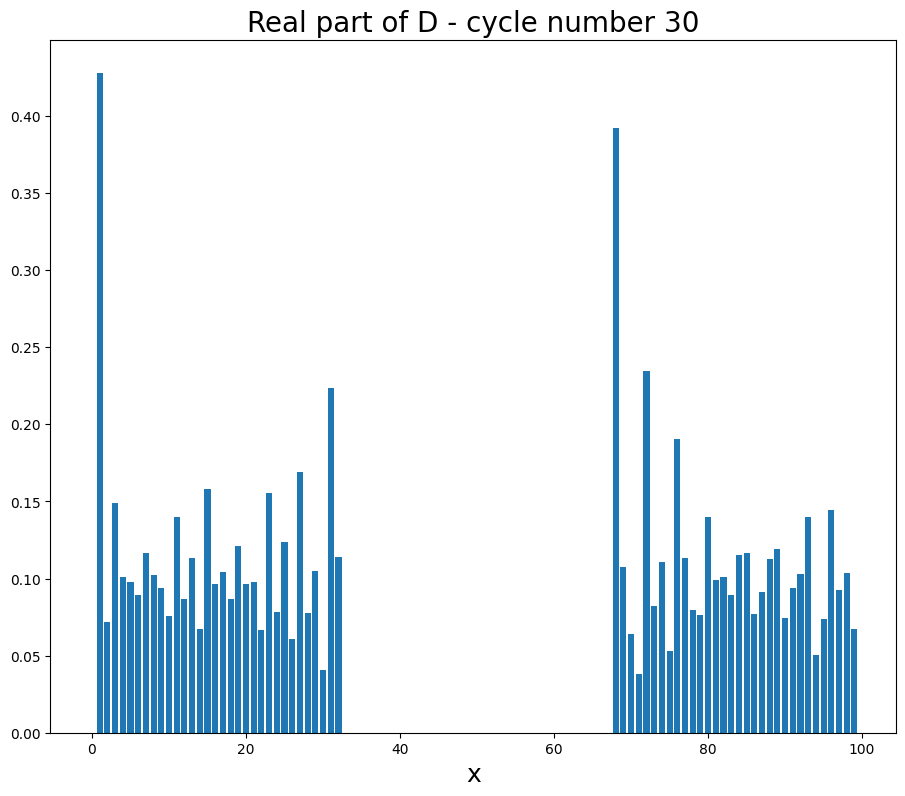} }}
        \subfloat[$\mathfrak{R}(D)$ at t=60\centering]{{\includegraphics[width=.3\linewidth]{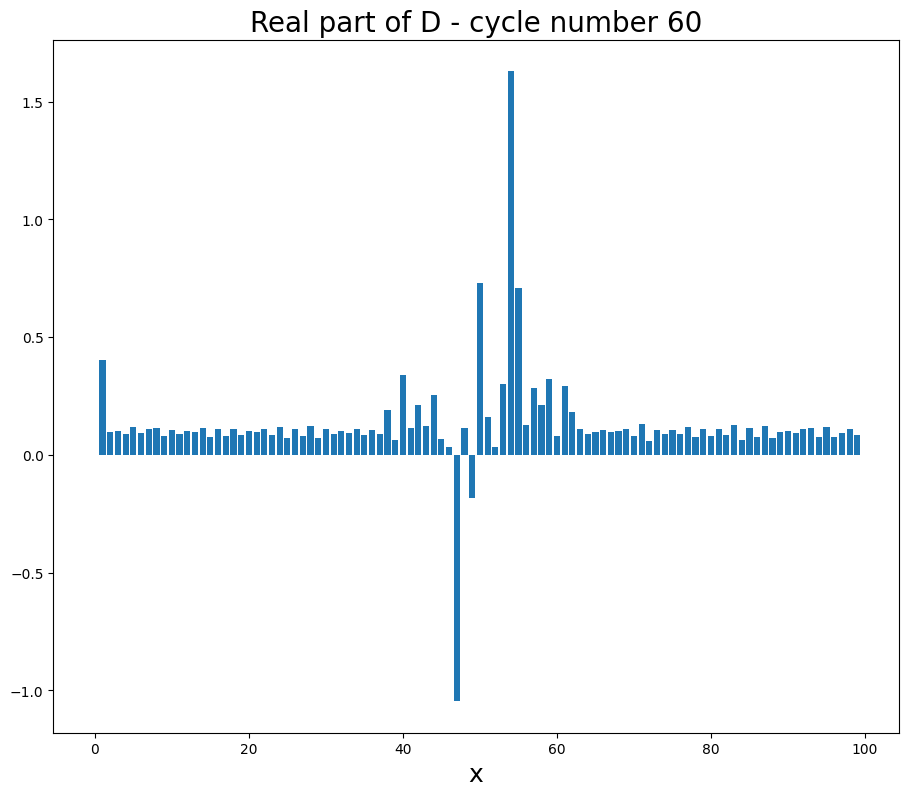} }}
        \subfloat[$\mathfrak{R}(D)$ at t=150\centering]{{\includegraphics[width=.3\linewidth]{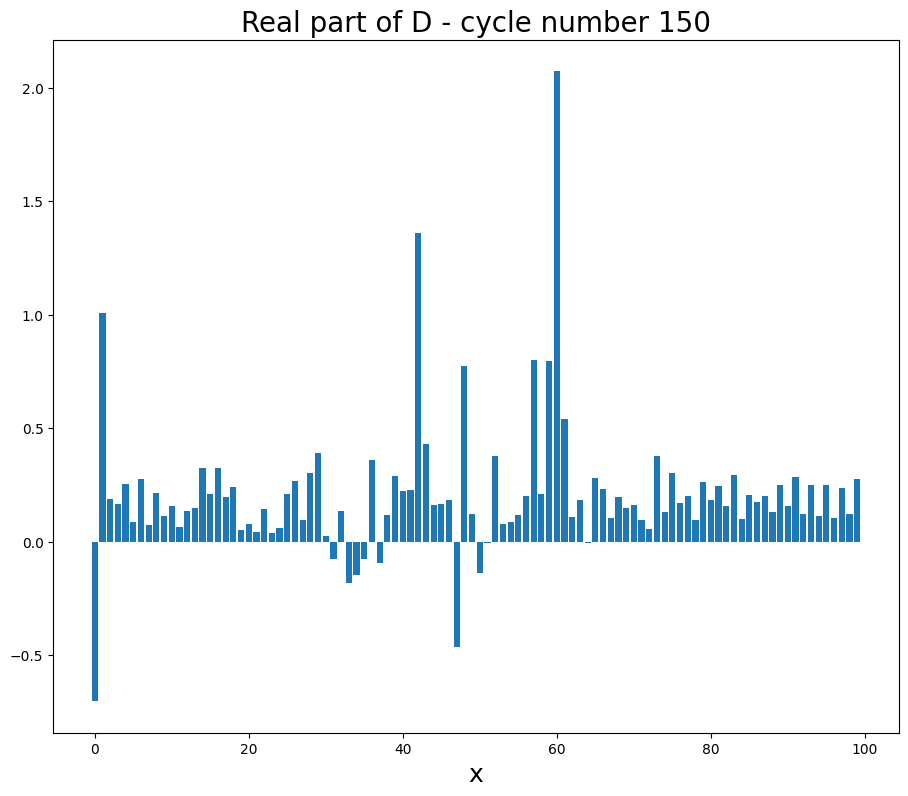} }}
        \caption{Anomalous dependence of diffusion coefficient in real value Conway Game of Life.}
        \label{fig:D_no_phase}
    \end{figure}
    \FloatBarrier

    \section{Concept of complex valued mass in Stochastic Conway Game of life}
    \label{sec:concept-of-complex-valued-mass-in-stochastic-conway-game-of-life}
\begin{figure}
        \centering
        \subfloat[\centering]{{\includegraphics[width=.25 \linewidth]{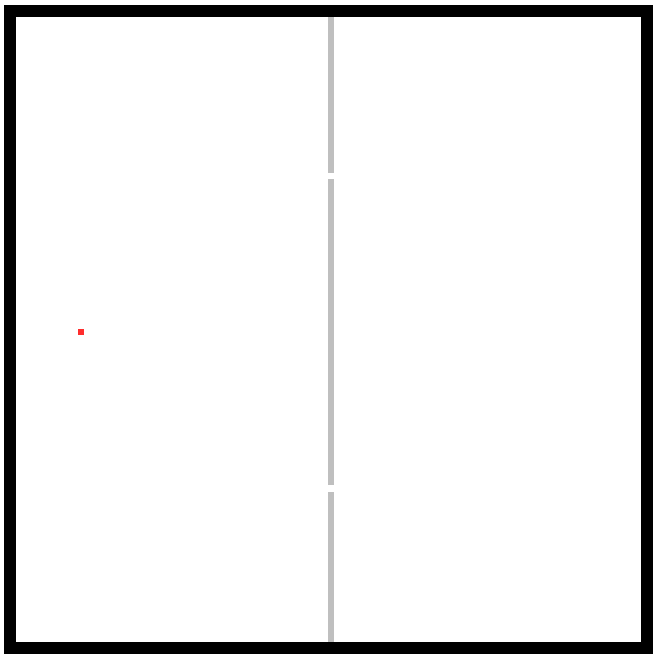} }}
        \qquad\qquad
        \subfloat[\centering]{{\includegraphics[width=.25 \linewidth]{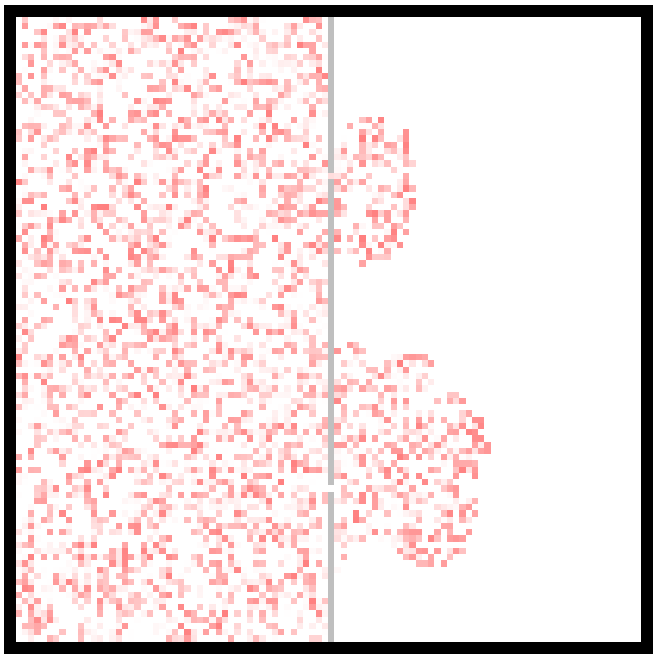} }}
        \caption{Diffusion process in the Stochastic Conway Game of Life of two weakly interconnected chambers connected by two small holes in the barrier (what express perturbative interactions between two reservoirs).}
        \label{fig:konf14}
    \end{figure}

\begin{figure}
        \centering
        \subfloat[\centering]{{\includegraphics[height=.3 \linewidth]{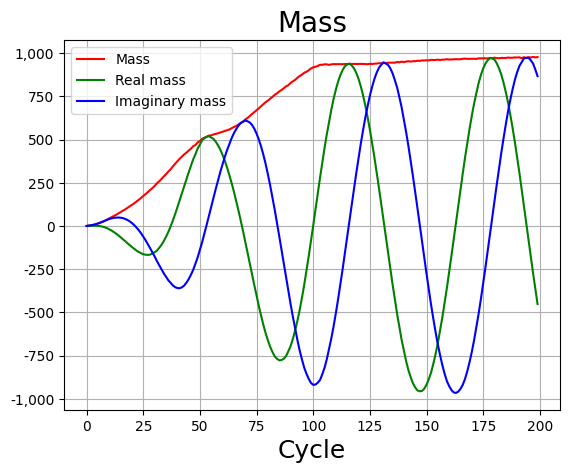} }}
        \quad
        \subfloat[\centering]{{\includegraphics[height=.3 \linewidth]{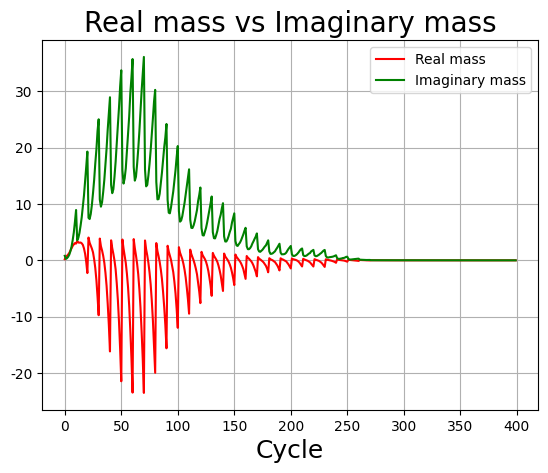} }}
        \caption{Evolution of mass in time with distinction of real and imaginary parts in the Stochastic Conway Game of Life (left picture). Evolution of mass over time with the distinction of real and imaginary parts in the Stochastic Game of Life with a 2 by 2 square averaging procedure every 10 time steps (right picture). Strong anti-correlation of the real and imaginary parts of mass of the cellular automata system is reported, which is quite analogous to the evolution of the wave function over time in the case of the Schr\"{o}dinger equation.}
        \label{fig:konf14_mass}
    \end{figure}

\begin{figure}
        \centering
        \subfloat[$\lvert mass\rvert$ at t=30\centering]{{\includegraphics[width=.2\linewidth]{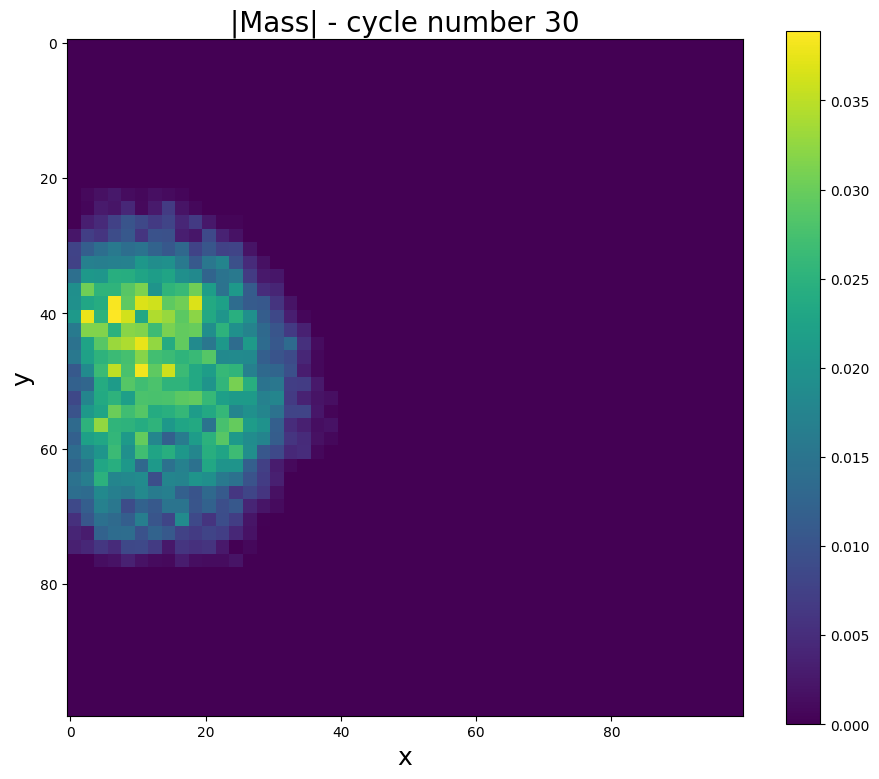} }}
        \subfloat[$\lvert mass\rvert$ at t=80\centering]{{\includegraphics[width=.2\linewidth]{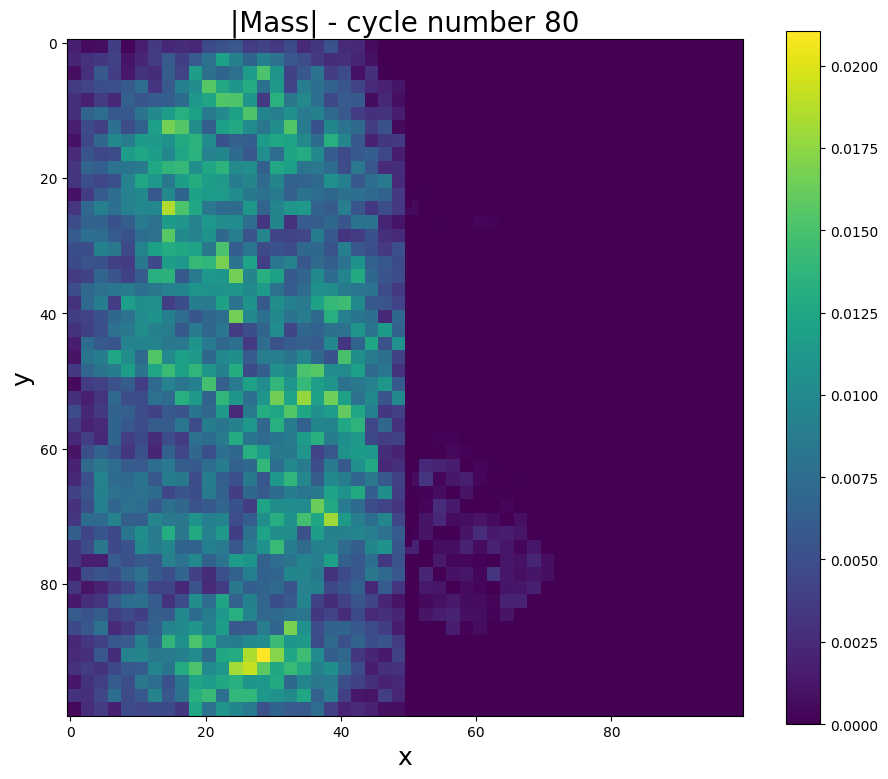} }}
        \subfloat[$\lvert mass\rvert$ at t=130\centering]{{\includegraphics[width=.2\linewidth]{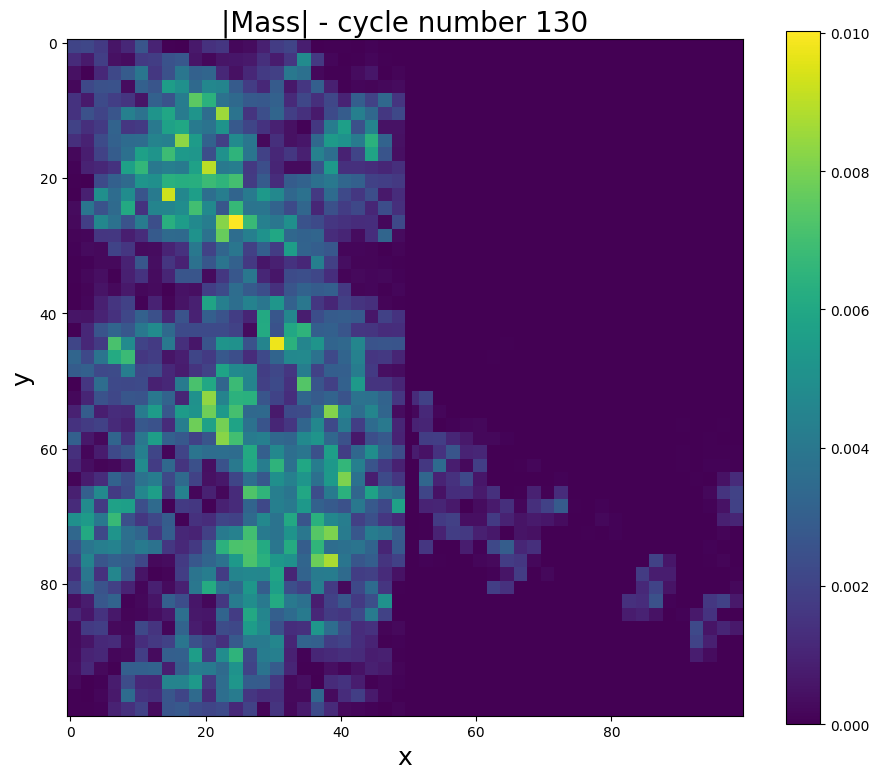} }}
        \qquad
        \subfloat[Real mass at t=30\centering]{{\includegraphics[width=.2\linewidth]{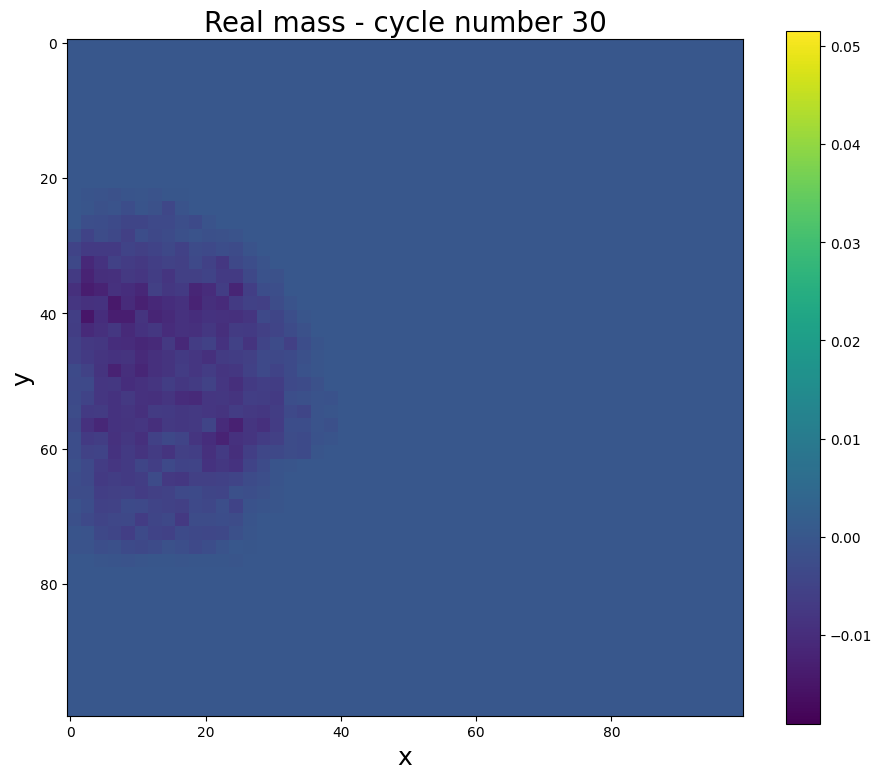} }}
        \subfloat[Real mass at t=80\centering]{{\includegraphics[width=.2\linewidth]{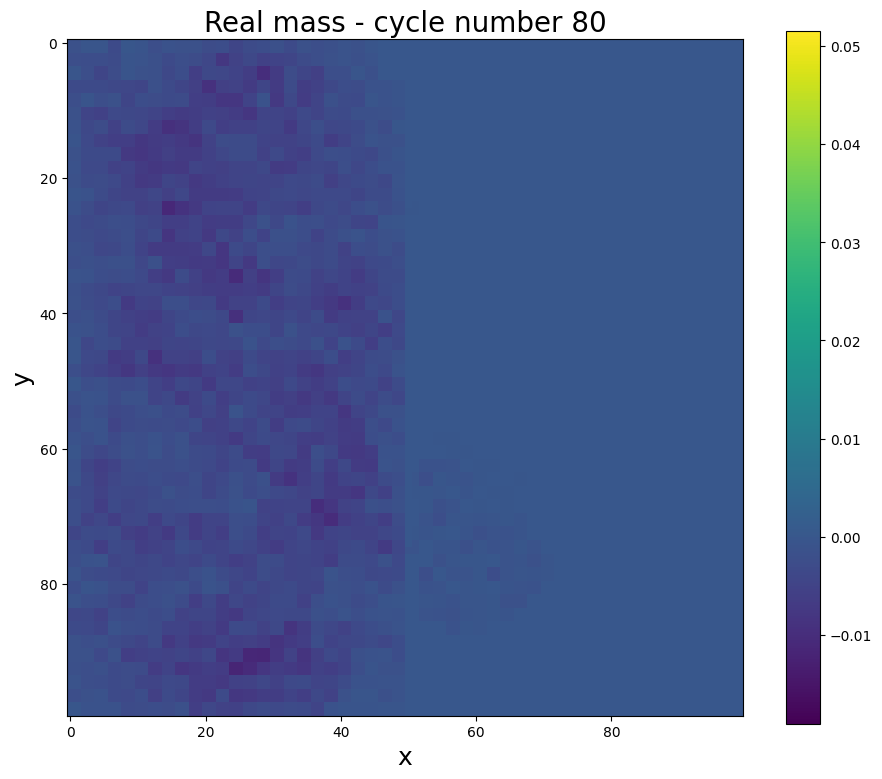} }}
        \subfloat[Real mass at t=130\centering]{{\includegraphics[width=.2\linewidth]{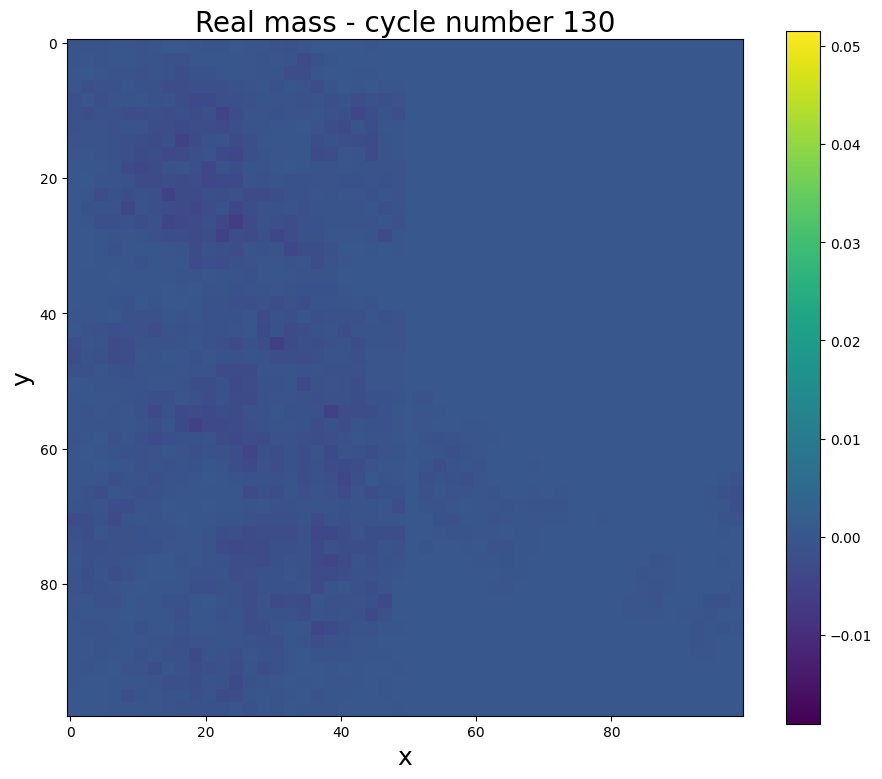} }}
        \qquad
        \subfloat[Imaginary mass at t=30\centering]{{\includegraphics[width=.2\linewidth]{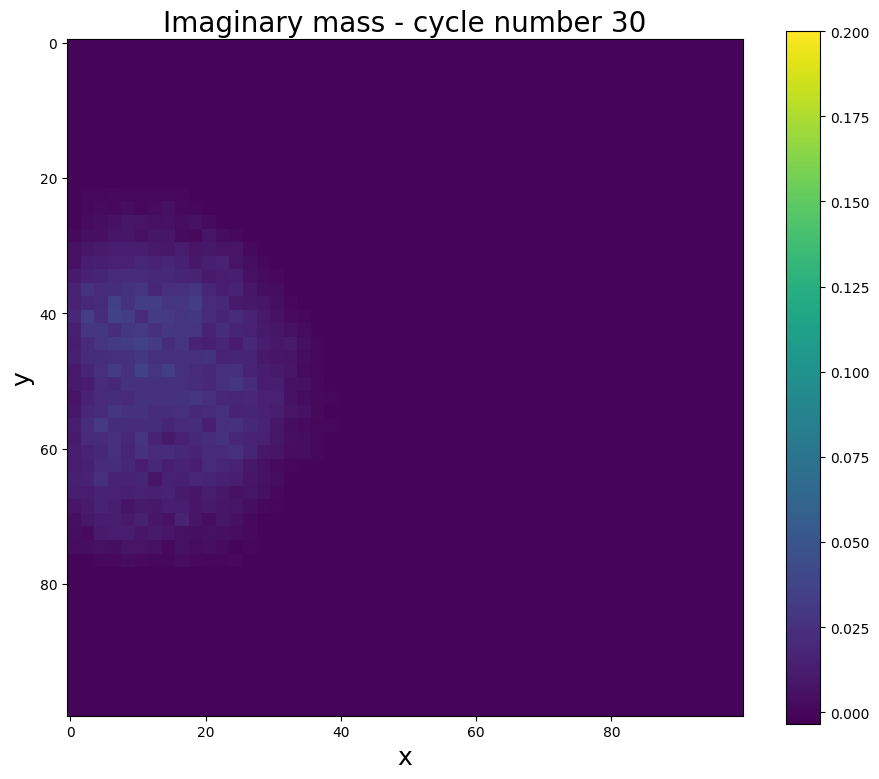} }}
        \subfloat[Imaginary mass at t=80\centering]{{\includegraphics[width=.2\linewidth]{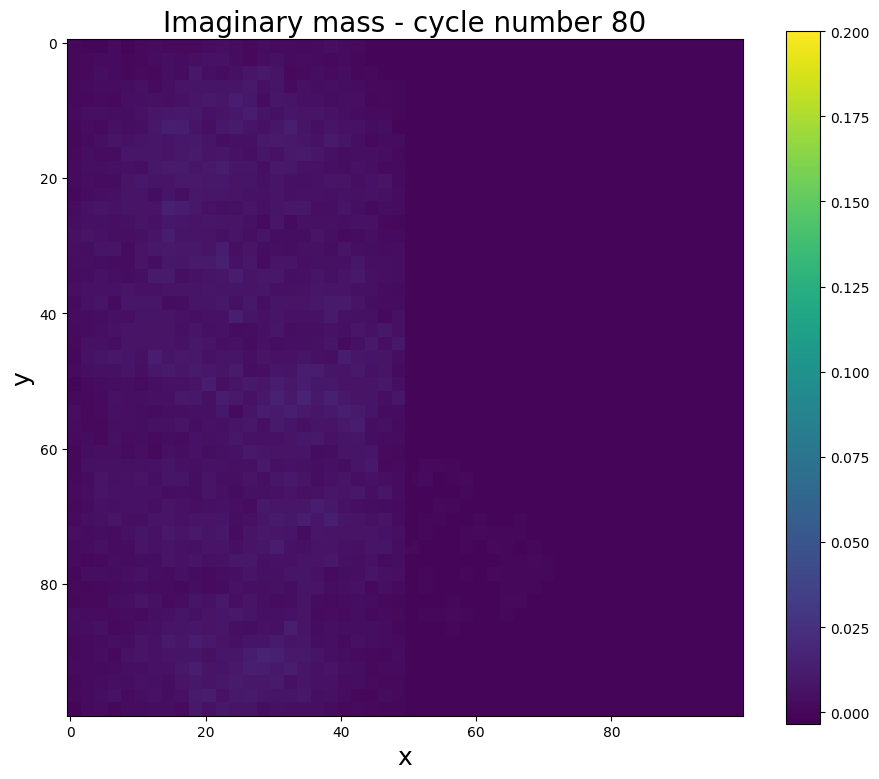} }}
        \subfloat[Imaginary mass at t=130\centering]{{\includegraphics[width=.2\linewidth]{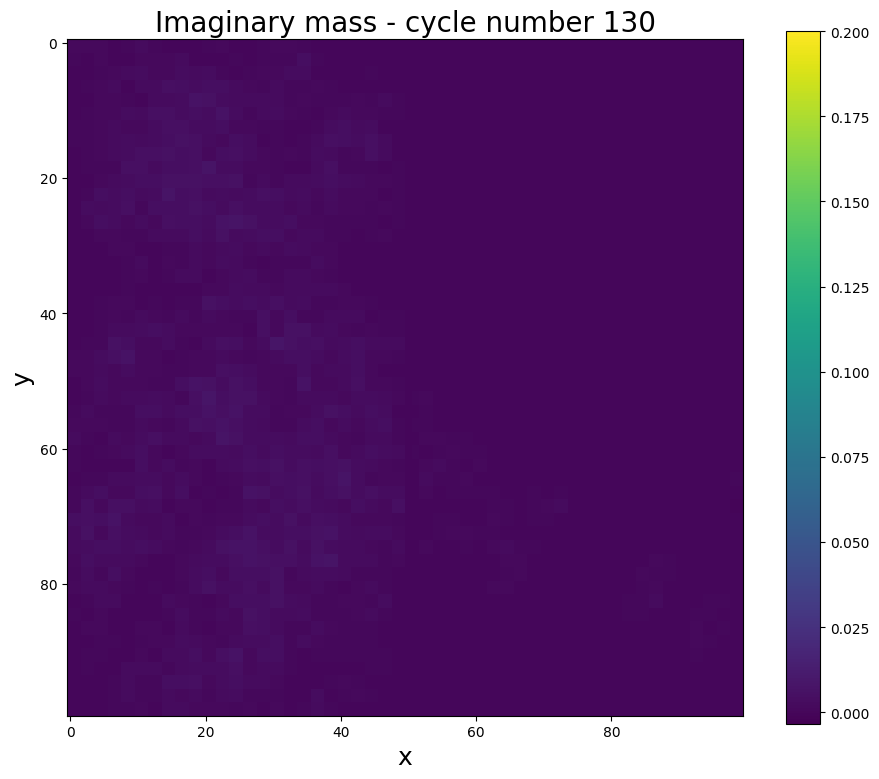} }}
        \qquad
        \subfloat[Phase at t=30\centering]{{\includegraphics[width=.2\linewidth]{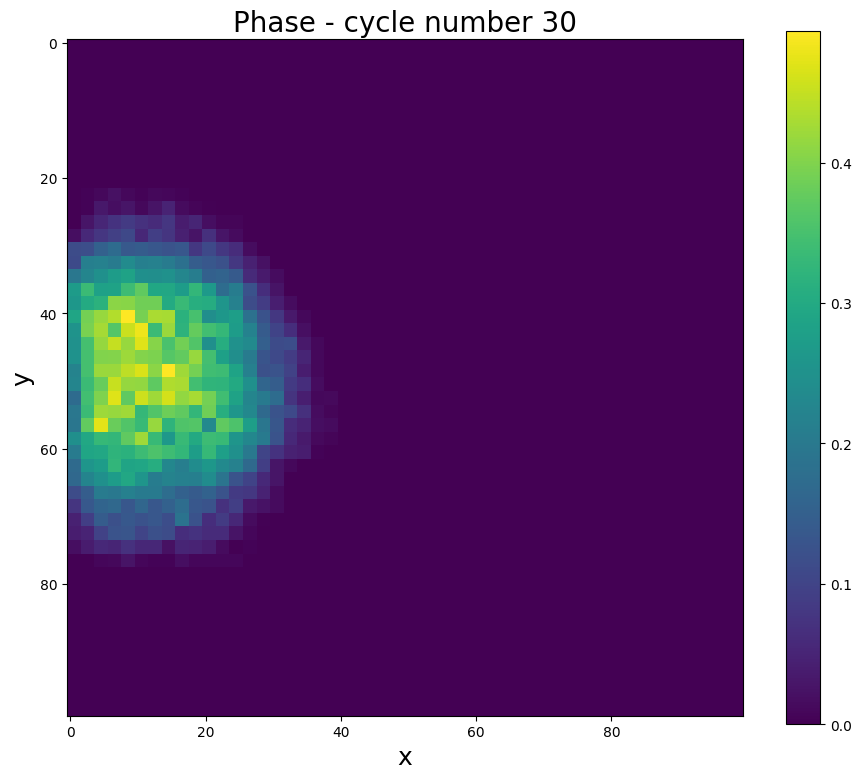} }}
        \subfloat[Phase at t=80\centering]{{\includegraphics[width=.2\linewidth]{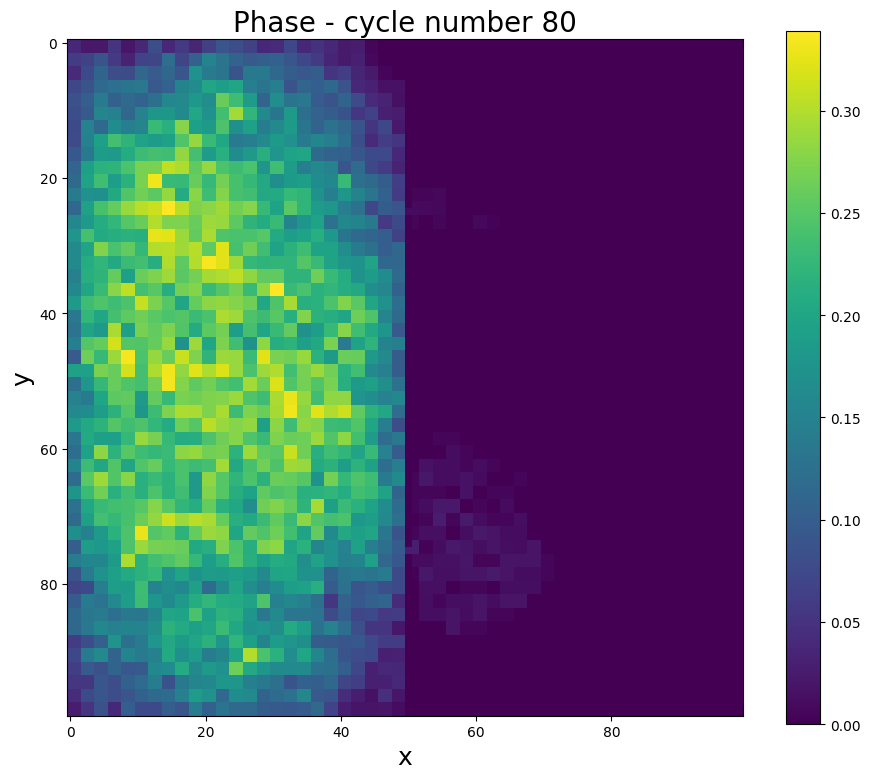} }}
        \subfloat[Phase at t=130\centering]{{\includegraphics[width=.2\linewidth]{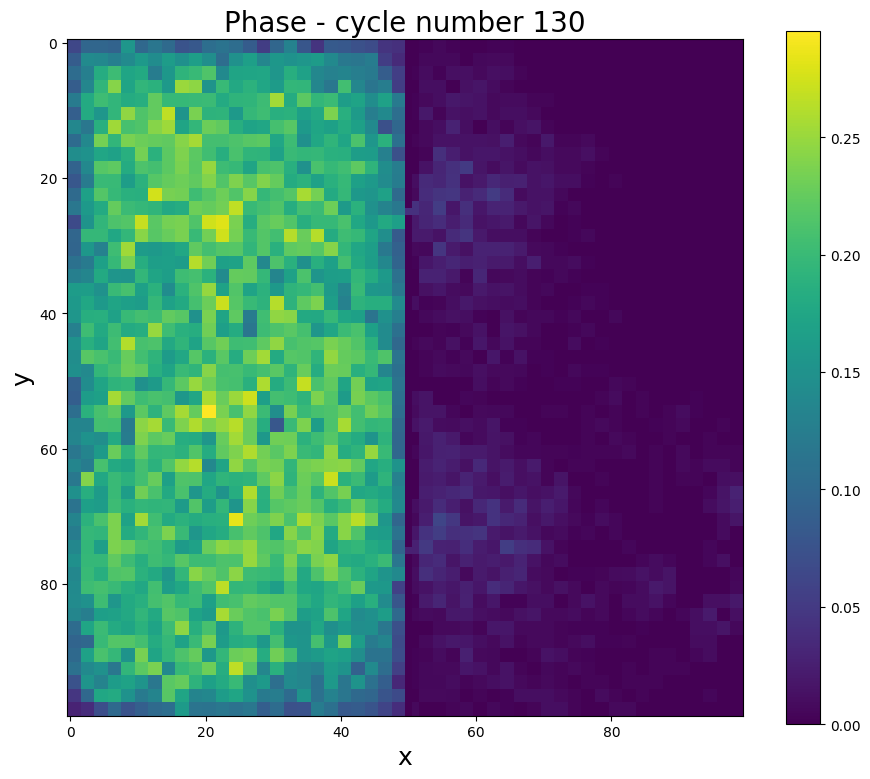} }}
        \qquad
        \subfloat[Pressure at t=30\centering]{{\includegraphics[width=.2\linewidth]{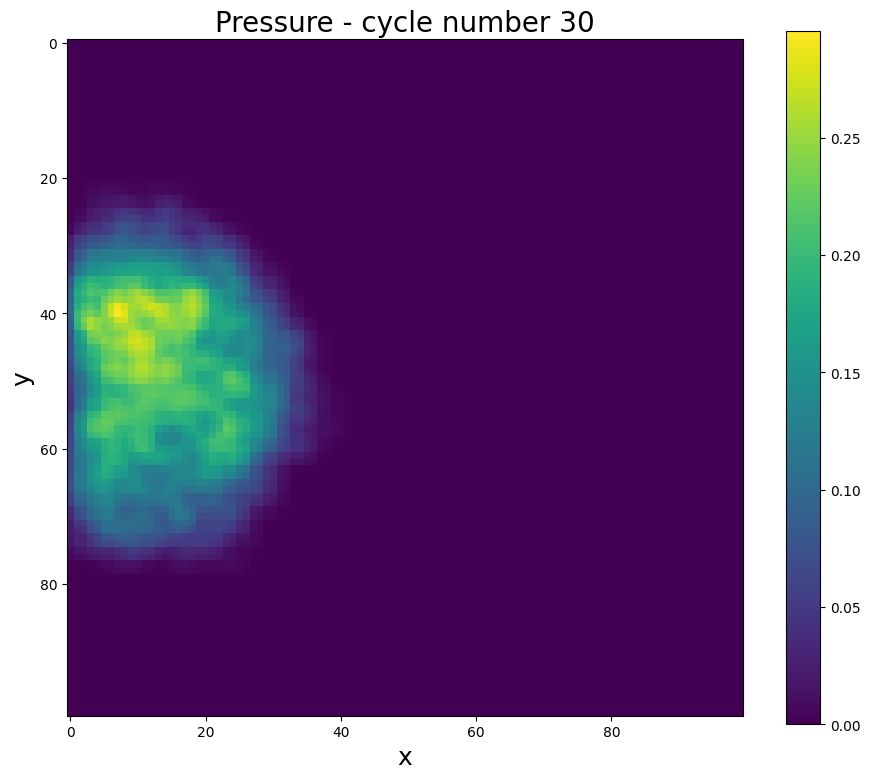} }}
        \subfloat[Pressure at t=80\centering]{{\includegraphics[width=.2\linewidth]{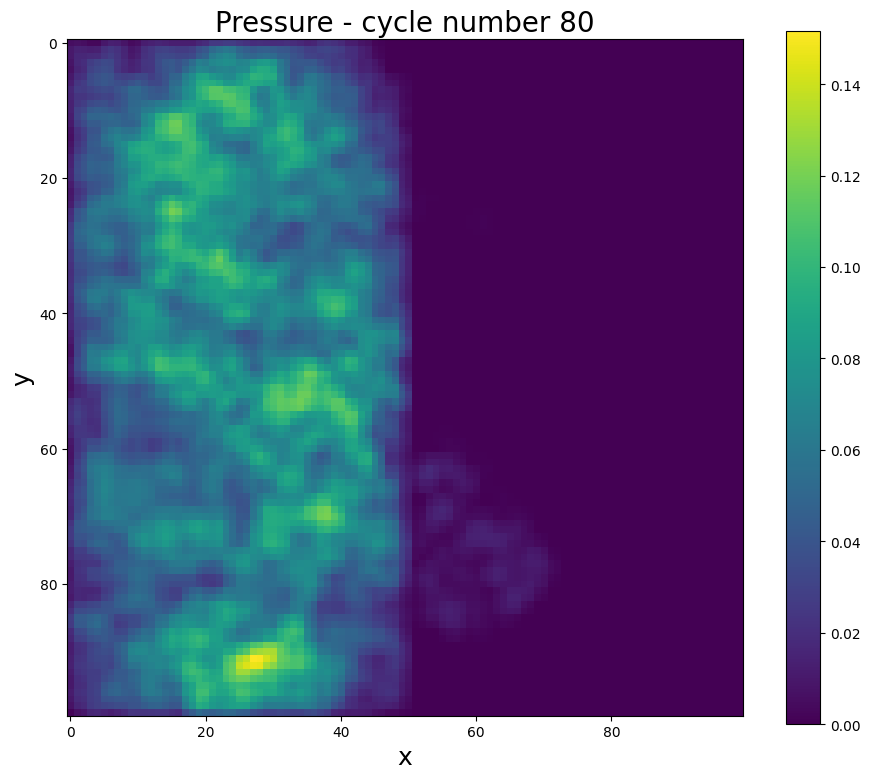} }}
        \subfloat[Pressure at t=130\centering]{{\includegraphics[width=.2\linewidth]{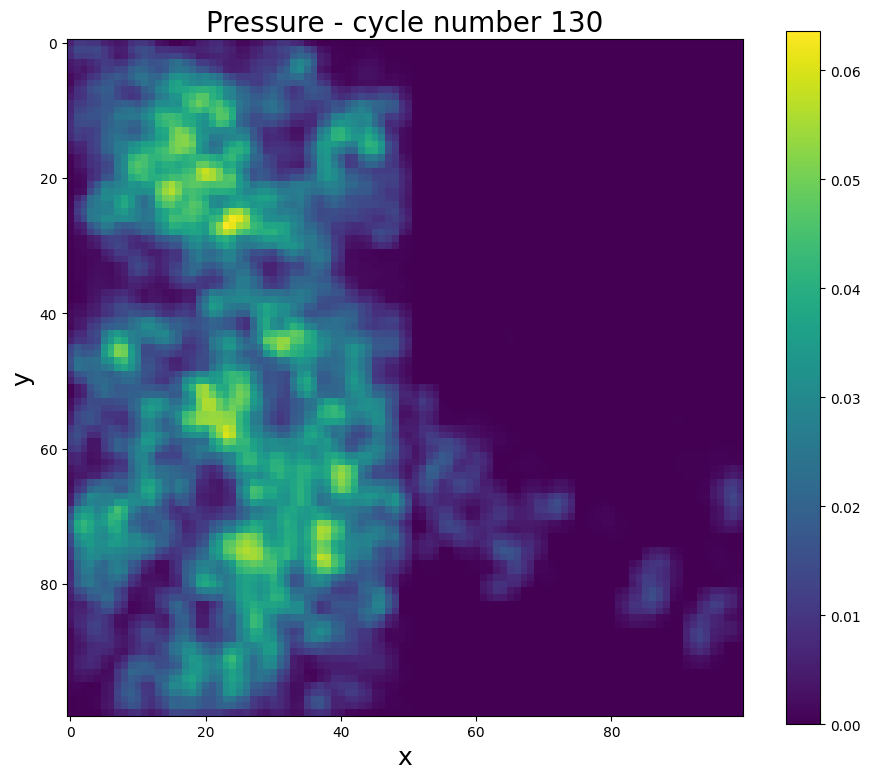} }}
        \caption{Dynamics of thermodynamic parameters with simulation time (mass, phase, pressure) of Stochastic Conway Game of Life in case of a system with one barrier and 2 by 2 square averaging procedure applied every 10 time step.}
        \label{fig:konf14_average_characteristics}
    \end{figure}

    In order to make the Stochastic Conway Game of Life similar to quantum mechanics, and to obtain the phenomenon of interference, we introduce a complex mass given by the formula
    \begin{align}
        \lvert m(x,y,t)\rvert=\sqrt{\mathfrak{R}\left(m(x,y,t)\right)^2+\mathfrak{I}\left(m(x,y,t)\right)^2}.
    \end{align}
    In each time step, the cell mass changes its value as follows
    \begin{align}
        m(x,y,t+1)=\lvert m(x,y,t)\rvert e^{i\phi(x,y,t)}=\lvert m(x,y,t)\rvert\left(cos\left(\phi(x,y,t)\right)+i\sin\left(\phi(x,y,t)\right)\right).
    \end{align}
    The conducted simulations for the complex mass were carried out for a system with a single barrier having two gaps through which cells can pass from one chamber to the other (as depicted in Figure~\ref{fig:konf14}).
    Diffusion process is observed and it has very different dynamics when we have course-graining (averaging procedure for nearest cells) or without that.
    Lack of averaging procedure results in system dynamics of full presence of cells with uniform mass, temperature and entropy distribution and lasting for very long time (tending to infinity).
    However in case of presence of averaging procedure we report that cell distribution is maximally delocalized in non-uniform way and there are pockets of lower and higher density (as depicted in Figure~\ref{fig:konf14_average_characteristics}).
    The obtained mass with separation into real and imaginary parts is shown in Figure~\ref{fig:konf14_mass}a.
    Evolution of mass over time with the distinction of real and imaginary parts in the Stochastic Game of Life with a 2 by 2 square averaging procedure every 10 time steps (right picture).
    Strong anti-correlation of the real and imaginary parts of mass of the cellular automata system is reported, which is quite analogous to the evolution of the wave function over time in the case of the Schr\"{o}dinger equation.
    We can add a modification of averaging the cells in a 2 by 2 square every tenth time step:
    \begin{align}
        m_{av}(k,l,t)=\frac{1}{4}(m(k,l,t)+m(k+1,l,t)+m(k,l+1,t)+m(k+1,l+1,t)
    \end{align}
    with assignment of values ($(m(k,l,t+1)\rightarrow m_{av}(k,l,t),(m(k+1,l,t+1)\rightarrow m_{av}(k,l,t),(m(k,l+1,t+1)\rightarrow m_{av}(k,l,t),(m(k+1,l+1,t+1)\rightarrow m_{av}(k,l,t)$).
    Such averaging of observables in nearest neighborhood is known in statistical physics as coarse-graining.
    Such procedure can be also seen as construction of mean-field theory in the framework of stochastic Conway Game of Life.
    Furthermore, complex value stochastic Conway Game of Life can be seen as attempt to reproduce features of quantum mechanics by cellular automata.
    Figure~\ref{fig:konf14_average_characteristics} shows the time evolution of such thermodynamic parameters as mass, phase and pressure in generalized stochastic Conway Game of Life that is introduced in this work and in accordance to our knowledge it is not yet represented in world literature.
    In particular, we pay attention to the pressure graph, which looks like a fuzzy wave function.
    \FloatBarrier

    \section{Thermodynamical cycles of Stochastic Conway Game of Life}
    \label{sec:thermodynamical-cycles-of-stochastic-conway-game-of-life}
\begin{figure}
        \centering
        \subfloat[\centering]{{\includegraphics[width=.3 \linewidth]{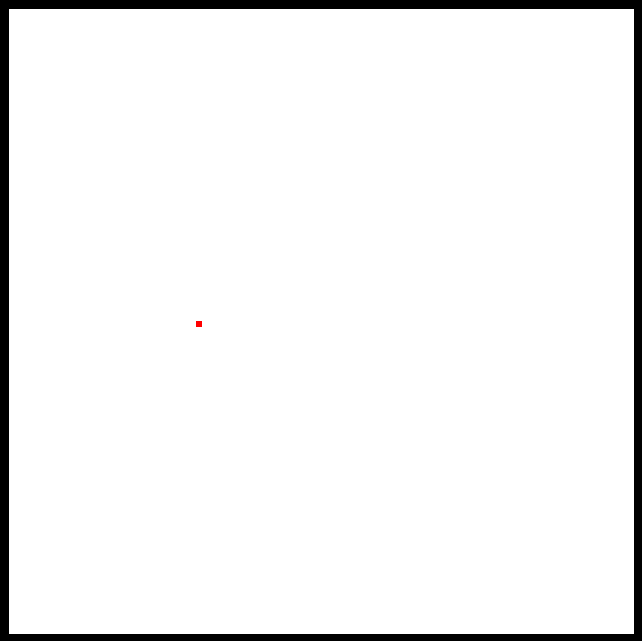} }}
        \qquad\qquad
        \subfloat[\centering]{{\includegraphics[width=.3 \linewidth]{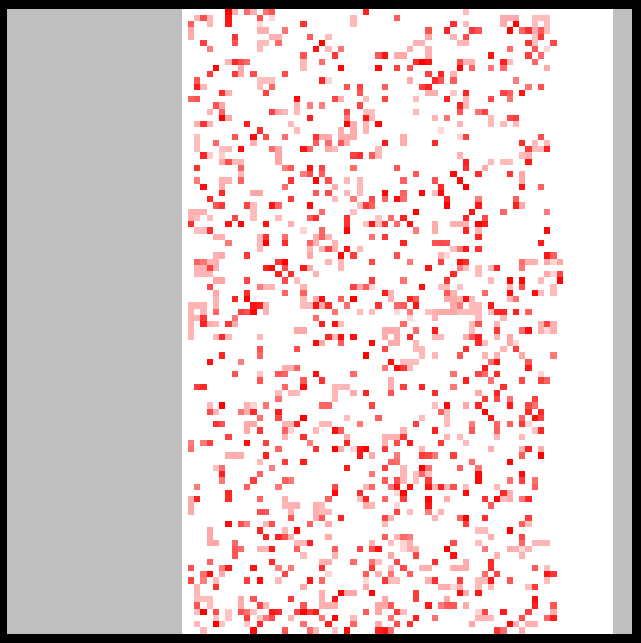} }}
        \qquad
        \subfloat[\centering]{{\includegraphics[height=.3 \linewidth]{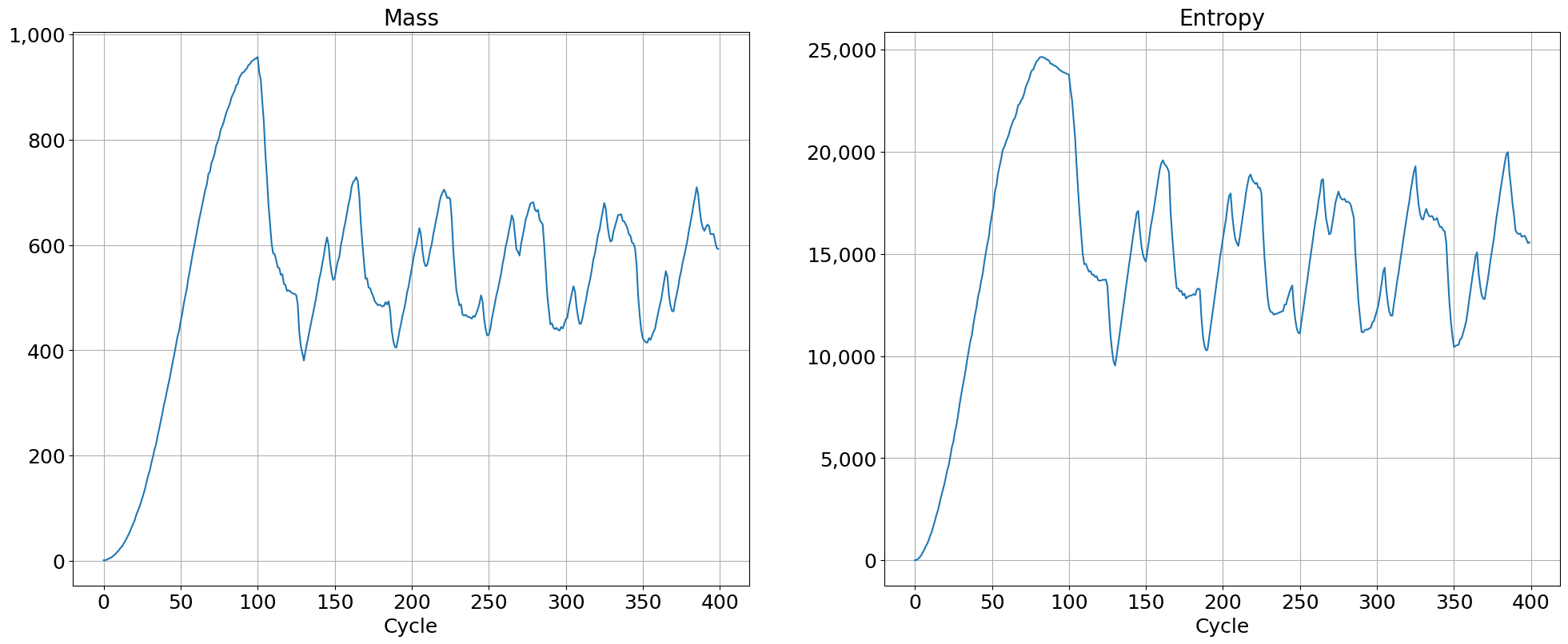} }}
        \caption{Dependence of mass and entropy in stochastic Conway game of Life population subjected to barriers moving in cyclic way. Quasi-periodicity of mass and entropy is observed. }
        \label{fig:moving_barriers}
    \end{figure}

\begin{figure}
        \centering
        \subfloat[Mass at t=60\centering]{{\includegraphics[width=.3\linewidth]{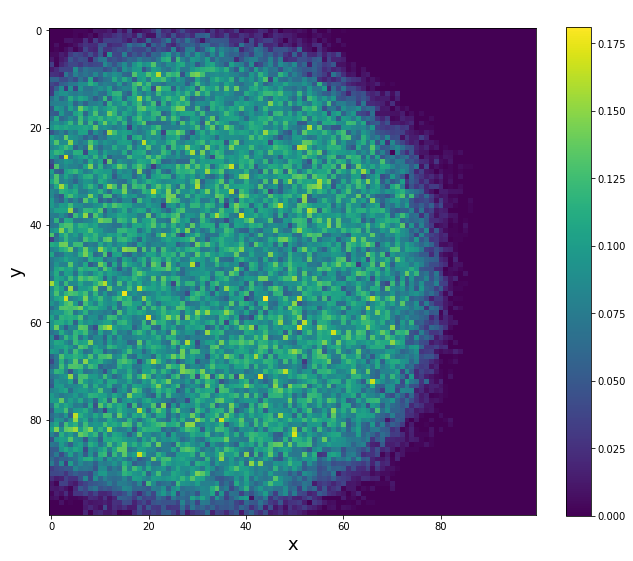} }}
        \subfloat[Mass at t=150\centering]{{\includegraphics[width=.3\linewidth]{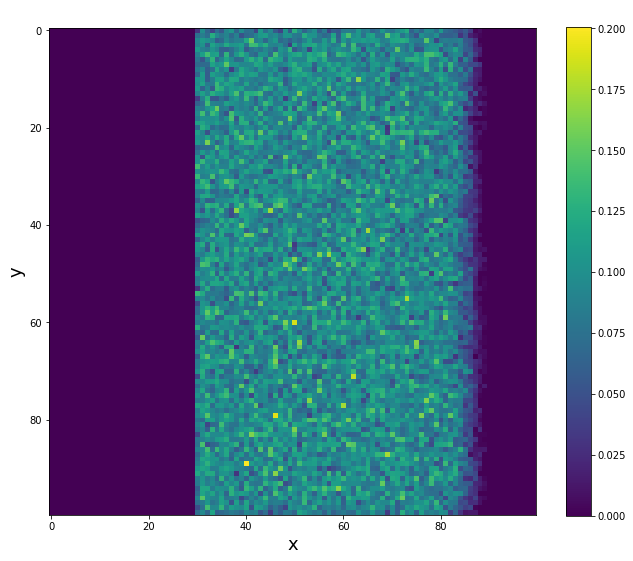} }}
        \subfloat[Mass at t=305\centering]{{\includegraphics[width=.3\linewidth]{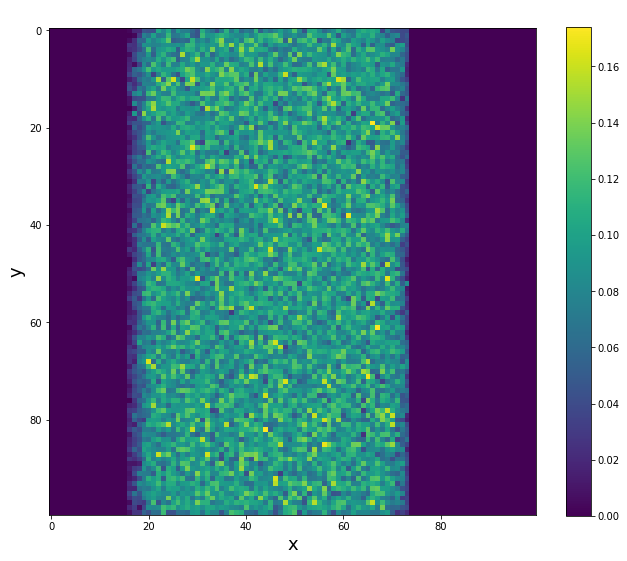} }}
        \qquad
        \subfloat[Entropy at t=60\centering]{{\includegraphics[width=.3\linewidth]{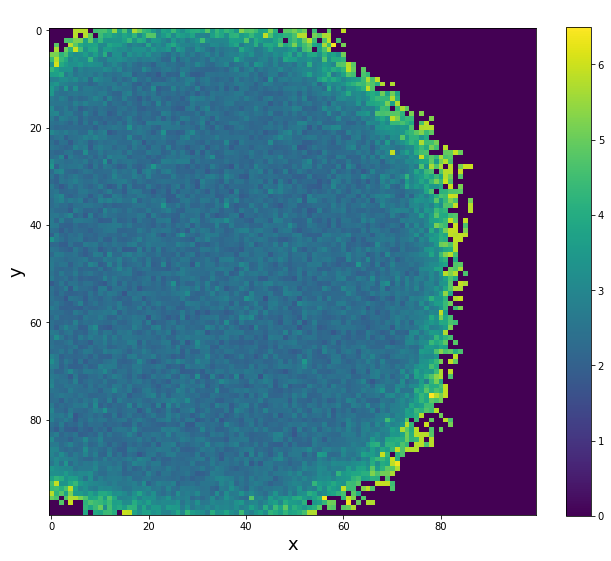} }}
        \subfloat[Entropy at t=150\centering]{{\includegraphics[width=.3\linewidth]{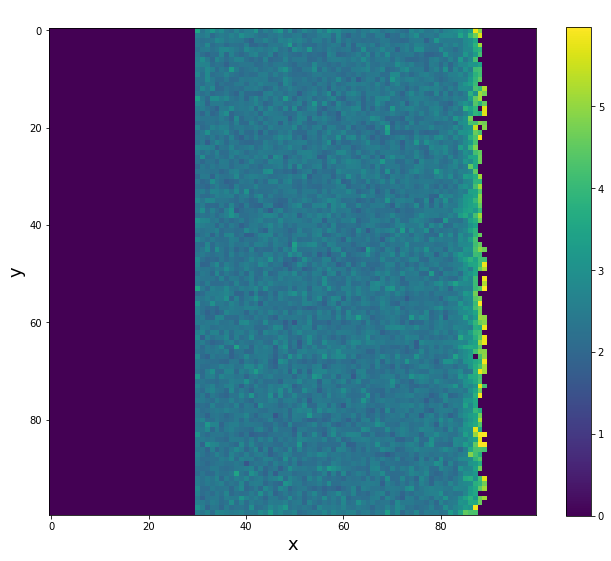} }}
        \subfloat[Entropy at t=305\centering]{{\includegraphics[width=.3\linewidth]{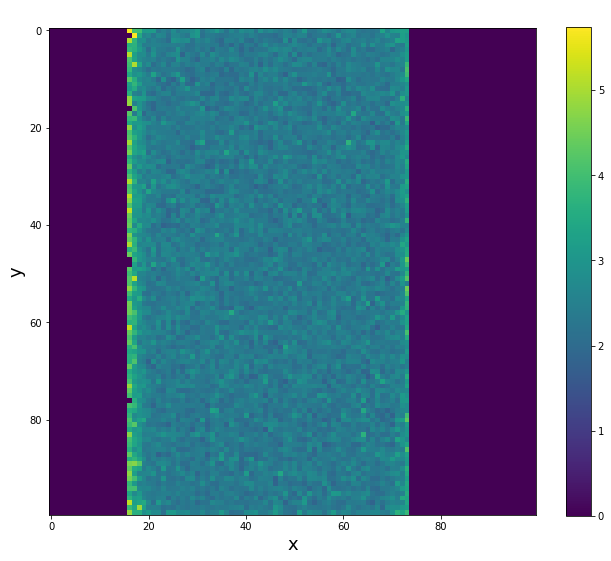} }}
        \qquad
        \subfloat[T(x,y,t=60)\centering]{{\includegraphics[width=.3 \linewidth]{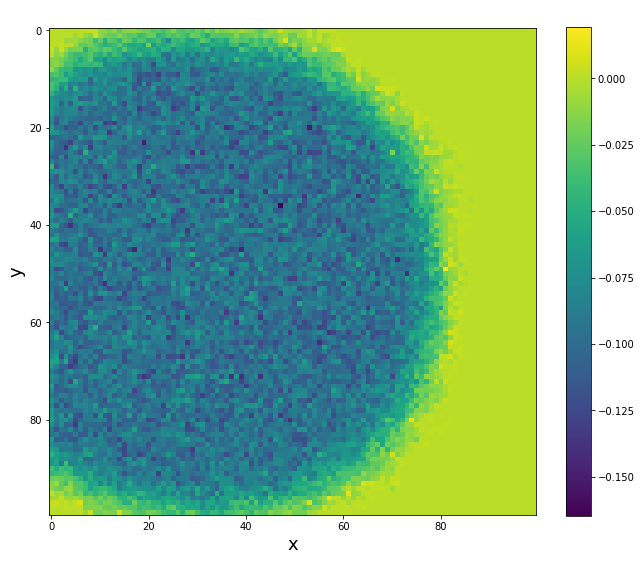} }}
        \subfloat[T(x,y,t=150)\centering]{{\includegraphics[width=.3 \linewidth]{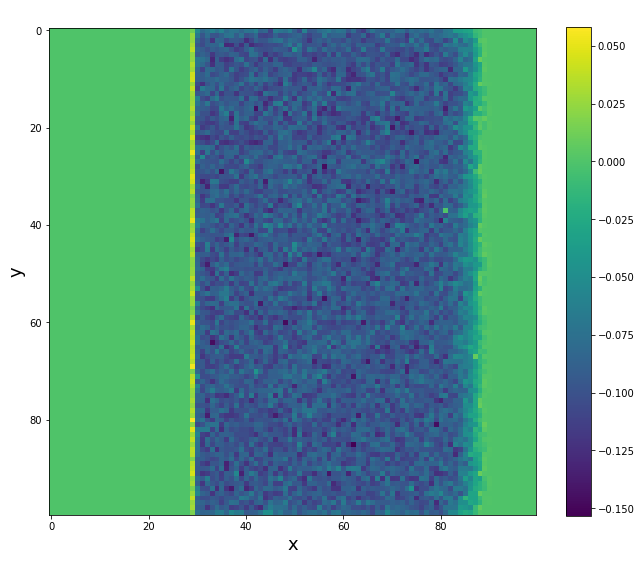} }}
        \subfloat[T(x,y,t=305)\centering]{{\includegraphics[width=.3 \linewidth]{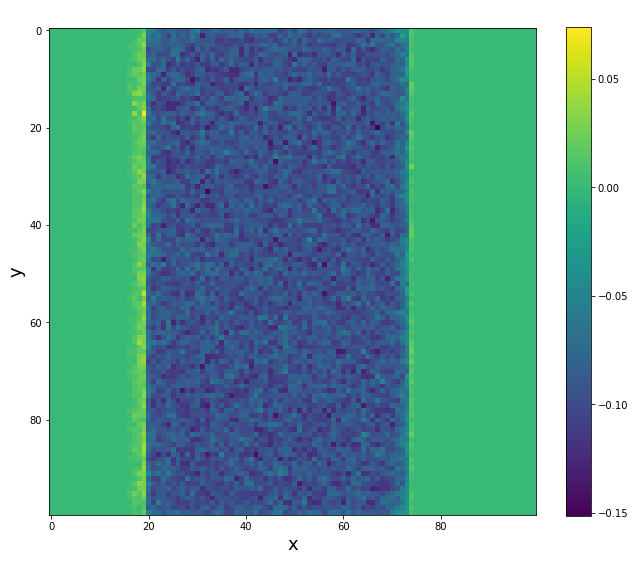} }}
        \caption{Snapshots of mass, entropy and temperature dependence over space at different simulation time in case of barriers moving in cyclic way as from Fig.~\ref{fig:moving_barriers}.}
        \label{fig:characteristics_moving_barriers}
    \end{figure}

\begin{figure}
        \centering
        \subfloat[$\frac{dm}{dt}$ with time\centering]{{\includegraphics[height=.2 \linewidth]{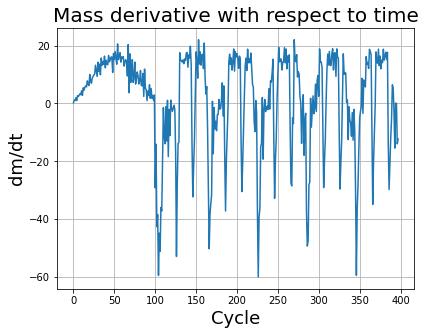} }}
        \subfloat[$\frac{dS}{dt}$ with time\centering]{{\includegraphics[height=.2 \linewidth]{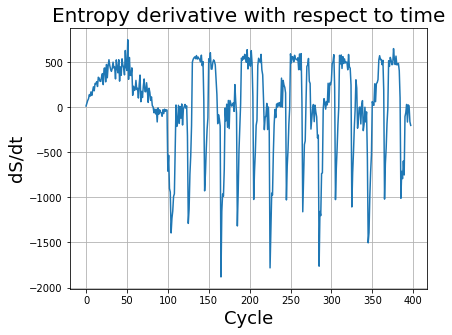} }}
        \subfloat[Temperature with time\centering]{{\includegraphics[height=.2 \linewidth]{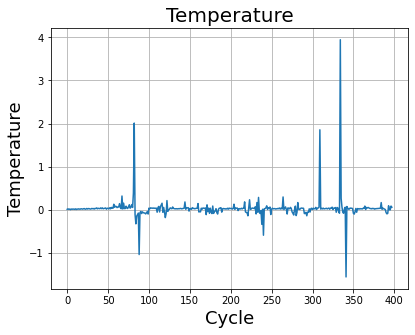} }}
        \caption{Snapshots of mass derivative with respect to simulation time, entropy derivative with respect to time and temperature dependence at different simulation time in case of barriers
        moving in cyclic way as from Fig. \ref{fig:moving_barriers}.}
        \label{fig:temperature_moving_barriers}
    \end{figure}

    There is common intuition that cellular automata in Stochastic or in deterministic Conway Game of Life behaves to certain degree as particles pumped into chamber from external unlimited source that could be regarded as a kind of diffusion process.
    Indeed we encounter diffusion process in case of cellular automata mass, entropy and temperature as given by Figures \ref{fig:characteristics_moving_barriers}, \ref{fig:temperature_moving_barriers}.
    Despite existing differences classical statistical physics and stochastic Conway Game of Life the quasi-periodic cycle of mass, entropy and temperature are reported as expressed by Fig. \ref{fig:characteristics_moving_barriers} what experimentally validates thermodynamics methodology in Conway Game of Life.
    Diffusion takes place in system with barriers and small holes or in empty space.
    In very real way we can consider some analogies of Game of Life with analogy to Grand Canonical Ensemble in classical thermodynamics.
    Cellular automata tend to propagate until certain density distribution is achieved and there is limiting distribution that cannot be crossed.
    We can even have the pressure concept if we start to compute the number of nearest neighbors for giving position of cellular automata.
    Due to existing analogies we can consider the gas of cellular automata (corresponding to particles) subjected to cyclic volume squeezing and expansion as it is depicted in Fig.~\ref{fig:moving_barriers}.
    It should be underlined that we have already introduced new concept of pressure in Conway Game of Life that is defined of numbers of nearest neighbors (or more precisely mass of them) given by formula
    \begin{align}
        press_{Conway}(k,l,t_{s})=m(k-1,l-1,t_s)+m(k,l-1,t_s)+m(k+1,l-1,t_s) \nonumber
    \end{align}
    \begin{align}
        +m(k-1,l,t_s)+m(k+1,l,t_s)+m(k-1,l+1,t_s)+m(k,l+1,t_s)+m(k+1,l+1,t_s).
    \end{align}
    In most cases pressure in Conway Game of Life behave the as in accordance of intuition in classical statistical physics.
    However anomalous behavior in pressure of Conway Game of Life was reported in two-dimensional complex value Conway Game of Life with coarse-graining (averaging on squares 2 by 2 every tenth time step) as depicted by two-dimensional pressure distribution in subsequent time steps presented in Figure \ref{fig:konf14_average_characteristics}(m)-(o).
    Final pressure distribution was not homogeneous pointing possible localization of Conway Game of automata in certain space areas.
    Situation of wavefunction localization is known from condensed matter physics as an example of crystals with local defects.
    In such case wave function is localized around defect.
    In conducted simulations we did not break space symmetry but still we observe localization phenomena.
    \FloatBarrier

    \section{Tight-binding model in description of Conway Game of Life and Hardware Reference}
    \label{sec:tight-binding-model-in-description-of-conway-game-of-life}
\begin{figure}[hbt!]
        \centering
        \includegraphics[scale=0.7]{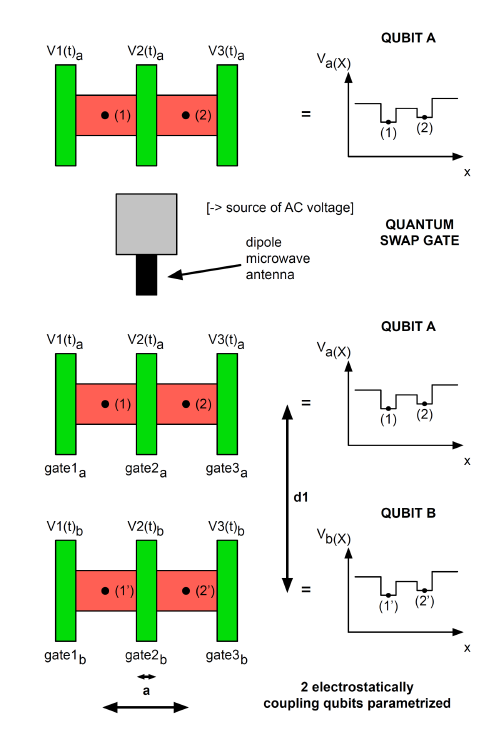}\includegraphics[scale=0.5]{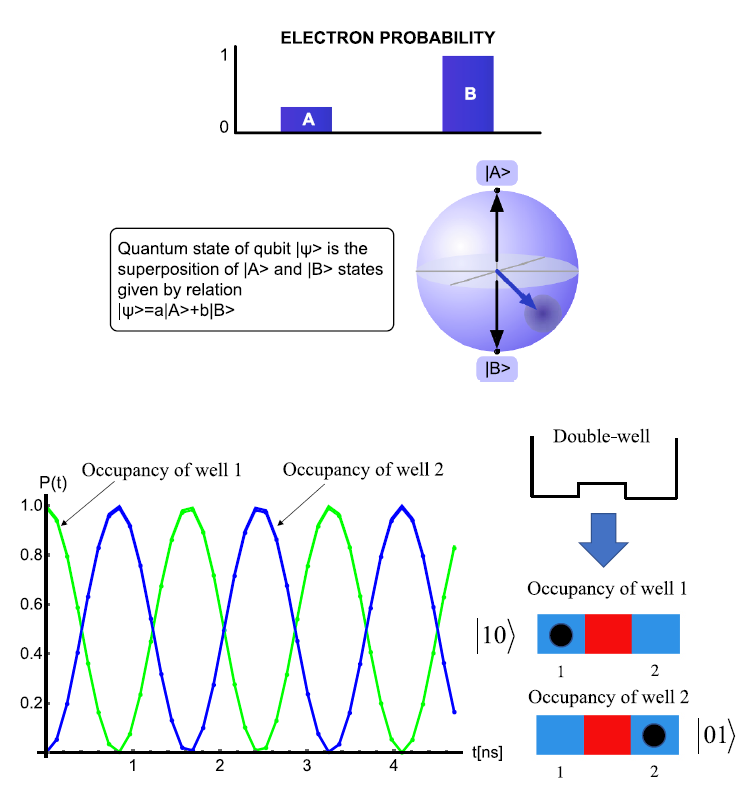}
        \caption{Position-based qubit implemented in the chain of coupled quantum dots as given by~\cite{Cryogenics} and by~\cite{Panos}. }
        \label{fig:WannierQ}
    \end{figure}

\begin{figure}
        \centering
        \includegraphics[scale=0.5]{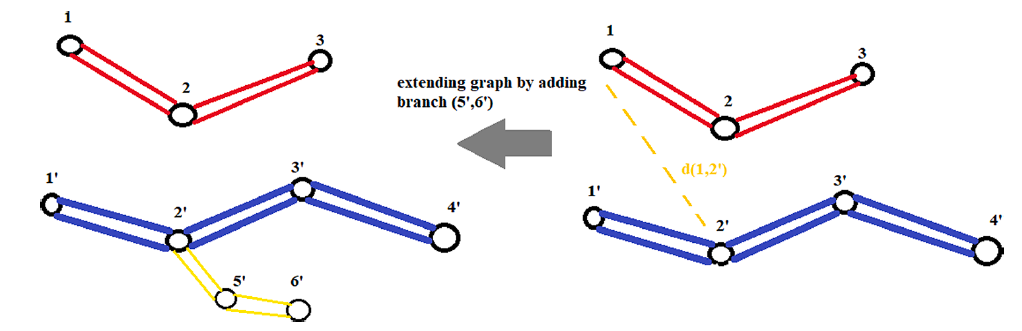}
        \caption{Electrostatically controlled graph of coupled quantum dots as by~\cite{Cryogenics} that can be also obtained from 2 dimensional model. }
        \label{fig:QGraph}
        \centering
        \includegraphics[scale=0.9]{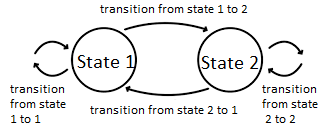}\includegraphics[scale=0.9]{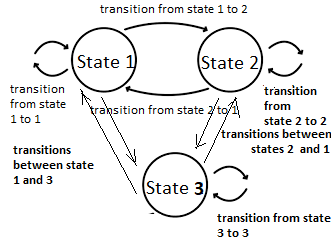}
        \caption{Illustration of epidemic model referring to stochastic finite state machine being 2 level system with 2 distinguished states 1 and 2. 4 possible transitions are characterized by 4 time-dependent coefficients $s_{1 \rightarrow 1}(t)=s_{11}(t)$ ,$s_{1 \rightarrow 2}(t)=s_{12}(t)$, $s_{2 \rightarrow 1}(t)=s_{21}(t)$, $s_{2 \rightarrow 2}(t)=s_{22}(t)$ (Left picture). Using induction reasoning we can extend 2 dimensional stochastic finite state machine to 3 state machine and N state machine (Right picture).}
        \label{fig:FiniteStates}
    \end{figure}

    Currently single-electron devices are becoming the more and more dominant trend in implementation of quantum technologies as given by Likharev~\cite{Likharev}, Fujisawa~\cite{Fujisawa}, Petta~\cite{Petta}, Leipold~\cite{Dirk}.
    The theory of operation on single-electron devices was developed in framework of tight-binding and Schr\"{o}dinger model by Pomorski~\cite{Spie,Cryogenics}, Giounanlis~\cite{Panos} and many others.
    Essentially one electron is injected into one among N coupled quantum dots and has oscillations of occupancy.
    Structures with such physical phenomena are shown by Fig.~\ref{fig:WannierQ} and in Fig.~\ref{fig:QGraph}.
    Probabilistic nature of this process implies hypothesis that electron occupancy of certain regions can be described by stochastic finite state machine.
    Indeed we can deal with reconfigurable quantum matter as pointed in Fig.~\ref{fig:QGraph}, where we can set quantum dot connectivity in electrical way.
    Quantum matter has features of superposition of many states at the same time, entanglement and is subjected to the strong or weak interaction during strong and weak measurement.
    Furthermore quantum matter is very sensitive towards external noise and decoherence processes.
    Quantum electrostatic entanglement emerges when we are dealing with 2 single-electron lines as pointed by~\cite{2SEL}.


    To create a suitable Hamiltonian, we must realize that the \textbf{GoL} has creationism - there is no mass conservation.
    We need to introduce the appropriate operator:\\
    $\ket{x+1}\bra{x}$ - creation of particle presence at $x+1$ point and annihilation of particle presence at x point.
    Mapping from $(k,l)$ to $(k+1,l+1)$ can be given by operator $\ket{k+1,l+1}\bra{k,l}$.
    Tight-binding Hamiltonian~\cite{cite23} of position based qubit~\cite{cite24}:
    \begin{align}
        \hat{H}=E_L\ket{L}\bra{L}+E_R\ket{R}\bra{R}+t_s(L\rightarrow R)\ket{R}\bra{L}+t_s(R\rightarrow L)\ket{L}\bra{R}
    \end{align}
    where $E_L$, $E_R$ are localized energies at $L$, $R$ nodes and $t_s(R\rightarrow L)$ describes hopping from $R$ to $L$ node, etc.
    In the equation of motion for a ``quantum state'' of \textbf{GoL} there is complex value time:
    \begin{align}
        i\hbar\frac{d}{dt}\ket{\Psi}_t=\hat{H}\ket{\Psi}_t
    \end{align}
    \begin{align}
        \hat{H}=\sum_{k=-\infty}^{k=+\infty}\sum_{l=-\infty}^{l=+\infty}\sum_{m=-\infty}^{m=+\infty}\sum_{n=-\infty}^{n=+\infty}\ket{k,l}\bra{m,n}\cdot f(k,l,m,n)
        \label{eqn:tightbinding}
    \end{align}
    where $i$ is imaginary unit, $\hbar$ reduced Planck's constant, $\hat{H}$ Hamiltonian operator.
    The Hamiltonian of the system consists of the creation and annihilation operators, which gives us three operations: killing the quantum state by underpopulation, preserving the state, and killing the quantum state by the overpopulation.
    This is achieved by using the sum of the hyperbolic tangent functions, which can be shown as
    \begin{figure}[ht]
        \centering
        \includegraphics[width=.5 \linewidth]{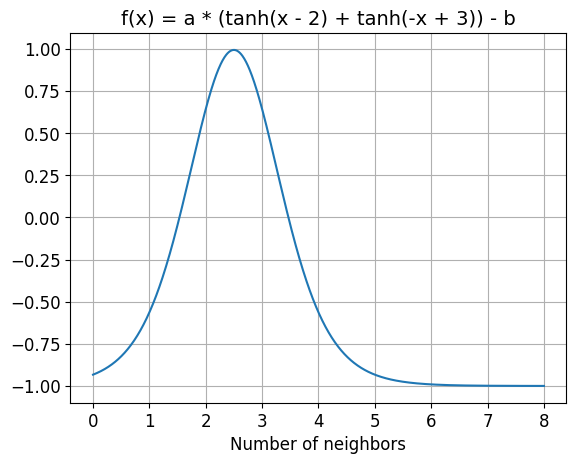}
        \caption{State function f dependence (with parameters $a$ set to 2.16 and $b$ set to 1.001) on the number of neighbors occurring in tight-binding Hamiltonian Conway Game of Life given by formula~\ref{eqn:tightbinding}.}
        \label{fig:tangenth}
    \end{figure}
    Using the above, we introduce dissipation in Hamiltonian \textbf{GoL} (responsible for cell appearance or disappearance) in artificial way:
    \begin{itemize}
        \item Killing of quantum state by underpopulation corresponds to non-Hermitian values of Hamiltonian
        \begin{align}
            \hat{H}(k,l)=\ket{k,l}\bra{k,l}(-i)[\tanh{(1-\lvert p_1\rvert+\cdots+\lvert p_8\rvert)}+\tanh{(\lvert p_1\rvert+\cdots+\lvert p_8\rvert)}]
        \end{align}
        \item Preservation of state corresponds to Hermitian values of Hamiltonian
        \begin{align}
            \hat{H}(k,l)=-\ket{k,l}\bra{k,l}\lambda\left[\tanh{(-3+\lvert p_1\rvert+\cdots+\lvert p_8\rvert)}+\tanh{(2-\lvert p_1\rvert-\cdots-\lvert p_8\rvert)}\right]
        \end{align}
        \item Dying from overpopulation corresponds to non-Hermitian values of polynomial
        \begin{align}
            \hat{H}(k,l)=-\ket{k,l}\bra{k,l}(-1)\left[e^{-(\lvert p_1\rvert+\cdots+\lvert p_8\rvert-3)}\right]
        \end{align}
    \end{itemize}

    \section{Summary of obtained results}
    \label{sec:summary-of-obtained-results}
    Creationism or annihilationism of cellular automata can be seen from physical perspective as reflected in the exchange of system energy and mass with the environment in a time dependent manner. Saying time-dependent complex value eigenenergies are similar to complex values of resonant frequencies in electromagnetic resonant cavities reflecting presence of dissipation that reflects energy leaving the system or being pumped into the system. Results allow for quantum chip design implementing the quantum Conway Game of Life. The concept of complex-valued mass was introduced into the Stochastic Classical Conway Game of Life as a tool mimicking behavior of certain quantum system behavior. The pointed research results show the possibilities of a conceptual research framework for classical systems governed by classical physics mimicking by its behavior the quantum systems. Thus it has the importance in development of programmable quantum matter.
    \newline
    
    We have presented an introduction to the Stochastic Conway Game of Life in one and in two dimensions (with very straightforward generalization to N dimensional case) confirming validity of usage of classical statistical physics methodology as temperature, entropy, energy. The complex value Stochastic Conway Game of Life was introduced in one and in two dimensions. The concept of phase associated with automata continuous mass has a deep impact on one dimensional dynamics of complex value Game of Life vs real value Game of Life. Coarse-graining applied to two dimensional complex Conway Game of Life has generated a non-uniform probability map that has a tendency to decay, while the same complex value two dimensional Game with no coarse graining generated uniform probability map at the very end. Effective tool used in description of dynamics one dimensional complex value Game of Live is by means of complex Schr\"{o}dinger potential. Therefore, the dissipative Schr\"{o}dinger equation is a good way to mimic stochastic complex value Conway Game of Life. Creationism and annihilationism is against preservation of normalization of a quantum state, what can be recognized as dissipative quantum mechanics. Due to the nature of originally Conway Game of Life it is instructive to consider dissipative tight-binding models that have special importance in description of single-electron devices. Hermitian features of this model will preserve the occupancy of quantum state, while non-Hermitian features will be responsible for state creation or annihilation. It is instructive to notice that dissipative quantum mechanics systems can mimic most classical systems, while very few classical statistical systems can mimic or partially mimic dissipative quantum mechanics systems.

    \section{Acknowledgment}
    We thank professor Adam Bednorz from University of Warsaw for helpful comments.
    All concepts presented in this work were introduced by Krzysztof Pomorski as complex value Conway Game of Life, Conway tight-binding model, parametrization of Conway Game of Life by Schr\"{o}inger complex value potential.
    All numerical computation was conducted by Dariusz Kotula and he has assisted in analytical computations as well. Krzysztof Pomorski and Dariusz Kotula contribution to this work is 60 percent vs 40 percent.
    Authors declare no conflict of interests.


\begin{thebibliography}{9}
\bibitem{cite1} M.Gardner, \textit{Mathematical Games},Scientific American, Vol.223, Nr.4, 1970.
\bibitem{cite4} S.Wolfram, \textit{Statistical mechanics of cellular automata},Review of Modern Physics, Vol.55, Issue.3, 1983.
\bibitem{cite5} H.Peitgen, H.Jurgens, D.Saupe, \textit{Chaos and Fractals: New Frontiers of Science: Pascal's Triangle: Cellular Automata and Attractors},Springer New York, 2004.
\bibitem{cite15} F.Berto, J.Tagliabue, \textit{Cellular Automata}, The Stanford Encyclopedia of Philosophy, 2012
\bibitem{cite16} S.Prasanta Bandyopadhyay, N.Grunska, D.Dcruz, M.C.Greenwood, Mark , \textit{Are Scientific Models of Life Testable? A Lesson from Simpson's Paradox}, Vol.3,Nr.1, Science, 2021
\bibitem{cite17} G.Aguilera, G.G.Jose, et al. , \textit{A probabilistic extension to Conway's Game of Life}, Vol.45,  Advances in Computational Mathematics, 2019
\bibitem{cite18} S.Vandevelde, J.Vennekens , \textit{Problife: a Probabilistic Game of Life},  Arxiv:2201.09521, 2022
\bibitem{cite23} K.Pomorski,\textit{Equivalence between finite state stochastic machine, non-dissipative and dissipative tight-binding and Schr\"{o}dinger model}, Vol.209, Mathematics and Computers in Simulation, 2023
\bibitem{cite24} K.Pomorski,\textit{Analytic view on N body interaction in electrostatic quantum gates and decoherence effects in tight-binding model}, Vol.19, Nr.4, International Journal of Quantum Information, 2021
\bibitem{Likharev} K.K.Likharev, \textit{Single-electron devices and their applications}, Vol.87, Nr.4, Proceedings of the IEEE, 1999   
\bibitem{Fujisawa} T.Fujisawa, T.Hayashi, H.D.Cheong, Y.H.Jeong, Y.Hirayama,  \textit{Rotation and phase-shift operations for a charge qubit in a double quantum dot}, Vol.21, Nr.2, Physica E: Low-dimensional Systems and Nanostructures, Proceedings of the Eleventh International Conference on Modulated Semiconductor Structures,  2004
\bibitem{Petta} K.D.Petersson, J.R.Petta, ,H.Lu A.C.Gossard, \textit{Quantum Coherence in a One-Electron Semiconductor Charge Qubit},
  Vol.105, Issue. 24, Physical Review Letters, 2010
\bibitem{Dirk} D.Leipold, \textit{Controlled Rabi Oscillations as foundation for entangled quantum aperture logic},  Seminar at UC Berkley Quantum Labs, 2018 
\bibitem{2SEL} K.Pomorski et al. , \textit{Analytic view on Coupled Single-Electron Lines}, Vol.34, Semiconductor Science and Technology, 2019
\bibitem{Cryogenics}, K.Pomorski, P.Peczkowski, R.B.Staszewski, \textit{Analytical solutions for N interacting electron system confined in graph of coupled electrostatic semiconductor and superconducting quantum dots in tight-binding model}, Vol.109, Cryogenics, 2020
\bibitem{Panos} P.Giounanlis, E.Blokhina, K.Pomorski, D.Leipold, R.B.Staszewski, \textit{Modeling of Semiconductor Electrostatic Qubits Realized Through Coupled Quantum Dots}
    Vol.7, IEEE Access, 2019
\bibitem{Spie} K.Pomorski et al. , \textit{From two types of electrostatic position-dependent semiconductor qubits to quantum universal gates and hybrid semiconductor-superconducting quantum computer}, Vol.11054, Spie, 2019 
\bibitem{MDPIKotula} K.Pomorski, D.Kotula,\textit{Thermodynamics in Stochastic Conway's Game of Life}, Vol.8, Nr.47, Condensed Matter, MDPI, 2023
\bibitem{London} F.London, H.London, \textit{The Electromagnetic Equations of the Supraconductor},Proceedings of the Royal Society A: Mathematical, Physical and Engineering Sciences,Vol.149,Nr.866, 1935.
\end{thebibliography}
\end{document}